\documentclass[journal]{IEEEtran}
\usepackage{amsmath,amsfonts}
\usepackage{algorithmic}
\usepackage{algorithm}
\usepackage[caption=false,font=footnotesize,labelfont=rm,textfont=rm]{subfig}
\usepackage{textcomp}
\usepackage{stfloats}
\usepackage{url}
\usepackage{verbatim}
\usepackage{graphicx}
\usepackage{array}
\usepackage{booktabs}
\usepackage{cite}
\usepackage{xcolor}
\usepackage{mathtools}
\usepackage{bm}
\usepackage{amsthm}
\usepackage[export]{adjustbox}
\usepackage[
    colorlinks=true,
    linkcolor=blue,
    citecolor=blue,
    urlcolor=blue
]{hyperref}

\newtheorem{remark}{\textbf{Remark}}

\newcommand{\be}{\bm{e}}

\newcommand{\bi}{\bm{i}}

\newcommand{\bu}{\bm{u}}
\newcommand{\bv}{\bm{v}}

\newcommand{\bx}{\bm{x}}
\newcommand{\by}{\bm{y}}

\newcommand{\ts}{\mathrm{s}} 
   
\newcommand{\tM}{\mathrm{M}}     
\newcommand{\tL}{\mathrm{L}}   
\newcommand{\tj}{\mathrm{j}}   
\newcommand{\tabc}{\mathrm{abc}}   
\newcommand{\tT}{\mathrm{T}}  
\newcommand{\tC}{\mathrm{C}}  
\newcommand{\ta}{\mathrm{a}}  
\newcommand{\tb}{\mathrm{b}}
\newcommand{\tc}{\mathrm{c}}  
\newcommand{\tm}{\mathrm{m}}  
  
\newcommand{\tf}{\mathrm{f}}  
\newcommand{\tPLL}{u} 
\begin{document}

\title{A Unified Framework for Hybrid Grid-Forming 
and Grid-Following Inverter Control}

\author{Xiaoyang Wang,~\IEEEmembership{Graduate Student Member,~IEEE,}
        Xin Chen,~\IEEEmembership{Member,~IEEE}

    \thanks{X. Wang and X. Chen are with the Department of Electrical and Computer Engineering, Texas A\&M University,  College Station, TX 77840 USA (emails: wangxy@tamu.edu, xin\_chen@tamu.edu).} 
        \thanks{This work was supported in part by NSF AMPS 2523934, in part by NSF CAREER 2541998, and in part by the Consortium on AI and Large Flexible Load (CALL) at Texas A\&M University. 
        \emph{(Corresponding author: Xin Chen)}.} 
}


\IEEEpubid{}


\maketitle

\begin{abstract}
This paper proposes a novel unified control framework for achieving hybrid grid-forming (GFM) and grid-following (GFL) inverter operation by integrating dispatchable virtual oscillator control with reference-following synchronization. The proposed inverter control method supports multiple operating modes within a unified structure, including voltage- and frequency-following ($PQ$ mode), voltage-forming and frequency-following ($PV$ mode), voltage-following and frequency-forming ($Qf$ mode), voltage- and frequency-forming ($Vf$ mode), and a hybrid mode with mixed GFM and GFL behaviors. 
In particular, the proposed method achieves smooth pre-synchronization and enables seamless transitions across a spectrum of inverter operating modes by tuning a small set of continuous control parameters, rather than relying on discrete controller switching. This framework provides a flexible and physically interpretable approach for adapting inverter dynamics to varying grid conditions and operational requirements. 
The small-signal stability and input-output frequency-domain characteristics are further analyzed under different control parameter settings. The effectiveness and robustness of the proposed unified control method are demonstrated through extensive electromagnetic transient (EMT) simulations and hardware-in-the-loop (HIL) experiments.
\end{abstract}





\section{Introduction}

\IEEEPARstart{P}{ower} electronic converters are increasingly deployed as interfaces for various types of renewable
generation, energy storage systems, and grid-supporting devices, leading to the large-scale integration of inverter-based resources (IBRs) into modern power grids~\cite{pogaku2007modeling,markovic2021understanding}. Depending on the synchronization and control mechanisms, the operation of voltage-source inverters in IBRs can be broadly classified into two main categories: grid-following (GFL) control and grid-forming (GFM) control~\cite{markovic2021understanding,gupta2025revisiting,11482770}. 
GFL inverters rely on an external grid voltage for synchronization, typically using a phase-locked loop (PLL) to estimate the grid voltage phase angle and frequency. Based on this measured grid reference, they regulate the injected current or power and thus behave like controllable current sources. In contrast, GFM inverters establish an internal voltage phasor and synchronize with the grid through their own voltage and frequency dynamics, thus behaving like controllable voltage sources. Typical GFM control methods include droop-based control~\cite{markovic2021understanding}, virtual synchronous machine (VSM) control~\cite{d2015virtualSM}, and dispatchable virtual oscillator control (dVOC)~\cite{lu2019grid}. These different GFL and GFM control mechanisms lead to distinct dynamic characteristics, stability properties, and grid-support capabilities~\cite{li2022revisiting,DuweiGFMANDGFL,9181463}.

Unlike synchronous generators, whose dynamics are largely governed by their physical structures and electromechanical characteristics, inverter dynamics are primarily determined by control algorithms. This control-dominated nature provides substantial flexibility in shaping inverter behavior. Therefore, inverter dynamic behavior is not  inherently constrained by the conventional GFM/GFL dichotomy, but can instead be designed to span a broad spectrum beyond these two categories~\cite{10499214,askarianMultiModeInvertersUnified2024}. In particular, the voltage- and frequency-forming/following responses of inverters can be blended and continuously adjusted according to operational requirements. 
Such flexibility is highly desirable for supporting power system stability and reliable operation under varying grid conditions~\cite{gaoSeamlessSwitchingMethod2025}. For example, an inverter may need to provide strong voltage and frequency support under weak-grid or islanded conditions, while behaving more like a power-tracking resource under stiff-grid conditions. 
This motivates the development of a unified inverter control framework that can continuously transition among different GFM/GFL operating modes through a small set of tunable control parameters.


\begin{table*}[t]
\caption{Summary and comparison of existing unified or hybrid GFM-GFL inverter control methods}
\label{tab:hybridclassification}
\centering
\renewcommand{\arraystretch}{1.2}
\begin{tabular}{
>{\raggedright\arraybackslash}p{1.9cm}
>{\raggedright\arraybackslash}p{1.6cm}
>{\raggedright\arraybackslash}p{3.4cm}
>{\raggedright\arraybackslash}p{3.3cm}
p{5.8cm}
}
\toprule
\textbf{Category} &
\textbf{Continuous Transition} &
\textbf{Achievable Modes} &
\textbf{Synchronization Mechanism} &
\textbf{Key Features} \\
\midrule

Switching-based \cite{gaoSeamlessSwitchingMethod2025,10811873,OptimalIBRTomas,zeng2025transient,odunlami2025dynamic} &
No &
Either GFL or GFM without hybrid modes &
PLL (GFL) or droop-based power synchronization (GFM) &
GFM and GFL controls coexist and are switched according to operating conditions; discontinuous switching; may introduce additional switching transients, state mismatch during mode transition, and stability concerns. \\
\midrule

Weighted-blending \cite{9264177,10339862,9956749,11DuWREGFM_C1,9714845} &
Yes &
Intermediate hybrid modes between GFM and GFL &
Weighted blending of PLL and power synchronization   &
Blending of current-control-loop references or voltage modulation signals using weighting factors; enable smooth transitions but lack a clear physical interpretation in intermediate operating modes. \\
\midrule

Auxiliary-loop-embedding \cite{askarianMultiModeInvertersUnified2024,6152189,gengUnifiedGridFormingFollowing2022,10499214,8011460,8825565,9793610,10478164} &
No\cite{gengUnifiedGridFormingFollowing2022,10499214,8011460,9793610,8825565,10478164}, 
Yes\cite{askarianMultiModeInvertersUnified2024,6152189} &
Frequency-following \& voltage-forming; or hybrid frequency forming-following &
Synchronization established through droop control, PLL, or DC-voltage regulation &
Incorporate supplementary control loops into a baseline GFM or GFL control; retain the original controller structure, while the embedding mechanisms are often heuristic. \\
\midrule
\textbf{Our proposed unified control} & Yes & $PQ$, $PV$, $Qf$, $Vf$, and hybrid modes (voltage- and frequency-forming/following) & Blending of measured grid frequency and self-sustained oscillator frequency for power synchronization & Enable multiple operating modes within a unified control structure and support smooth mode transitions through continuous parameter tuning. \\
\bottomrule
\end{tabular}
\begin{minipage}{\linewidth}
\footnotesize
\noindent
\end{minipage}
\end{table*}

Existing studies have explored several unified or hybrid GFM/GFL control frameworks. From a control-structure perspective, these approaches can be broadly categorized into three groups: switching-based methods, weighted-blending methods, and auxiliary-loop-embedding methods, as summarized in Table~\ref{tab:hybridclassification} and discussed below.

Switching-based methods~\cite{gaoSeamlessSwitchingMethod2025,10811873,OptimalIBRTomas,zeng2025transient,odunlami2025dynamic} realize GFM/GFL transitions by shifting the active control structure or signal path according to operating conditions. A common implementation is to switch the input signal of the current-control loop between the output of a voltage-control loop and that of a power-control loop. These methods typically rely on event detection, operating-condition assessment, or dedicated mode-transition logic. Although they can support transitions between GFM and GFL operation, the discontinuous switching process may introduce additional transient and stability concerns. Moreover, such methods generally switch among discrete operating modes and do not naturally provide intermediate hybrid operating characteristics.

Weighted-blending methods~\cite{9264177,10339862,9956749,11DuWREGFM_C1,9714845} combine GFM and GFL control actions by introducing weighting factors at certain control layers. For example, existing studies have blended the power-synchronization loop used in GFM control with the PLL-based synchronization loop used in GFL control~\cite{9264177,10339862,9956749}, e.g., introducing weighted coupling at the input of the current-control loop~\cite{11DuWREGFM_C1}, or directly blending the inverter voltage modulation signals~\cite{9714845}. Compared with switching-based methods, weighted-blending approaches can enable smoother transitions and partially hybridized dynamic responses. Their performance depends strongly on the control layer at which the blending is introduced and on the specific choice of the weighting mechanism. In addition, the resulting intermediate hybrid modes may lack a clear physical interpretation.

Auxiliary-loop-embedding approaches~\cite{askarianMultiModeInvertersUnified2024,6152189,gengUnifiedGridFormingFollowing2022,10499214,8011460,8825565,9793610,10478164} achieve hybrid dynamic behavior by incorporating supplementary control loops into an existing baseline 
GFM or GFL control architecture. For instance, multiple control loops can be combined through gain coefficients~\cite{askarianMultiModeInvertersUnified2024,6152189}; a power-synchronization loop~\cite{gengUnifiedGridFormingFollowing2022} or a DC-voltage-synchronization loop can be embedded into a GFL controller to enhance synchronization performance~\cite{10499214,8011460,8825565}. Other approaches integrate voltage-control loops into GFL frameworks to enable active-power tracking together with voltage-forming capability, or linearly combine DC-voltage-frequency and AC-power-frequency droop relationships to blend frequency-following and frequency-forming behaviors~\cite{9793610,10478164}.

Overall, existing unified or hybrid GFM/GFL control methods provide important mechanisms for bridging the two control paradigms. Nevertheless, many of them rely on discrete mode switching, architecture-specific weighted blending, or additional auxiliary-loop augmentation. These limitations motivate the development of a unified inverter control framework in which voltage- and frequency-forming/following behaviors can be continuously shaped through a small set of tunable control parameters, enabling seamless mode transitions with clear physical interpretations and performance guarantees.

From the perspective of oscillator theory, GFM and GFL control can be interpreted as two fundamentally different synchronization mechanisms. A representative GFM control approach is virtual oscillator control (VOC) \cite{johnson2015synthesizing,johnsonSynchronizationParallelSinglePhase2014}, in which each inverter is modeled as a self-sustained nonlinear oscillator. By incorporating power-dispatch feedback into the oscillator dynamics, dispatchable virtual oscillator control (dVOC) further enables synchronization while regulating active and reactive power according to prescribed setpoints \cite{dVoCDorfler,lu2019grid,luVirtualOscillatorGridForming2022,li2020inverter} (see Section~\ref{section:dVOC} for a detailed introduction). In contrast, GFL control relies on reference-following synchronization, where the inverter synchronizes to the external grid by tracking the grid voltage phase angle and frequency, while regulating the injected current for power tracking. Although these two synchronization mechanisms exhibit complementary and dual characteristics \cite{li2022revisiting}, a unified framework that continuously bridges them remains lacking.

Motivated by these complementary mechanisms of frequency formation and grid synchronization, we propose a unified GFM-GFL inverter control framework by integrating dVOC with reference-following synchronization. The proposed controller blends an internally generated self-sustained oscillation frequency for frequency-forming operation with the externally measured grid frequency for frequency-following operation. In parallel, it blends self-regulated voltage-magnitude control for voltage-forming operation with power-reference-tracking dynamics for voltage-following operation. This blending is governed by a small set of continuous control parameters, enabling smooth transitions among different operating modes and intermediate hybrid behaviors within a unified framework. As a result, the proposed unified controller can flexibly shape voltage- and frequency-forming/following inverter behaviors.
Moreover, the proposed framework allows continuous optimization of GFM and GFL capabilities among interconnected inverters, providing additional flexibility for system-level coordination. 

The key contributions of this paper are summarized below: 
\begin{itemize} 
\item [1)] A novel inverter control framework is proposed to unify GFM, GFL, and hybrid GFM/GFL control within a single controller structure. The proposed framework supports smooth pre-synchronization and achieves multiple operating modes, including $PQ$, $PV$, $Qf$, $Vf$, and various intermediate hybrid modes, bridging the conventional dichotomy between GFM and GFL control.

\item [2)] 
Seamless transitions across a broad spectrum of inverter operating modes are achieved by tuning a small set of continuous control parameters, without requiring discrete controller switching or structural reconfiguration. This provides a flexible and physically interpretable mechanism for adapting inverter dynamic behavior to varying grid conditions. The continuity of these parameters also facilitates grid-level IBR optimization, e.g., the optimal siting and allocation of GFM and GFL capacities.


\item [3)] The dynamic characteristics of the proposed controller are systematically studied through small-signal stability analysis and input-output frequency-domain analysis. Extensive electromagnetic transient (EMT) simulations and hardware-in-the-loop (HIL) experiments further validate the effectiveness and robustness of the proposed framework under different grid scenarios.

\end{itemize}




The remainder of this paper is organized as follows. Section~\ref{Section:IIPreliminaries} introduces the preliminaries on inverter-based systems and dVOC. Section~\ref{Section:UnifiedController} develops the proposed unified control framework and presents its multiple operating modes. Section~\ref{section:IV} provides the small-signal stability analysis. Section~\ref{Section:EMTsimulation} presents the EMT simulation results, and Section~\ref{Section:HIL} reports the HIL test results. Finally, Section~\ref{sec:conclusion} concludes the paper.

\section{Preliminaries on Inverter-Based Systems and Virtual Oscillator Control}
\label{Section:IIPreliminaries}

In this section, we first present the configuration and notation for the inverter-based power system, and then introduce the preliminaries on (dispatchable) virtual oscillator control.

\subsection{System Description and Notation}

\begin{figure}[ht]
    \centering
    \includegraphics[width=1\linewidth]{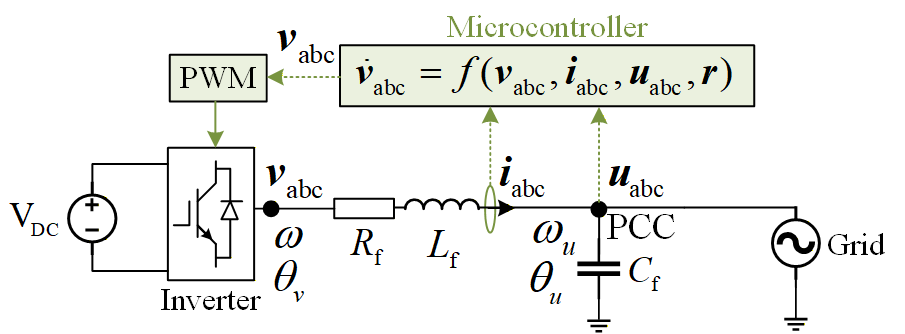}
    \caption{Typical structure of a voltage-source inverter system. 
    }
    \label{fig:SystemStructure}
\end{figure}

Consider a balanced three-phase system with a voltage-source inverter, as shown in Figure~\ref{fig:SystemStructure}. The DC voltage is assumed to be a constant value $\text{V}_\text{DC}$. The inverter is connected to the grid through an LC filter, 
where the resistance, inductance, and capacitance are denoted by $R_\tf$, $L_\tf$, and $C_\tf$, respectively. 
The three-phase current through the inductor is $\bi_\text{abc}$. The three-phase voltage at the point of common coupling (PCC) is denoted by $\bu_\text{abc}$, which is the same as the grid-side voltage.
The nominal reference angular frequency and voltage magnitude are denoted by $\omega_0$ and $V_0$, respectively. The PCC frequency measured by the PLL is denoted by $\omega_{u}$, while $\omega$ represents the frequency of the inverter terminal voltage. The corresponding phase angles are $\theta_u$ and $\theta_v$, respectively. Based on the measured PCC voltage $\bu_{abc}$, inductor current $\bi_{abc}$, and the preset reference vector $\bm{r}$, the microcontroller generates the inverter terminal voltage reference $\bv_{abc}$ according to a control law $f(\bv_{abc}, \bi_{abc}, \bu_{abc}, \bm{r})$, which is subsequently supplied to the PWM generator.
The high-frequency switching dynamics of the inverter PWM control are neglected. Hence, the inverter terminal voltage is represented by the controlled voltage $\bv_{abc}$. 



Using the Clarke transformation\cite{clarke1943ac}, three-phase values can be represented in the stationary $\alpha\beta$ reference frame as: 
\begin{align} \label{eq:clarkeTransformation} \bx = x_\alpha + \tj x_\beta := \begin{bmatrix}x_\alpha\\ x_\beta \end{bmatrix} = \underbrace{\frac{2}{3} \begin{bmatrix} 1 & -\tfrac{1}{2} & -\tfrac{1}{2}\\ 0 & \tfrac{\sqrt{3}}{2} & -\tfrac{\sqrt{3}}{2} \end{bmatrix}}_{\coloneqq \tT_\tC}  \underbrace{\begin{bmatrix}
x_\ta\\ x_\tb \\ x_\tc
\end{bmatrix}}_{\coloneqq \bx_{abc}}, 
\end{align} 
where $\tj=\sqrt{-1}$ denotes the imaginary unit and $\tT_\tC$ is the Clarke transformation matrix. Accordingly, the inverter terminal voltage $\bv_{\tabc}$, PCC voltage $\bu_{\tabc}$, and inductor current $\bi_{\tabc}$ are converted into the $\alpha\beta$ frame as $\bv$, $\bu$, and $\bi$, respectively.

The voltage vectors are expressed as
\begin{align}
\bv &\coloneqq [v_\alpha, v_\beta]^\top = v_\alpha+\tj v_\beta= V_\tm e^{\tj \theta_v}, \\
\bu &\coloneqq [u_\alpha, u_\beta]^\top =u_\alpha+\tj u_\beta= U_\tm e^{\tj \theta_u},
\end{align}
where $V_\tm\coloneqq ||\bv||$ and $U_\tm\coloneqq ||\bu||$ denote their voltage magnitudes.   
Define $\delta=\theta_v-\theta_u$ as the phase angle difference between the inverter terminal voltage 
and the PCC voltage. 

The active and reactive powers at the inverter terminal are calculated in the $\alpha\beta$ reference frame as 
\begin{align} P+\tj Q = \frac{3}{2}\bv\bar{\bi}, 
\end{align} 
where $\bar{\bi}$ denotes the complex conjugate of $\bi$.

Define the local inverter $dq$ frame \cite{park1929} as the rotating reference frame with angular frequency $\omega$ and phase angle $\theta_v$. 
The Park transformation  \cite{park1929} from the stationary $\alpha\beta$ frame to the local inverter $dq$ frame is given by \eqref{eq:dqtrans}:
\begin{equation} \label{eq:dqtrans}
\bx_{dq}\coloneqq
\begin{bmatrix}
x_d\\
x_q
\end{bmatrix}
=
R(-\theta_v)
\begin{bmatrix}
x_\alpha\\
x_\beta
\end{bmatrix} = R(-\theta_v)\,\bx,
\end{equation}
where the rotation matrix $R(\theta)$ is defined by \eqref{eq:rotation}:
\begin{equation} \label{eq:rotation}
R(\theta)\coloneqq
\begin{bmatrix}
\cos\theta & -\sin\theta\\
\sin\theta & \cos\theta
\end{bmatrix}.
\end{equation}
Similarly,
define the local grid-side $d_gq_g$ frame as the rotating reference frame with angular frequency $\omega_u$ and phase angle $\theta_u$. The transformation between the local inverter $dq$ frame and the local grid-side $d_gq_g$ frame is given by \eqref{eq:dq2todq1}:
\begin{align} \label{eq:dq2todq1}
\bx_{d_gq_g} = R(\delta)\bx_{dq}, \quad
\bx_{dq} = R(-\delta)\bx_{d_gq_g},
\end{align}
where $\delta = \theta_v-\theta_u$ denotes the phase angle difference.



\subsection{(Dispatchable) Virtual Oscillator Control}
\label{section:dVOC}

\subsubsection{Linear Harmonic Oscillator}

\begin{figure}
    \centering
    \includegraphics[width=1\linewidth]{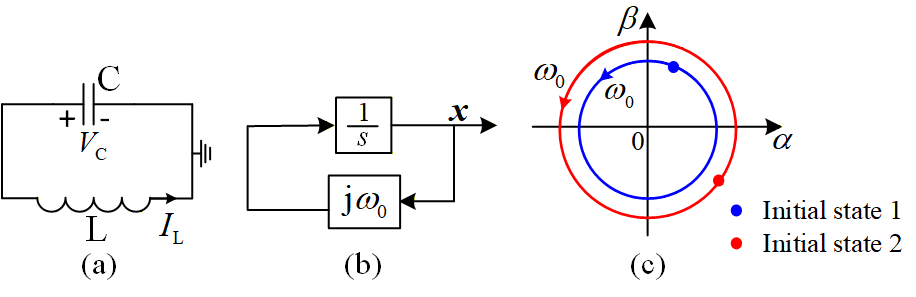}
    \caption{Linear harmonic oscillator. (a) LC oscillator circuit. (b) Control block diagram. (c) Oscillation trajectories with different initial states. 
    }
    \label{fig:LinearOscillator}
\end{figure}

A simple linear harmonic oscillator system is given by \eqref{eq:Harmonicoscillator}: 
\begin{align} \label{eq:Harmonicoscillator}
    \dot{\bx}=\tj \omega_0 \bx,
\end{align}
where $\bx \coloneqq x_\alpha + \tj x_\beta$, and $\dot{\bx}\coloneqq [\frac{d x_\alpha}{dt}, \frac{d x_\beta}{dt}]^\top$. It can be represented by a passive LC oscillator system, as shown in Figure \ref{fig:LinearOscillator}(a). The dynamics of the inductor current $I_L$ and the capacitor voltage $V_C$ are given by \eqref{eq:LCosci}: 
\begin{subequations}\label{eq:LCosci}
    \begin{align}
        \dot{V}_\tC&=-\omega_0 Z_0 I_\tL,\\
        Z_0\dot{I}_\tL&=\omega_0 V_\tC,
    \end{align}
\end{subequations} 
where $\omega_0=\frac{1}{\sqrt{LC}}$ denotes the natural resonant frequency, and $Z_0=\sqrt{{L}/{C}}$ is the characteristic impedance. The LC oscillator dynamics \eqref{eq:LCosci} become \eqref{eq:Harmonicoscillator} when defining $x_\alpha\coloneqq V_\tC$ and $x_\beta \coloneqq Z_0I_\tL$. 
The corresponding control block diagram is shown in Figure \ref{fig:LinearOscillator}(b), where $s$ denotes the Laplace variable. Since the LC oscillator  in Figure \ref{fig:LinearOscillator}(a) has no external source or damping, the oscillation magnitude is not controllable but depends on the initial state, as shown in Figure \ref{fig:LinearOscillator}(c). 

\begin{figure}
    \centering
    \includegraphics[width=1\linewidth]{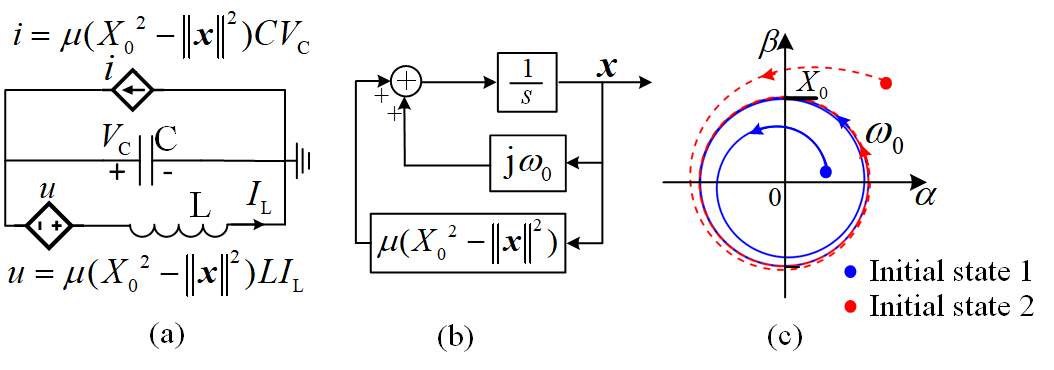}
    \caption{Andronov-Hopf oscillator. (a) LC oscillation circuit with a controlled current source and a voltage source to regulate oscillation magnitude. (b) Control block diagram. (c) Oscillation trajectories with different initial states.
    }
    \label{fig:AHO}
\end{figure}

\subsubsection{Andronov-Hopf Oscillator (AHO)} To actively regulate the magnitude of the linear oscillator, the difference between the magnitude and its reference can be fed back into the oscillator dynamics. A prominent approach is the 
nonlinear Andronov-Hopf oscillator (AHO)~\cite{lu2019grid,luVirtualOscillatorGridForming2022,li2020inverter}, which offers the advantage of generating harmonic-free oscillations via symmetric feedback. As shown in Figure~\ref{fig:AHO}(b), AHO introduces an additional nonlinear feedback term $\mu (X_0^2-||\bx||^2)\bx$, and its dynamics is given by \eqref{eq:AHO}:
\begin{align} \label{eq:AHO}
    \dot{\bx}=\tj \omega_0 \bx + \mu (X_0^2-||\bx||^2)\bx,
\end{align}
where $X_0$ is the reference magnitude of the oscillator, and $\mu$ is the feedback gain that governs the convergence rate to $X_0$. The corresponding LC oscillator realization of AHO is shown in Figure \ref{fig:AHO}(a), where a controlled current source $i$ is added in parallel with the capacitor C and a controlled voltage source $u$ is added in series with the inductor L. The current source $i$ and voltage source $u$ are controlled according to \eqref{eq:AHO-LC}:
\begin{subequations} \label{eq:AHO-LC}
    \begin{align}
        i&=\mu(X_0^2-||\bx||^2)CV_\tC,\\
        u&=\mu(X_0^2-||\bx||^2)LI_\tL.
    \end{align}
\end{subequations}
These two sources inject energy when the oscillation magnitude $||\bx||^2$ is less than $X_0$ and absorb energy when $||\bx||^2$ is larger than $X_0$. As a result, the AHO converges to a limit cycle with frequency $\omega_0$ and magnitude $X_0$, independent of the initial states, as shown in Figure~\ref{fig:AHO}(c).

The AHO can generate sinusoidal waveforms with a constant frequency $\omega_0$ and a constant magnitude $X_0$. Therefore, it can be used for GFM inverter control by using $x_\alpha$ and $x_\beta$ as modulation signals for single-phase \cite{li2020inverter} or three-phase \cite{lu2019grid} systems. Nevertheless, the basic AHO lacks inherent self-synchronization capability, as its frequency and magnitude are prescribed by $\omega_0$ and $X_0$, while its phase is not adjusted according to grid conditions.



\subsubsection{Dispatchable Virtual Oscillator Control (dVOC)}

\begin{figure}[hb]
    \centering
    \includegraphics[width=1\linewidth]{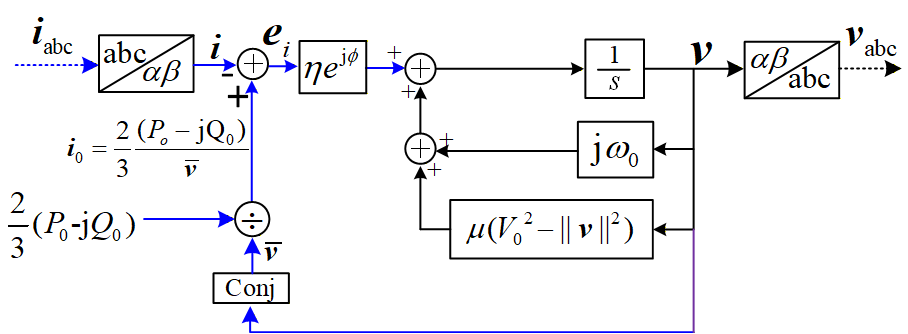}
    \caption{The AHO-based dispatchable virtual oscillator for grid-forming inverter control. (The blue lines represent the current-feedback term introduced to achieve power dispatch.)
    }
    \label{fig:dAHO}
\end{figure}


To enable global synchronization among multiple oscillators, synchronizing feedback based on the output current can be introduced, where the output current mediates the coupling among oscillators. 
In this way, AHO-based dispatchable virtual oscillator control was proposed in~\cite{lu2019grid} for GFM inverter control by incorporating current feedback. The term ``virtual" indicates that the oscillator is not
realized by a physical LC circuit but is implemented through software or digital control.
The term ``dispatchable"
means that the oscillator-based inverter can regulate its voltage waveform while adjusting its active and reactive power outputs according to prescribed setpoints~\cite{dVoCDorfler,8722028}. The rectangular-coordinate dynamics of the AHO-based dVOC are formulated as \eqref{eq:dispatchableAHO}:
\begin{align}\label{eq:dispatchableAHO}
& \textbf{(Rectangular Dynamics of AHO-Based dVOC)}:\nonumber \\
  & \qquad  \dot{\bv}=\tj\omega_0\bv+\mu(V_0^2-||\bv||^2)\bv+\eta(\bi_0-\bi)e^{\tj\phi},
\end{align}
where $\phi \in [0,\frac{\pi}{2}]$ is the rotation angle parameter that determines the coupling between voltage magnitude/phase and active/reactive power dynamics. 
$\eta$ is the current-feedback gain. $\bi_0$ denotes the reference current calculated from \eqref{eq:i_0} based on the active and reactive power references $P_0, Q_0$. $\bi$ is the measured current that depends on the inverter terminal voltage $\bv$ and the actual inverter active and reactive power $P, Q$ according to \eqref{eq:i_PQ}.
\begin{subequations} \label{eq:iandi0_PQ}
    \begin{align} 
    \bi_0=&\frac{2}{3}\frac{P_0-\tj Q_0}{\bar{\bv}}=\frac{2}{3}\frac{(P_0-\tj Q_0)}{||\bv||^2}\bv, \label{eq:i_0}\\
    \bi=&\frac{2}{3} \frac{P-\tj Q}{\bar{\bv}}=\frac{2}{3}\frac{P-\tj Q}{||\bv||^2}\bv. \label{eq:i_PQ}
\end{align}
\end{subequations}
Figure~\ref{fig:dAHO} illustrates the control block diagram of the dVoC.

Substituting \eqref{eq:iandi0_PQ} into \eqref{eq:dispatchableAHO} eliminates the current terms and it yields \eqref{eq:vdot_PQ}:
\begin{align}\label{eq:vdot_PQ}
    &\dot{\bv}=\tj \left[\omega_0+\eta \frac{2[(P_0\!-\!P) \sin \phi-(Q_0\!-\!Q)\cos \phi]}{3||\bv||^2}\right]\bv \nonumber \\
    &+\!\left[\mu(V_0^2\!-\!||\bv||^2)+\eta\frac{2[(P_0\!-\!P) \cos \phi+(Q_0\!-\!Q)\sin \phi]}{3||\bv||^2}\right]\bv.
\end{align}
In terms of the inverter terminal voltage $\bv\!=\!V_\tm(t) e^{\tj \theta_v(t)}$, its derivative with respect to time $t$ is given by \eqref{eq:dvdt}:
\begin{align}\label{eq:dvdt}
  \dot{\bv}=  \frac{d \bv}{dt}=\tj \frac{d\theta_v(t)}{dt}\bv+ \frac{1}{V_\tm(t)}\frac{d V_\tm(t)}{dt}\bv.
\end{align}
By comparing \eqref{eq:dvdt} and \eqref{eq:vdot_PQ} term by term and using the definition $V_m \coloneqq ||\bv||$, the polar-coordinate dynamics of the AHO-based dVOC are obtained as \eqref{eq:VthetadynamicsdVoC}:
\begin{subequations} \label{eq:VthetadynamicsdVoC}
  \begin{align}
  &\hspace{-22pt} \textbf{(Polar Dynamics of AHO-Based dVOC)}:\nonumber \\
    \dot{V}_\tm=&\,\mu V_\tm(V_0^2\!-\!V_\tm^2)\!+\!\eta \frac{2[(P_0\!-\!P) \cos \phi+(Q_0\!-\!Q)\sin \phi]}{3V_\tm},\\
   \dot{\theta}_v=&\, \omega=\omega_0+\eta\frac{2[(P_0\!-\!P) \sin \phi-(Q_0\!-\!Q)\cos \phi]}{3V_\tm^2}.
\end{align}
\end{subequations}

From \eqref{eq:VthetadynamicsdVoC}, both $V_\tm$ and $\theta_v$ (or $\omega$) are controlled according to the reference values $V_0$ and $\omega_0$ as well as the active and reactive power errors $P_0-P$ and $Q_0-Q$. The rotation angle parameter $\phi$ determines how active and reactive power participate in the voltage magnitude and phase dynamics. For example, when $\phi=\frac{\pi}{2}$, the polar dynamics of the AHO-based dVOC \eqref{eq:VthetadynamicsdVoC} reduce to \eqref{eq:Vthteadynamicphipi2}:
\begin{subequations}\label{eq:Vthteadynamicphipi2}
    \begin{align}
        \dot{V}_\tm&=\mu V_\tm(V_0^2-V_\tm^2)+\frac{2\eta}{3V_\tm}(Q_0-Q), \label{eq:Vdynamicphipo2}\\
        \dot{\theta}_v&=\omega=\omega_0+\frac{2\eta}{3V_\tm^2}(P_0-P),
    \end{align}
\end{subequations}
where reactive power $Q$ primarily contributes to voltage magnitude $V_m$ regulation, and active power $P$ primarily contributes to phase $\theta_v$ or frequency $\omega$ regulation. Moreover, in steady states, \eqref{eq:Vthteadynamicphipi2} exhibits nonlinear droop-like \cite{simpson2013synchronization} characteristics for $V_\tm$-$Q$ and $\omega$-$P$, as formulated in \eqref{eq:dVoCPfQVdroop}:
    \begin{align} \label{eq:dVoCPfQVdroop}
        V_\tm^2=V_0^2\!+\!\frac{2\eta}{3\mu V_\tm^2}(Q_0\!-\!Q),\ \ 
        \omega=\omega_0\!+\!\frac{2\eta}{3V_\tm^2}(P_0\!-\!P).
    \end{align}
By setting $\mu=0$ in \eqref{eq:Vthteadynamicphipi2}, a unified VOC framework was proposed in \cite{9203895}, in which active-power ($\phi=0$) or reactive-power tracking ($\phi=\frac{\pi}{2}$) can be achieved. This corresponds to voltage-following and frequency-forming operation. Nevertheless, the controller remains fundamentally based on self-sustained oscillator dynamics and therefore cannot represent frequency-following operation.

\section{Unified GFM and GFL Control Design}\label{Section:UnifiedController}

In this section, we develop the unified GFM-GFL inverter control method based on a virtual oscillator structure, which integrates self-sustained oscillation dynamics with grid-synchronization mechanisms. The proposed unified GFM-GFL controller is formulated in both the $\alpha\beta$ and $dq$ frames, and its multiple operating modes, including $PQ$, $PV$, $Qf$, $Vf$, and hybrid modes, are derived and discussed. Rather than switching discretely among operating modes, the unified controller enables continuous transitions by smoothly tuning a small set of control parameters. The grid-synchronization mechanisms, pre-synchronization capability, and current-limiting strategies are presented as well.



\subsection{Design Motivation} 

As introduced in Section~\ref{section:dVOC}, dVOC achieves GFM control through self-sustained oscillator dynamics with an internal nominal frequency. It further incorporates current-feedback terms parameterized by active and reactive power references to enable synchronization and power dispatch, together with voltage-magnitude feedback for voltage regulation. As a result, dVOC enables autonomous voltage and frequency formation while supporting dispatchable power sharing. In contrast, GFL control typically relies on a PLL to estimate the grid voltage phase angle and frequency for synchronization, and it follows the externally measured grid frequency to construct a grid-synchronized reference frame and regulate the injected current for power reference tracking~\cite{DuweiGFMANDGFL}.

Motivated by these complementary mechanisms of frequency formation and grid synchronization, we propose a unified GFM-GFL inverter control framework. The proposed controller blends an internally generated self-sustained oscillation frequency for frequency-forming operation with the externally measured grid frequency for frequency-following operation. In parallel, the proposed controller combines self-regulated voltage-magnitude control for voltage-forming operation with power-reference-tracking dynamics for voltage-following operation. This blending is governed by a small set of continuous control parameters, allowing smooth transitions among different operating modes. It also enables hybrid control modes that bridge GFM and GFL control within a unified framework. The detailed control design and operating mode analysis are presented in the following subsections.

\subsection{Unified GFM-GFL Inverter Control in \texorpdfstring{$\alpha\beta$}{alpha beta} Frame} \label{section:IIIA}

\begin{figure}[ht]
    \centering
    \includegraphics[width=1\linewidth]{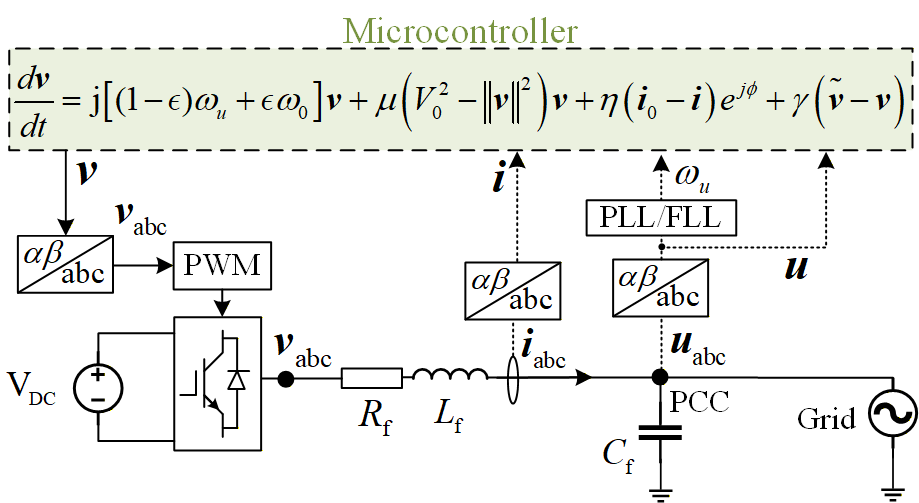}
    \caption{The proposed unified GFM-GFL controller for a three-phase inverter. 
    }
    \label{fig:System}
\end{figure}

Building upon the AHO-based dVOC \eqref{eq:dispatchableAHO} and PLL-based GFL control, we propose a unified GFM-GFL inverter control framework, as formulated in \eqref{eq:hybridDynamics}:
\begin{align} \label{eq:hybridDynamics}
& \textbf{(Rectangular Dynamics of Unified GFM-GFL Control)}:\nonumber \\
  &\quad  \dot{\bv}=\tj \big[\epsilon \omega_0 + (1-\epsilon) \omega_{u}\big]\bv + \mu(V_0^2-||\bv||^2)\bv \nonumber\\
    &\qquad\qquad\qquad+\eta(\bi_0-\bi)e^{\tj \phi} +\bm{\gamma} (\tilde{\bv}-\bv),
\end{align}
where $\omega_u$ denotes the PCC frequency measured by the PLL or frequency-locked loop (FLL)\cite{5446347}, and $\epsilon\in[0,1]$ is a weighting parameter. In addition, $\tilde{\bv}$ denotes the target value of $\bv$, and $\gamma$ is a nonnegative gain parameter. The term 
$\bm{\gamma} (\tilde{\bv}-\bv)$ is incorporated to pin the inverter voltage vector $\bv$ to the target value $\tilde{\bv}$.
Prior to grid connection, $\tilde{\bv}$ is set to the grid voltage vector $\bu$ to achieve pre-synchronization. 


The implementation of the proposed unified controller \eqref{eq:hybridDynamics} is illustrated in Figure~\ref{fig:System}. It integrates both self-sustained oscillation and frequency-following dynamic behavior. Specifically, the proposed control method \eqref{eq:hybridDynamics} integrates four mechanisms: 
\begin{itemize}
    \item [1)] A hybrid forming-following oscillation frequency:  
    \begin{align} \label{eq:combinnaturalfrequency}
     \omega_\epsilon=(1-\epsilon)\omega_\tPLL+\epsilon \omega_0.
\end{align}
\item [2)] Voltage magnitude feedback regulation.
\item [3)] Current/power tracking feedback control.  
\item [4)] Voltage reference tracking for pre-synchronization.  
\end{itemize}

Compared with the dVOC method \eqref{eq:dispatchableAHO}, the proposed unified GFM-GFL control \eqref{eq:hybridDynamics} replaces the constant natural oscillation frequency $\omega_0$ with a hybrid frequency $\omega_\epsilon$, which reduces to $\omega_0$ when setting $\epsilon\!=\!1$. In addition, a voltage-reference tracking term $\bm{\gamma}(\tilde{\bv}\!-\!\bv)$ is introduced to enable pre-synchronization prior to grid connection. During normal operation, $\gamma$ is set to zero, while $\gamma \neq 0$ is applied only during the pre-synchronization process. 

Denote the tracking errors in \eqref{eq:hybridDynamics} as $e_V \coloneqq V_0^2-||\bv||^2$, $\be_i\coloneqq\bi_0-\bi=e_{i\alpha}+\tj e_{i \beta}$, and $\be_{\bv}\coloneqq \tilde{\bv}-\bv=e_{v \alpha}+\tj e_{v \beta}$. Denote the active and reactive power tracking errors as:
 \begin{align}
     \be_S \coloneqq e_P+\tj e_Q=\frac{2}{3}(P_0-P)+\tj \frac{2}{3}(Q_0-Q).
 \end{align}

\begin{remark}
    In the following analysis, we focus on the
    normal operation with $\gamma=0$, and thus the unified GFM-GFL
    control dynamics \eqref{eq:hybridDynamics} become \eqref{eq:Maindynamic}:
    \begin{align} \label{eq:Maindynamic}
    \dot{\bv}=&\tj \omega_\epsilon \bv + \mu e_V \bv+\eta \be_{\bi} e^{\tj \phi},
\end{align}
    while the pre-synchronization process with $\gamma\neq 0$ is discussed in Section \ref{sec:presyn}. Moreover, the power references $P_0$ and $Q_0$ are assumed to be specified at the inverter terminal, i.e., before the filter inductor. The alternative case in which $P_0$ and $Q_0$ are specified at the PCC, i.e., after the filter inductor, is presented in Appendix~\ref{Appendix:MultiModedefinedbehindinductor}. \qed
\end{remark}

By definition \eqref{eq:iandi0_PQ}, we have $\be_i=\frac{\bar{\be}_S}{||\bv||^2}\bv$, and then the unified control dynamics \eqref{eq:Maindynamic} can be rewritten as \eqref{eq:dynamicvbye}:
 \begin{align} \label{eq:dynamicvbye}
     \dot{\bv}=&\tj [\omega_\epsilon + \eta \frac{(e_P \sin \phi - e_Q \cos \phi)}{||\bv||^2}] \bv \nonumber\\
   &  \qquad +[\mu e_V + \eta \frac{(e_P \cos \phi +e_Q \sin \phi)}{||\bv||^2} ] \bv.
 \end{align}
By comparing \eqref{eq:dynamicvbye} and \eqref{eq:dvdt} term by term, the polar-coordinate dynamics of the unified GFM-GFL controller \eqref{eq:Maindynamic} can be obtained as \eqref{eq:hybridthetaV}:
\begin{subequations} \label{eq:hybridthetaV}
    \begin{align}
        &\hspace{-30pt}\textbf{(Polar Dynamics of Unified GFM-GFL Control):}\nonumber\\   
        &\dot{V}_\tm=\mu V_\tm e_V + \eta_1 \frac{(e_P \cos \phi +e_Q \sin \phi)}{V_\tm},\\
&\dot{\theta}_v=\omega=\omega_\epsilon + \eta_2 \frac{(e_P \sin \phi - e_Q \cos \phi)}{V_\tm^2}. \label{eq:thetadyna_hybridthetaV}
    \end{align}
\end{subequations}
Here, we split the control gain $\eta$ into two independent gains, $\eta_1$ and $\eta_2$, 
which regulate the voltage magnitude and phase angle, respectively, to provide greater control flexibility.

\subsection{Unified GFM-GFL Inverter Control in dq-Frame}

Compared with the stationary $\alpha\beta$ frame, the local inverter $dq$ frame transforms balanced sinusoidal variables into constant steady-state quantities, which facilitate PI-based control, active/reactive power decoupling, and small-signal stability analysis around a time-invariant operating point. 
Hence, based on the polar dynamics \eqref{eq:hybridthetaV}, this subsection presents the unified GFM-GFL inverter control method in the $dq$ frame.

In the local inverter $dq$ frame, the active and reactive power $P,Q$ and their reference values $P_0,Q_0$ are represented as:
\begin{subequations}\label{eq:PQdqframe}
    \begin{align}
    P&=\frac{3}{2}V_\tm i_{d}, \ \  Q=-\frac{3}{2}V_\tm i_{q}, \\
    P_{0}&=\frac{3}{2}V_\tm i_{0, d},  \ \ Q_{0}=-\frac{3}{2}V_\tm i_{0, q},
\end{align}
\end{subequations}
Here, $\bi_{dq}\coloneqq [i_d,i_q]^\top$ and $\bi_{0,dq}\coloneqq [i_{0,d},i_{0,q}]^\top$ are the inductor current and its reference in the local $dq$ frame, which can be obtained by \eqref{eq:itrans}:
\begin{equation}\label{eq:itrans}
\bi_{dq}
=
R(-\theta_v)
\bi_{\alpha\beta}
=
R(-\theta_v)\tT_\tC
\bi_{abc}.
\end{equation}
Denote the current tracking error in the local {\it{dq}}-frame as:
\begin{equation}
    e_{id}\coloneqq i_{0,d}-i_d, \qquad e_{iq}\coloneqq i_{0,q}-i_q.
\end{equation}

Based on \eqref{eq:hybridthetaV} and \eqref{eq:PQdqframe}, the proposed unified GFM-GFL control is formulated as \eqref{eq:Vthetaidiq} in the local inverter $dq$ frame:
\begin{subequations} \label{eq:Vthetaidiq}
    \begin{align}
    &\hspace{-20pt}\textbf{(Unified GFM-GFL Control in \textit{dq} Frame):}\nonumber\\ 
     \dot{V}_\tm&= \mu V_\tm(V_{0}^2-V_\tm^2) + \eta_{1}\big( e_{id} \cos \phi - e_{iq} \sin \phi \big) , \label{eq:Vidiq}\\
\dot{\theta}_{v}&=\omega=\omega_{\epsilon} +  \eta_{2}(e_{id}\sin \phi + e_{iq} \cos  \phi)  \label{eq:thetaidiq}.
    \end{align}
\end{subequations}

\begin{figure}[ht]
    \centering
    \includegraphics[width=0.9\linewidth]{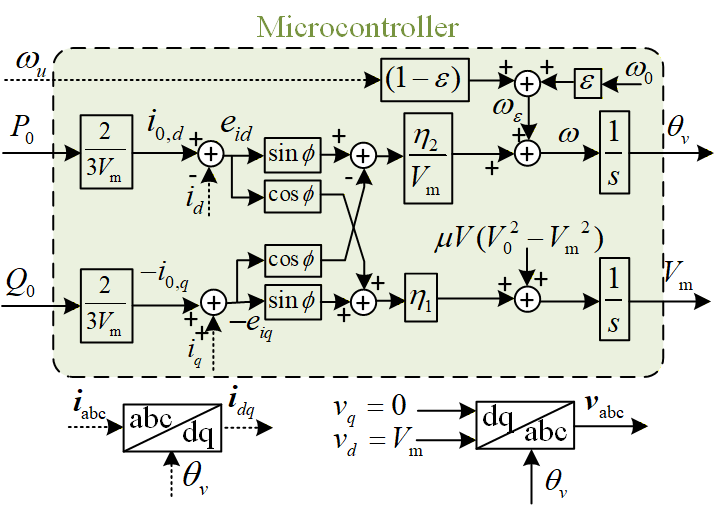}
    \caption{The unified GFM-GFL control in local inverter $dq$ frame.
    }
    \label{fig:dqframecontrolinverterside}
\end{figure}

The control block diagram for implementing the proposed unified GFM-GFL inverter control \eqref{eq:Vthetaidiq} in the local $dq$ frame is shown in Figure \ref{fig:dqframecontrolinverterside}. In addition, the control dynamics \eqref{eq:Vthetaidiq} can be transformed into the local grid-side $d_gq_g$ frame via \eqref{eq:dq2todq1}.


\subsection{Multiple Operating Modes Enabled by the Unified GFM-GFL Control Framework}
\label{section:modecontrol}

\begin{figure}
    \centering
    \includegraphics[width=0.8\linewidth]{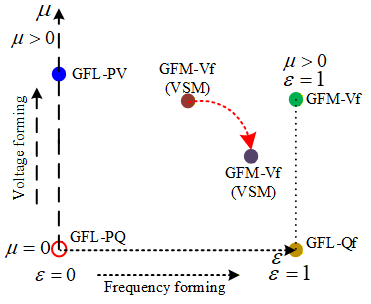}
    \caption{Multiple operating modes enabled by the proposed unified GFM-GFL control framework. (The parameter $\mu$ continuously tunes the voltage behavior from voltage-following to voltage-forming, while $\epsilon$ continuously tunes the frequency behavior from frequency-following to frequency-forming. For $\epsilon \in (0,1)$, different hybrid operating modes with corresponding steady-state $V$-$Q$ and $f$-$P$ droop characteristics can be obtained, enabling continuous tuning and seamless transitions among these modes.)}
    \label{fig:fourmodes}
\end{figure}

This subsection presents the multiple operating modes enabled by the proposed unified GFM-GFL control framework. 
As illustrated in Table \ref{Table:Modes} and Figure~\ref{fig:fourmodes},
by properly selecting the control parameters $\epsilon$, $\mu$, $\eta_1$, and $\eta_2$, the proposed framework can realize $PQ$ mode, $PV$ mode, $Qf$ mode, $Vf$ mode, and hybrid mode, allowing flexible transitions between GFM and GFL dynamic behaviors. 
Since the reactance typically dominates the resistance in the inverter LC filter, i.e., $X \gg R$, the active power $P$ and reactive power $Q$ are strongly coupled with the phase angle $\theta$ and voltage magnitude $V$, respectively. Thus, we consider $\phi=\pi/2$ when presenting the operating modes below for simplicity and clarity, while similar analysis can be performed for other $\phi$ values.



\begin{table*}[t]
\centering
\caption{Operating modes with different parameters enabled by the unified GFM-GFL control framework.}
\label{Table:Modes}

\renewcommand{\arraystretch}{1.15}
\setlength{\tabcolsep}{5pt}

\begin{tabular*}{\textwidth}{@{\extracolsep{\fill}}ccccccccccc}
\toprule
Control Mode &
$\epsilon$ &
$\mu$ &
$\eta_1$ &
$\eta_2$ &
$\omega_\epsilon$ &
$\gamma$ &
Frequency &
Voltage &
$P$ &
$Q$
\\
\midrule

GFL-PQ
& 0 & 0 & $>0$ & $>0$
& $\omega_{u}$
& 0
& Following
& Following
& Tracking $P_0^\dagger$
& Tracking $Q_0^\ddagger$
\\
\midrule

GFL-PV
& 0 & $>0$ & $\ge0$ & $>0$
& $\omega_{u}$
& 0
& Following
& Forming
& Tracking $P_0^\dagger$
& $V$-$Q$ Droop
\\
\midrule

GFL-Qf
& 1 & 0 & $>0$ & $\ge0$
& $\omega_0$
& 0
& Forming
& Following
& $f$-$P$ Droop
& Tracking $Q_0^\ddagger$
\\
\midrule

GFM-Vf
& 1 & $>0$ & $\ge0$ & $\ge0$
& $\omega_0$
& 0
& Forming
& Forming
& $f$-$P$ Droop
& $V$-$Q$ Droop
\\
\midrule

Hybrid
& $(0,1)$ & $\ge 0$ & $\ge0$ & $\ge0$
& $(1-\epsilon)\omega_{u}+\epsilon\omega_0$
& 0
& Forming
& Forming
& $f$-$P$ Droop
& $V$-$Q$ Droop
\\
\midrule
Pre-Synchronization
& 0 & * & * & *
& *
& $\gg 0$
& Following
& Following
& 0
& 0
\\

\bottomrule
\end{tabular*}

\vspace{2mm}

\begin{minipage}{\linewidth}
\footnotesize
\textit{*: The value depends on the expected mode after synchronization.\\
$\dagger$: An outer primary $P$-$f$ droop loop can be incorporated.\\
$\ddagger$: An outer primary $Q$-$V$ droop loop can be incorporated.
}
\end{minipage}
\end{table*}
\subsubsection{Frequency and Voltage Following - PQ Mode}
\begin{figure}[ht]
    \centering
    \includegraphics[width=0.9\linewidth]{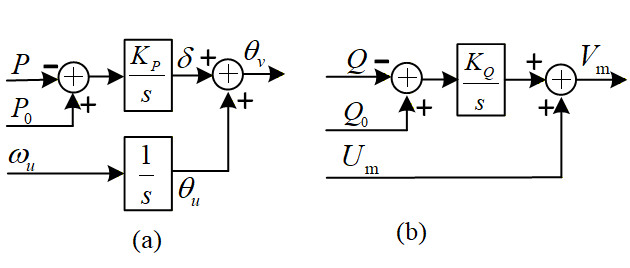}
    \caption{The control block diagram of typical power synchronization methods.}
    \label{fig:Power-Synchronization}
\end{figure}

By setting $\epsilon=0$, $\mu=0$, $\eta_1>0$, and $\eta_2>0$, the unified GFM-GFL control dynamics \eqref{eq:Maindynamic} become \eqref{eq:PQ_vdot}:
\begin{align} \label{eq:PQ_vdot}
     \dot{\bv}=&\tj \omega_{u} \bv +\eta \tj (\bi_0 -\bi),
\end{align}
where the natural frequency of the oscillator \emph{follows} the PCC frequency measured by the PLL, i.e., $\omega_\epsilon=\omega_\tPLL$. As the phase angle difference $\delta$ between the inverter and the PCC follows $\dot{\delta}=\omega-\omega_\tPLL$, the polar dynamics \eqref{eq:hybridthetaV} of the unified GFM-GFL control become \eqref{eq:PQmode_polar_phi=pai2}:
\begin{subequations} \label{eq:PQmode_polar_phi=pai2}
    \begin{align}
     \dot{V}_\tm&=\frac{\eta_1}{V_\tm}e_Q,\\
        \dot{\delta}&= \frac{\eta_2}{V_\tm^2}e_P,
    \end{align}
\end{subequations}
where the active-power tracking error regulates the phase (or frequency) dynamics, and the reactive-power tracking error regulates the voltage magnitude. This indicates a power synchronization mechanism\cite{powersynchronization}, as illustrated in Figure ~\ref{fig:Power-Synchronization}, by letting the integral gains $K_P=\frac{\eta_2}{{V_\tm}^2}$ and $K_Q=\frac{\eta_1}{V_\tm}$.

The steady-state solutions of \eqref{eq:PQmode_polar_phi=pai2} satisfy:
    \begin{align} \label{eq:PQmode_steadystate}
        e_P=0,\quad 
        e_Q=0,
\end{align}
which indicates that the inverter regulates its active and reactive power outputs to closely track the references $P_0$ and $Q_0$, achieving zero tracking error at steady state. Hence, this operating mode is referred to as the $PQ$ mode.

\begin{remark}[Droop control in PQ mode]
\label{reamrk2} 
    To add $P$-$f$ and $Q$-$V$ primary droop control in the $PQ$ operating mode, the reference power $P_0$ and $Q_0$ can be set according to \eqref{eq:PQdroop}:
    \begin{subequations} \label{eq:PQdroop}
        \begin{align}
            P_0 &= P_\ts + D_{P-f} (\omega_0-\omega_\tM), \label{eq:PfdroopPQ}\\
            Q_0 &= Q_\ts + D_{Q-V} (V_0-V_\tM) \label{eq:QVdroopPQ},
        \end{align}
    \end{subequations}
where $P_\ts$ and $Q_\ts$ are the droop power setpoints, 
$D_{P-f}$ and $D_{Q-V}$ are the droop coefficients, $\omega_\tM$ and $V_\tM$ are the measured grid frequency and voltage magnitude at the PCC.
 \qed
\end{remark}
\subsubsection{Frequency Following and Voltage Forming - PV Mode}
By setting $\epsilon=0, \mu>0, \eta_1 \ge 0$, and $\eta_2>0$, the unified GFM-GFL control dynamics \eqref{eq:Maindynamic} become \eqref{eq:PVDynamics_i0i}:
\begin{align} \label{eq:PVDynamics_i0i}
    \dot{\bv}=&\tj \omega_{u}\bv + \mu(V_0^2-||\bv||^2)\bv+ \eta \tj (\bi_0 -\bi),
\end{align}
where the natural frequency of the oscillator \emph{follows} the PCC frequency with $\omega_\epsilon=\omega_\tPLL$, while the voltage regulation term with $\mu>0$ serves to \emph{form} the inverter voltage magnitude.
The polar dynamics \eqref{eq:hybridthetaV} of the unified control become \eqref{eq:PVmode_polarhalfpi}:
\begin{subequations} \label{eq:PVmode_polarhalfpi}
    \begin{align} 
      \dot{V}_\tm&=\mu V_\tm e_V  +\frac{\eta_1}{V_\tm}e_Q,\\
    \dot{\delta}&=\frac{\eta_2}{V_\tm^2}e_P,  
\end{align}
\end{subequations}
which implies that the voltage magnitude tracking error $e_V$, together with the reactive power tracking error $e_Q$, is used to regulate the voltage, while
the active power tracking error $e_P$ regulates the phase (or frequency) dynamics. 

The steady-state solutions of \eqref{eq:PVmode_polarhalfpi} satisfy:
    \begin{align} \label{eq:PVsteadyPQV}
     V_\tm^2=V_0^2+\eta_1 \frac{2(Q_0-Q)}{3\mu V_\tm^2}, \quad  P=P_0. 
\end{align}
This indicates that, in this operating mode, the inverter tracks the active power reference $P_0$ by regulating the phase angle difference $\delta$, while following the grid frequency $\omega_u$. Meanwhile, the voltage magnitude is formed through a nonlinear $V$-$Q$ droop mechanism based on the voltage reference $V_0$ and the reactive power tracking error $Q_0-Q$. In particular, when setting $\eta_1=0$, the steady-state voltage magnitude becomes $V_m=V_0$. Hence, this operating mode is referred to as the $PV$ mode. Similarly, to incorporate $P$-$f$ primary droop control,
the active power reference $P_0$ can be adjusted according to \eqref{eq:PfdroopPQ}.





\subsubsection{Frequency Forming and Voltage Following - Qf Mode}

By setting $\epsilon=1, \mu=0, \eta_1>0$, and $\eta_2 \ge 0$, the unified GFM-GFL control dynamics \eqref{eq:Maindynamic} become \eqref{eq:QfDynamics_i0i}:
\begin{align} \label{eq:QfDynamics_i0i}
    \dot{\bv}=&\tj \omega_0 \bv + \eta \tj (\bi_0 -\bi),
\end{align}
where the natural frequency of the oscillator becomes $\omega_\epsilon=\omega_0$ and the inverter exhibits frequency-forming capability. 
The polar dynamics \eqref{eq:hybridthetaV} of the unified control become \eqref{eq:QfthetaV_hatpi2}: 
\begin{subequations} \label{eq:QfthetaV_hatpi2}
   \begin{align}
        \dot{V}_\tm=& \frac{\eta_1}{V_\tm}e_Q,\\
          \dot{\theta}_v=& \,\omega=\omega_0 + \frac{\eta_2}{V_\tm^2}e_P.
    \end{align}
\end{subequations}


The steady-state solutions of \eqref{eq:QfthetaV_hatpi2} satisfy: 
   \begin{align} \label{eq:QfthetaV_hatpi2_steadystate}
      Q=Q_0,\quad  \omega=\omega_0 + \frac{2\eta_2}{3 V_\tm^2}(P_0-P), 
\end{align}
which indicates that, in this operating mode, the inverter tracks the reactive power reference $Q_0$ by regulating the voltage magnitude. Meanwhile, the inverter forms the terminal frequency through an $f$-$P$ droop control centered at the natural frequency $\omega_0$. In particular, when setting $\eta_2=0$, the steady-state inverter terminal frequency remains fixed at $\omega=\omega_0$. Hence, this operating mode is referred to as the $Qf$ mode. Similarly, to add $Q$-$V$ primary droop control,
the active power reference $Q_0$ can be adjusted according to \eqref{eq:QVdroopPQ}.




\subsubsection{Frequency and Voltage Forming - Vf Mode}
\label{section:VfmodeVtheta}
By setting $\epsilon = 1$, $\mu > 0$, $\eta_1 \ge 0$, and $\eta_2 \ge 0$, the unified GFM-GFL control dynamics \eqref{eq:Maindynamic} become \eqref{eq:VfDynamics_i0i}:
\begin{align} \label{eq:VfDynamics_i0i}
    \dot{\bv} = \tj \omega_0 \bv 
    + \mu (V_0^2 - \|\bv\|^2) \bv 
    + \eta \tj (\bi_0 - \bi),
\end{align}
which is exactly the AHO-based dVOC \eqref{eq:dispatchableAHO} with $\phi=\pi/2$. Therefore, this operating mode ($Vf$ mode) fully preserves the grid-forming dynamic behavior of dVOC, in which both the frequency and voltage dynamics are established by the oscillator with self-synchronization capability. 
The polar dynamics \eqref{eq:hybridthetaV} of the unified GFM-GFL control become \eqref{eq:VFmodepolar}:
\begin{subequations}\label{eq:VFmodepolar}
    \begin{align}
        \dot{V}_\tm&=\mu V_\tm(V_0^2-V_\tm^2)+\frac{2\eta_1}{3V_\tm}(Q_0-Q), \\
\dot{\theta}_v&=\omega=\omega_0+\frac{2\eta_2}{3V_\tm^2}(P_0-P),
    \end{align}
\end{subequations}
with the steady-state solutions satisfying
\eqref{eq:PfQVdroopVfmode}:
\begin{subequations} \label{eq:PfQVdroopVfmode}
    \begin{align}
        V_\tm^2&=V_0^2+\eta_1 \frac{2(Q_0-Q)}{3\mu V_\tm^2}, \label{eq:VQdroopVfmode}\\
        \omega&=\omega_0+\eta_2 \frac{2(P_0-P)}{3V_\tm^2},\label{eq:fPdroopVfmode}
    \end{align}
\end{subequations}
which exhibit nonlinear $V$-$Q$ and $f$-$P$ droop control characteristics. Here, we split the feedback gain $\eta$ in \eqref{eq:Vthteadynamicphipi2} and \eqref{eq:dVoCPfQVdroop} into two independent parameters: $\eta_1$ for $V$-$Q$ control and $\eta_2$ for $f$-$P$ control, offering greater control flexibility than the standard dVOC formulation. For example, when setting $\eta_1=0$, the steady-state voltage magnitude satisfies $V_\mathrm{m} = V_0$, indicating a constant voltage regulation without $V$-$Q$ droop control. When setting $\eta_2=0$, the inverter terminal frequency becomes $\omega = \omega_0$, indicating a  
constant frequency regulation without $f$-$P$ droop.

\subsubsection{Hybrid Mode}
\label{section:hybridmode}

By setting $\epsilon \in (0,1)$, $\mu \geq 0$, $\eta_1 \ge 0$, and $\eta_2 \ge 0$, the inverter operates in a hybrid mode 
that exhibits both grid-following and grid-forming dynamic behaviors in the unified control formulation \eqref{eq:Maindynamic}. In this operating mode, the natural frequency of the oscillator is $\omega_\epsilon = (1-\epsilon)\omega_{{u}} + \epsilon \omega_0$, which is a convex combination of the measured grid frequency $\omega_{{u}}$ and the internal reference frequency $\omega_0$. 

From the polar dynamics \eqref{eq:hybridthetaV} with $\phi=\pi/2$, 
the voltage magnitude is formed according to the voltage reference $V_0$ and a nonlinear $V$-$Q$ droop feedback \eqref{eq:VQdroopVfmode}, which is 
the same as dVOC \eqref{eq:Vdynamicphipo2}.  The phase (or frequency) dynamics \eqref{eq:thetadyna_hybridthetaV} can be reformulated as \eqref{eq:reformufreq}:
\begin{align} \label{eq:reformufreq}
    (1-\epsilon)(\omega-\omega_\tPLL) + \epsilon (\omega - \omega_0) = \frac{2\eta_2}{3 V_\tm^2}(P_0-P).
\end{align}

\begin{figure}[t]
    \centering
    \includegraphics[width=0.8\linewidth]{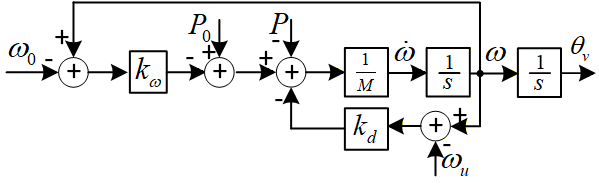}
    \caption{Equivalent frequency/phase control loop of the proposed unified GFM-GFL controller in the hybrid mode, exhibiting the same structure as a virtual synchronous machine \cite{d2015virtualSM}.}
    \label{fig:VSM}
\end{figure}

When applying a low-pass filter $\frac{1}{1+sT_P}$ to the power-tracking error $P_0-P$ and neglecting $\dot{\omega}_\tPLL$, the frequency dynamics \eqref{eq:reformufreq} become \eqref{eq:reforVSM}:
\begin{align}\label{eq:reforVSM}
    M\dot{\omega}=  P_0-P-k_d (\omega - \omega_\tPLL) - k_\omega (\omega-\omega_0) ,
\end{align}
which is exactly the virtual synchronous machine control introduced in \cite{d2015virtualSM} with $M:= \frac{3V_\tm^2}{2\eta_2}T_P$, $k_d:=\frac{3(1-\epsilon)V_m^2}{2\eta_2}$ and $k_\omega := \frac{3\epsilon V_\tm^2}{2\eta_2}$.
The corresponding control block diagram is shown in Figure~\ref{fig:VSM}. This implies that the inverter frequency dynamics are partially synchronized to the grid frequency $\omega_u$ and partially formed by the internal oscillator frequency $\omega_0$, depending on the parameter $\epsilon$. In addition, in the steady states when $\dot{\delta}=0$ (or $\omega=\omega_u$), the inverter frequency satisfies:
\begin{align}
   \epsilon (\omega - \omega_0) &=  \frac{2\eta_2}{3V_\mathrm{m}^2} (P_0 - P). \label{eq:hybridfPdroop}
\end{align}
This indicates that decreasing $\epsilon$ from 1 to 0 enables a smooth transition from frequency-forming control with an $f$-$P$ droop mechanism to active-power-following control.

As a result, the hybrid operating mode blends the $Vf$ mode, i.e., dVOC, with the $PQ$ mode when $\mu=0$ (or the $PV$ mode when $\mu>0$), depending on the choice of $\epsilon$. As $\epsilon$ decreases from 1 to 0, the inverter smoothly transitions from a frequency-forming source to a frequency-following unit. 


In addition, the implementations of impedance-mode control and virtual impedance control in the proposed unified GFM-GFL control framework are discussed in Appendix \ref{Appendix:Impedance}.

\subsection{Pre-Synchronization and Current Limiting Strategies} \label{sec:presyn}

\subsubsection{Pre-Synchronization}
Before connecting to the grid, pre-synchronization is required to align the inverter terminal voltage with the grid-side voltage to reduce switching transients and avoid large inrush currents during connection.  
To achieve pre-synchronization, $\gamma$ can be set to a sufficiently large value while enforcing $\tilde{\bv}=\bu$ and setting $\epsilon=0$, $\mu=0$, and $\eta=0$. Under these settings, the proposed unified control dynamics  \eqref{eq:hybridDynamics} reduce to \eqref{eq:presyn}:
\begin{align} \label{eq:presyn}
\dot{\bv}= \tj \omega_u \bv + \gamma (\bu-\bv).
\end{align}
For a given grid-side voltage oscillation vector $\bu$ with frequency $\omega_u$, the solution to \eqref{eq:presyn} is given by:
 \begin{equation}
    \label{eq:presolution}
\bv(t)=\bu(t) + \big(\bv(0)-\bu(0)\big) e^{-\gamma t + \tj \omega_u t},
\end{equation} 
where $\bv(0)$ and $\bu(0)$ are the initial values of $\bv$ and $\bu$. Equation \eqref{eq:presolution} shows that the inverter voltage vector converges exponentially to the grid voltage vector at a convergence rate of $\gamma$.
As a result, the inrush current during grid connection can be effectively avoided. Immediately after the inverter is connected to the grid, its power output remains zero. Then, the parameter $\gamma$ can be reduced to zero, allowing the inverter to transition to the prescribed operating mode determined by the selected values of $\epsilon$, $\mu$, and $\eta$.


\subsubsection{Current Limiting}

For the proposed unified GFM-GFL inverter control framework, current limiting can be implemented through several strategies at different control layers, as discussed in~\cite{10603443}. A direct current-limiting approach is applied at the switching level, where the PWM modulation is replaced by a hysteresis current controller once the current exceeds the predefined limit. Alternatively, the reference power $P_0+\tj Q_0$ or the reference current $\bm{i}_0$ can also be constrained. This enables accurate and fast current limiting in the $PQ$ mode. Nevertheless, in other operating modes, such as grid-forming operation, these quantities serve as droop-based reference power or current signals rather than strict tracking references, making precise current limiting difficult to achieve. In such cases, the virtual impedance control introduced in Appendix~\ref{AppendixB_B:VirtualImpedance} can be employed for current limiting; see \cite{10603443} for more details.

\section{Stability Analysis}\label{section:IV}

In this section, we analyze the small-signal stability of the proposed unified GFM-GFL inverter control method, study the impacts of the control parameters $\mu$ and $\eta$ as well as the series resistance $R_\tf$, and then discuss the input-output frequency-domain characteristics.





\subsection{Small-Signal Linearized Model for Infinite-Bus Connection}

Consider an individual inverter connected to an infinite bus system with frequency $\omega_u $ and voltage $U_\tm$. 
 The unified GFM-GFL control dynamics \eqref{eq:Vthetaidiq} can be reformulated in the local grid-side $d_gq_g$ frame as \eqref{eq:SmallsignaldeltaV}:
\begin{subequations}
\label{eq:SmallsignaldeltaV}
    \begin{align}
    \hspace{-5pt} \dot{V}_\tm&= \mu V_\tm(V_{0}^2\!-\!V_\tm^2) + \eta_{1} (\frac{2 Q_{0}}{3V_\tm}\!-\!i_{d_g} \sin \delta \!+\! i_{q_g} \cos \delta), \label{eq:Vidgiqg}\\
    \dot{\delta}&=  \epsilon (\omega_0 - \omega_u)+  \frac{\eta_{2}}{V_\tm} (\frac{2P_{0}}{3V_\tm} - i_{d_g} \cos \delta  - i_{q_g} \sin \delta). 
\end{align}
\end{subequations}

The inverter current dynamics in the $\alpha\beta$ frame are \eqref{eq:linealphabeta}:
\begin{align}\label{eq:linealphabeta}
    \dot{\bi} = - \frac{R_{\tf}}{L_{\tf}} \bi + \frac{\bv - \bu}{L_{\tf}}.
\end{align}
Using the Park transformation \eqref{eq:dqtrans}, dynamics \eqref{eq:linealphabeta} can be transformed into the local grid-side $d_g q_g$ frame as \eqref{eq:RLfdynamicinlocalbusframe}:
\begin{align} \label{eq:RLfdynamicinlocalbusframe}
\begin{bmatrix}
    {\dot i}_{d_g} \\ {\dot i}_{q_g}
\end{bmatrix}
 = \begin{bmatrix}
      - \frac{R_{\rm{f}}}{L_{\rm{f}}} & \omega_u \\ - {\omega_u} & - \frac{R_{\rm{f}}}{L_{\rm{f}}}
 \end{bmatrix} 
\begin{bmatrix}
    {i}_{d_g} \\ {i}_{q_g}
\end{bmatrix} + \frac{1}{L_{\rm{f}}} \begin{bmatrix}
    {V_\tm}\cos {\delta} - {U_\tm} \\ {V_\tm}\sin {\delta}
\end{bmatrix}.
\end{align}


For the inverter connected to an infinite-bus system, we define the state variables as $\bx=[\delta, V_\tm, i_{d_g}, i_{q_g}]^\top$, the control inputs as $\bm{u}=[U_\tm, \omega_u]^\top$, and the output variables as $\bm{y}=[i_{d_g}, i_{q_g}]^\top$.
Combining \eqref{eq:RLfdynamicinlocalbusframe} with \eqref{eq:SmallsignaldeltaV} yields the fourth-order nonlinear state-space dynamic model \eqref{eq:generalnonlineardynamic}:
\begin{subequations} \label{eq:generalnonlineardynamic}
    \begin{align}
    \dot{\bm{x}}&=\bm{f}(\bm{x}, \bm{u}), \\
    \dot{\bm{y}}&=C \bm{x},
    \end{align}
\end{subequations}
where
$C \coloneqq
\begin{bmatrix}
0 & 0 & 1 & 0\\
0 & 0 & 0 & 1
\end{bmatrix}$.
Based on a given operation point $\bm{x}^*=[\delta^*, V_\tm^*, i_{d_g}^*, i_{q_g}^*]^\top$ and $\bu^*=[U_\tm^*, \omega_u^*]^\top$, one can linearize \eqref{eq:generalnonlineardynamic} and obtain the linearized dynamic model \eqref{eq:DynamicsgeneralAxbc}:
\begin{subequations} \label{eq:DynamicsgeneralAxbc}
    \begin{align}
        \dot{\bm{x}}&= A \bx + B \bu,\\
        \by &=C \bx.
    \end{align}
\end{subequations}
As formulated in \eqref{eq:AandB}, $A = \left.\frac{\partial \bm{f}}{\partial \bx}\right|_{(\bx^*,\bu^*)}$ and $B = \left.\frac{\partial \bm{f}}{\partial \bu}\right|_{(\bx^*,\bu^*)}$ denote the Jacobian matrices evaluated at the equilibrium point $(\bx^*,\bu^*)$. In \eqref{eq:AandB}, $P^*$ and $Q^*$ denote the steady-state inverter output active and reactive powers, respectively, given by:
\begin{subequations}
    \begin{align}
        P^*&=\frac{3}{2}V_\tm^* i_d^*= \frac{3}{2}V_\tm^* (i_{d_g}^* \cos \delta^*+i_{q_g}^* \sin \delta^*), \\
        Q^*&=-\frac{3}{2}V_\tm^* i_q^* =\frac{3}{2}V_\tm^* (i_{d_g}^* \sin \delta^*-i_{q_g}^* \cos \delta^*). 
    \end{align}
\end{subequations}

\begin{figure*}[t]
\centering
\begin{equation}
\label{eq:AandB}
\begin{aligned}
A &=
\begin{bmatrix}
\frac{2\eta_2 Q^*}{3 (V_\tm^*)^2}
& \frac{2\eta_2 (P^* - 2P_0)}{3 (V_\tm^*)^3}
& - \frac{\eta_2}{V_\tm^*}\cos\delta^*
& - \frac{\eta_2}{V_\tm^*}\sin\delta^*
\\[6pt]
\frac{2\eta_1 P^*}{3 V_\tm^*}
& \mu (V_0^2 - 3 (V_\tm^*)^2)
- \frac{2\eta_1 Q_0}{3 (V_\tm^*)^2}
& - \eta_1 \sin\delta^*
& \eta_1 \cos\delta^*
\\[6pt]
- \frac{V_\tm^* \sin\delta^*}{L_{\rm f}}
& \frac{\cos\delta^*}{L_{\rm f}}
& - \frac{R_{\rm f}}{L_{\rm f}}
& \omega_{u}^*
\\[6pt]
\frac{V_\tm^* \cos\delta^*}{L_{\rm f}}
& \frac{\sin\delta^*}{L_{\rm f}}
& - \omega_{u}^*
& - \frac{R_{\rm f}}{L_{\rm f}}
\end{bmatrix},
\qquad
B =
\begin{bmatrix}
0 & -\varepsilon\\
0 & 0\\
-\dfrac{1}{L_{\rm f}} & i_{q_g}^*\\
0 & - i_{d_g}^*
\end{bmatrix}.
\end{aligned}
\end{equation}
\end{figure*}

The reference active and reactive powers are set to $P_0=0.333$ pu and $Q_0=0.267$ pu, respectively. The reference voltage is $V_0=1.0138$ pu, and the reference frequency is $\omega_0=120\pi$ rad/s. The inverter filter parameters are selected as $R_\tf=0.01$ pu and $L_\tf={0.04}$ pu (i.e., $X_\tf=0.04$ pu). The grid-side voltage and frequency are set to $U_\tm^*=1$ pu and $\omega_u^*=120\pi$ rad/s, respectively. 
With these settings, the operating point coincides with the reference point of the $V$-$Q$ and $f$-$P$ droop characteristics, i.e., $V_\tm^*=V_0$ and $\omega^*=\omega_0=\omega_u^*$. Therefore, the equilibrium point is independent of the control parameters $\epsilon$, $\eta_1$, $\eta_2$, and $\mu$. The resulting equilibrium point is presented in Table~\ref{tab:equilibrium}. When the resistance $R_\tf$ changes, the equilibrium point is obtained by solving the steady-state equations under the same prescribed references.

\begin{table}[ht]
    \centering
     \caption{Equilibrium of state, control, and computed values}
    \begin{tabular}{c|c|c|c|c|c|c|c} 
    \hline
     \multicolumn{4}{c|}{State} & \multicolumn{2}{c|}{Control} & \multicolumn{2}{c}{Computed}\\
     \hline
        $\delta^*$ & $V_\tm^*$ & $i_{d_g}^* $  & $i_{q_g}^* $ &$U_\tm^*$ & $\omega_u^*$ & $P^*$ & $Q^*$\\
        \hline
        0.0105 & 1.0138 & 0.3316& -0.2596 & 1& $120\pi$ &0.333 &0.267 \\
        \hline
    \end{tabular}
    \label{tab:equilibrium}
\end{table}

\subsection{Small-Signal Stability Results}


Based on the linearized model \eqref{eq:DynamicsgeneralAxbc}, we assess small-signal stability by analyzing the trajectories of the eigenvalues (or modes) in the complex plane as the controller parameters vary.

In addition, participation factors \cite{PerezArriaga1982} are used to quantify the contribution of each state variable to different eigenmodes. The participation factor $\chi_{ki}$ of the $k$th state variable in the $i$th mode is defined as \eqref{eq:parfact}:
\begin{equation}\label{eq:parfact}
\chi_{ki} = \ell_{ik} r_{ki},
\end{equation}
where $r_{ki}$ and $\ell_{ik}$ denote the $k$th entries of the right and left eigenvectors associated with the eigenvalue $\lambda_i$, respectively. 
The participation factor $\chi_{ki}$ quantifies the extent to which the state variable $x_k$ contributes to, and is affected by, the dynamic mode associated with $\lambda_i$. A large value of $|\chi_{ki}|$ indicates that $x_k$ plays a significant role in the corresponding mode, whereas a small value indicates weak participation.



\subsubsection{From Voltage Following to Voltage Forming (Increasing \texorpdfstring{$\mu$}{mu})}

Fixing $\eta_1=\eta_2=1$, Table~\ref{tab:Eignevalueoperationmode} presents four sets of eigenvalues corresponding to four different operating modes ($PQ,Qf,PV,Vf$). Since $\omega_u$ is a control variable and $\epsilon$ does not appear in the state matrix $A$, the frequency-forming modes ($\epsilon=1$) and frequency-following modes ($\epsilon=0$) share the same state matrix $A$ and thus exhibit identical eigenvalues. 
The modes near 60~Hz (i.e., imaginary parts are close to $\pm 377$~rad/s) are mainly associated with the grid-current states $i_{d_g}$ and $i_{q_g}$. In the voltage-following mode ($PQ,Qf$), a pair of eigenvalues with imaginary parts of approximately $\pm 4.56$~rad/s corresponds to a 0.73~Hz subsynchronous transient. This subsynchronous dynamic is eliminated in the 
voltage-forming modes ($PV,Vf$).

\begin{table}[ht]
    \centering
     \caption{Eigenvalues under different operating modes}
    \begin{tabular}{c|c|c|c|c}
      \hline
     \multicolumn{2}{c|}{Mode ($\epsilon$, $\mu$)} & \multicolumn{3}{c}{Four Eigenvalues ($\lambda_1,\lambda_2,\lambda_3,\lambda_4$)}   \\
      \hline
       $PQ$ (0, 0)  & $Qf$ (1,0) & $-69.80 \pm$ $\tj 372.41$ & \multicolumn{2}{c}{$-24.45 \pm \tj 4.56$}   \\
       \hline
       $PV$ (0, 30) & $Vf$ (1,30) & $-69.51 \pm \tj 374.46$ & $-24.23 $ & $-86.91$\\
       \hline
    \end{tabular}
    \label{tab:Eignevalueoperationmode}
\end{table}

\begin{figure}[ht]
    \centering
\includegraphics[width=0.9\linewidth]{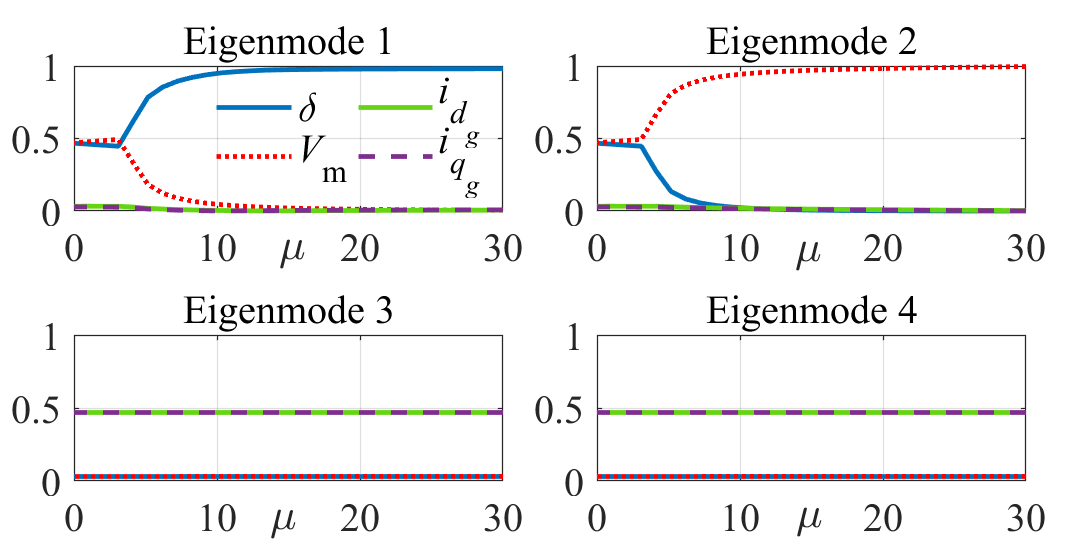}
    \caption{Participation factors under varying voltage feedback gain $\mu$.}
    \label{fig:PFmu030}
\end{figure}

\begin{figure}[ht]
    \centering
    \includegraphics[width=1\linewidth]{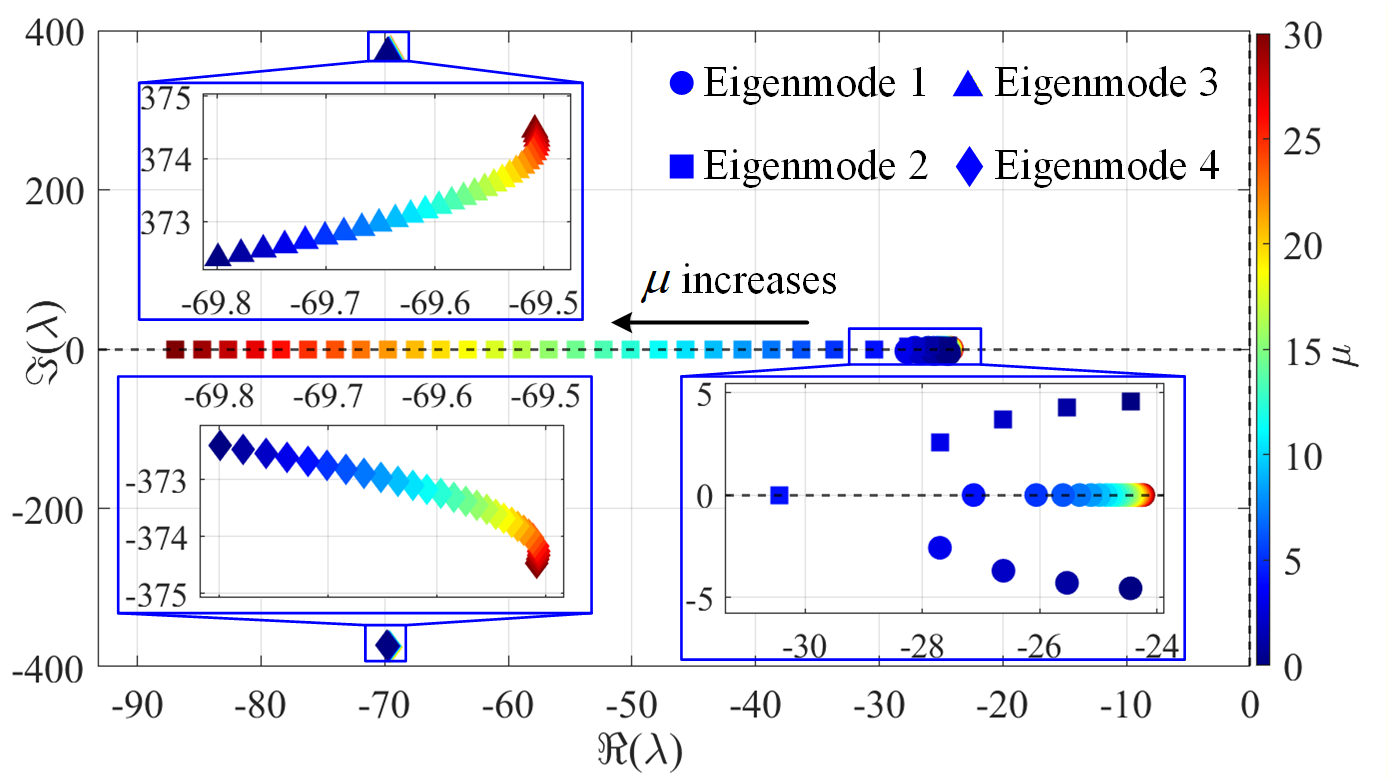}
    \caption{Eigenmode trajectories under varying voltage feedback gain $\mu$.}
    \label{fig:Mu0to30}
\end{figure}

Under varying $\mu$,
the participation factors of each state variable for different eigenmodes are shown in Figure~\ref{fig:PFmu030}. 
The eigenmode trajectories with varying $\mu$ are illustrated in Figure~\ref{fig:Mu0to30}. As $\mu$ increases, the voltage-forming capability is strengthened, and the inverter exhibits greater voltage-magnitude regulation, corresponding to enhanced damping in the $V$-$Q$ droop dynamics. Consequently, the real eigenvalue of eigenmode 2, which is mainly associated with the voltage magnitude $V_\tm$ (see Figure~\ref{fig:PFmu030}), significantly shifts further to the left in the complex plane, as shown in Figure~\ref{fig:Mu0to30}, indicating faster voltage convergence. In contrast, other eigenmodes mainly associated with the phase angle $\delta$ and the filter-current states $i_{d_g}$ and $i_{q_g}$ remain nearly unchanged.

\subsubsection{Enhanced Power Tracking (Increasing \texorpdfstring{$\eta$}{eta})}
\label{section:etaincreasebeforeinductor}

\begin{figure}[!t]
    \centering
    \includegraphics[width=0.9\linewidth]{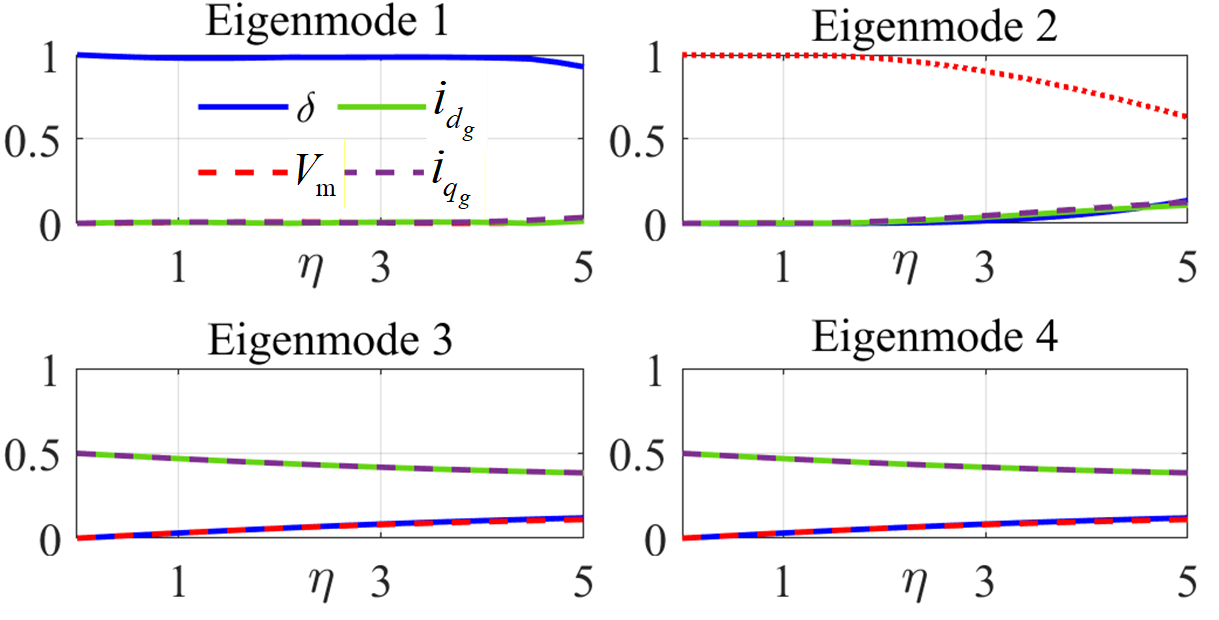}
    \caption{Participation factors under varying current feedback gain $\eta$.}
    \label{fig:PFeta0to5}
\end{figure}

\begin{figure}[!t]
    \centering
    \includegraphics[width=1\linewidth]{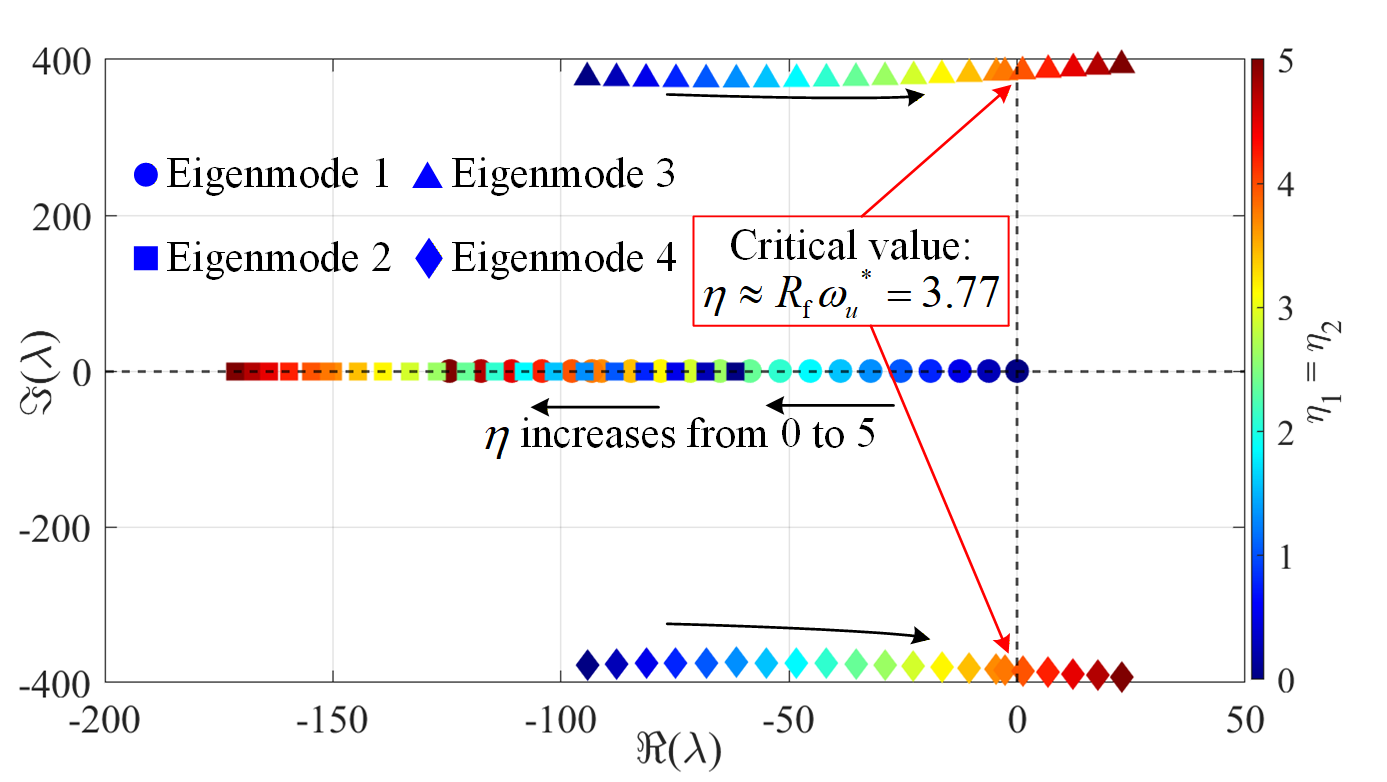}
    \caption{Eigenmode trajectories under varying current feedback gain $\eta$ and when $R_\tf=0.01$ pu.}
    \label{fig:Etaincrease}
\end{figure}

To investigate the impact of the current feedback gain, we evaluate the system trajectories by varying $\eta=\eta_1=\eta_2$ from $0$ to $5$ while holding $\mu=0$,  the participation factors and the eigenvalue trajectories are illustrated in Figure~\ref{fig:PFeta0to5} and Figure~\ref{fig:Etaincrease}.

When the current-feedback gain $\eta$ is very small, the phase angle $\delta$ is mainly associated with eigenmode 1, as shown in Figure~\ref{fig:PFeta0to5}. This mode is a slow eigenmode with a real part close to zero, as observed in Figure~\ref{fig:Etaincrease}. This behavior occurs because a small $\eta$ provides only weak feedback to the phase dynamics. In the extreme case of $\eta=0$, no control objective regulates $\delta$, and the corresponding eigenmode 1 is located at the origin.
As $\eta$ increases, the coupling between the inverter voltage dynamics and current dynamics becomes stronger. 
Consequently, eigenmodes 1 and 2, which are dominated by the phase angle $\delta$ and voltage magnitude $V_\tm$, respectively, move further to the left in the complex plane, indicating improved stability and faster dynamic response. 
Meanwhile, the oscillatory eigenmodes with large imaginary parts (i.e., eigenmodes 3 and 4) remain dominated by the filter current states $i_{d_g}$ and $i_{q_g}$, as confirmed by the participation factors in Figure~\ref{fig:PFeta0to5}. When $\eta$ exceeds a certain threshold, a complex-conjugate pair of eigenvalues crosses the imaginary axis, resulting in small-signal instability, as shown in Figure~\ref{fig:Etaincrease}.

To identify the physical mechanism driving this instability and determine the critical gain $\eta_\text{crit}$, we analyze the interaction between the voltage-frequency loop and the current dynamics. The small-signal filter current dynamics in the local $d_g q_g$ frame can be derived from \eqref{eq:RLfdynamicinlocalbusframe} and compactly expressed in complex form as \eqref{eq:current_dyn_complex}: 
\begin{equation} 
L_\tf \Delta \dot{\bi}
= -R_\tf \Delta \bi
- \mathrm{j}\omega_u^* L_\tf \Delta \bi
+ \Delta {\bv},
\label{eq:current_dyn_complex}
\end{equation}
where $\Delta \bi = \Delta i_{d_g} + \mathrm{j}\Delta i_{q_g}$ and
$\Delta {\bv} = \Delta v_{d_g} + \mathrm{j}\Delta v_{q_g}$ denote the small-signal current and voltage, respectively.
Around the operating point $\delta^* \approx 0$, the voltage perturbation can be approximated as \eqref{eq:v_complex}:
\begin{equation}
\Delta {\bv}
\approx \Delta V_\tm + \mathrm{j}V_\tm^*\Delta\delta.
\label{eq:v_complex}
\end{equation}

\begin{figure}
    \centering
    \includegraphics[width=1\linewidth]{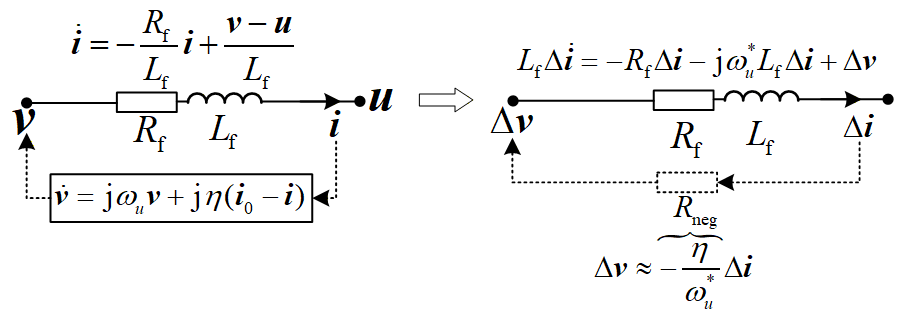}
    \caption{Control dynamics in the $\alpha \beta$ frame and the corresponding small-signal linearized model in the $d_g q_g$ frame, with $\bi_0$ treated as a constant during linearization.}
    \label{fig:virtualresistor}
\end{figure}

Under $\mu=0$ and $\omega_u^*=\omega_0$, the polar dynamics \eqref{eq:SmallsignaldeltaV} yield:
\begin{equation}
\Delta\dot{V}_{\tm} \approx \eta\,\Delta i_{q_g},
\qquad
\Delta\dot{\delta} \approx -\frac{\eta}{V_\tm^*}\Delta i_{d_g}.
\label{eq:outer_loop_lin}
\end{equation}
Transforming \eqref{eq:outer_loop_lin} into the frequency domain gives:
\begin{equation}
\Delta V_\tm \approx \eta\frac{1}{s}\Delta i_{q_g},
\qquad
\Delta\delta \approx -\frac{\eta}{V_\tm^*}\frac{1}{s}\Delta i_{d_g}.
\label{eq:outer_loop_freq}
\end{equation}
Substituting \eqref{eq:outer_loop_freq} into \eqref{eq:v_complex} yields
\begin{equation}
\Delta {\bv}
\approx - \tj \frac{\eta}{s}\Delta \bi.
\label{eq:neg_resistance_general}
\end{equation}
For the current-dominated oscillatory mode, the critical condition occurs near
$s \approx \tj \omega_u^*$. Hence, it leads to
\begin{equation}
\Delta {\bv}
\approx -\frac{\eta}{\omega_u^*}\Delta \bi.
\label{eq:neg_resistance}
\end{equation}
\begin{remark}[Physical interpretation of stability boundary under high-gain current feedback]
Equation \eqref{eq:neg_resistance} indicates that at frequency $\omega_u^*$, the current feedback behaves like a virtual negative resistance injected into the system:
\begin{equation}
R_{\mathrm{neg}} = -\frac{\eta}{\omega_u^*},
\end{equation}
Consequently, the total effective damping resistance seen by the filter current becomes:
\begin{equation}
R_{\mathrm{eff}} = R_{\mathrm{f}} - \frac{\eta}{\omega_u^*},
\end{equation}
Small-signal instability occurs when this virtual negative resistance completely cancels out the physical filter damping ${R_\mathrm{eff}} < 0$.
This yields the critical current feedback gain \eqref{eq:critaleta} that determines the small-signal stability boundary:
\begin{equation}
\eta_{\mathrm{crit}} \approx R_{\mathrm{f}} \omega_u^*.
\label{eq:critaleta}
\end{equation}
\end{remark}
Figure \ref{fig:virtualresistor} illustrates the control dynamics ($\mu = 0$) in the $\alpha \beta$ frame and the corresponding small-signal linearized model in the $d_g q_g$ frame, where $\bi_0$ is assumed approximately constant during linearization. The small-signal model explicitly reveals the equivalent negative resistance introduced by the current-feedback loop, providing further physical insight into and validation of the derived stability boundary.


This stability boundary result also explains the eigenvalue migration observed in Figure~\ref{fig:Etaincrease}, i.e.:
\begin{equation}
    \eta_\mathrm{crit}=R_\tf \omega_u^*=0.01 \times 377=3.77.
\end{equation}

\begin{figure}
    \centering
    \includegraphics[width=1\linewidth]{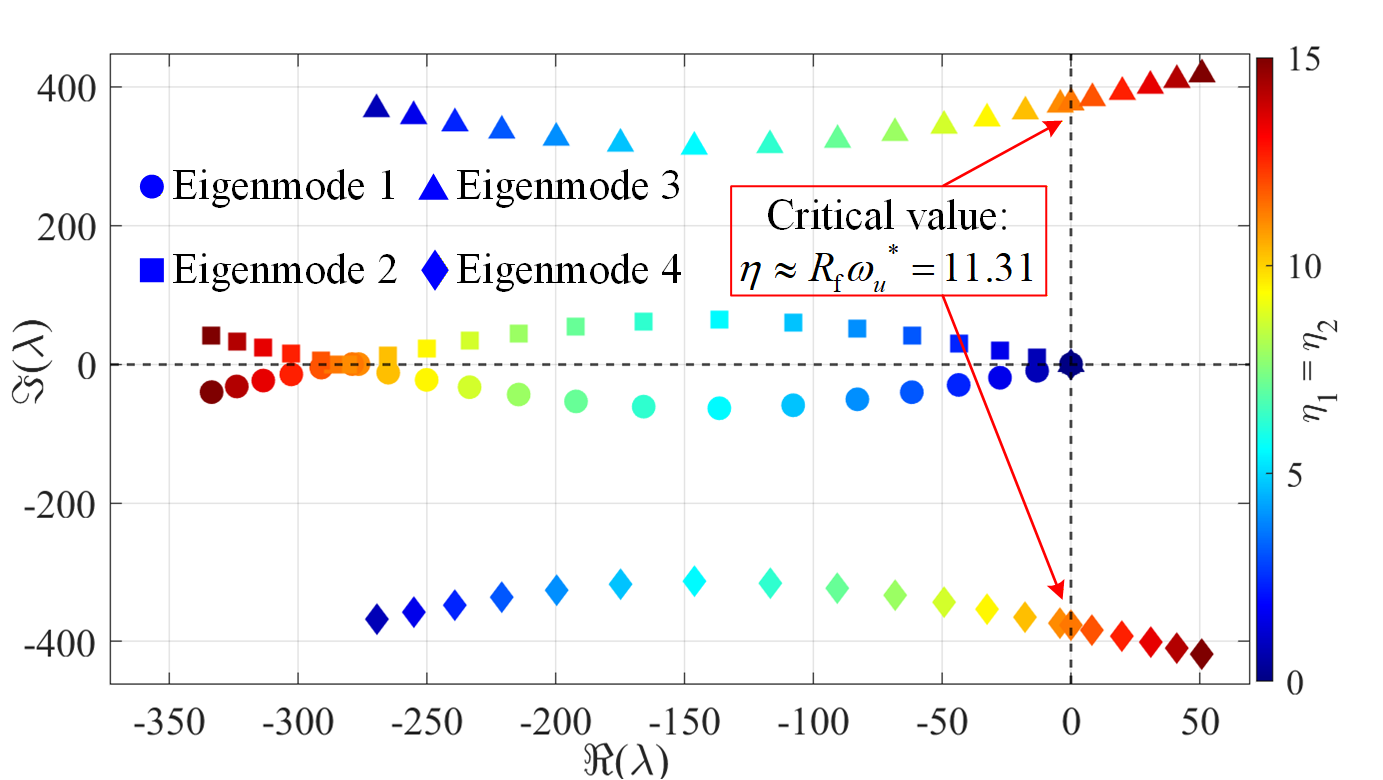}
    \caption{Eigenmodes trajectories under varying current feedback gain $\eta$ and when $R_\tf=0.03 \text{ pu}$.}
    \label{fig:R003}
\end{figure}

To further validate this critical value, $R_\tf$ is increased to 0.03 pu, the eigenmode trajectories varying with $\eta$ from 0 to 15 are shown in Figure \ref{fig:R003}, the critical value becomes:
\begin{equation}
    \eta_\mathrm{crit}=R_\tf \omega_u^*=0.03 \times 377= 11.31.
\end{equation}
It further validates the derived stability boundary criterion \eqref{eq:critaleta}.

\subsubsection{Effect of LC Filter Series Resistance (Increasing \texorpdfstring{$R_\tf$}{Rf})} 

\begin{figure}[hb]
    \centering
    \includegraphics[width=1\linewidth]{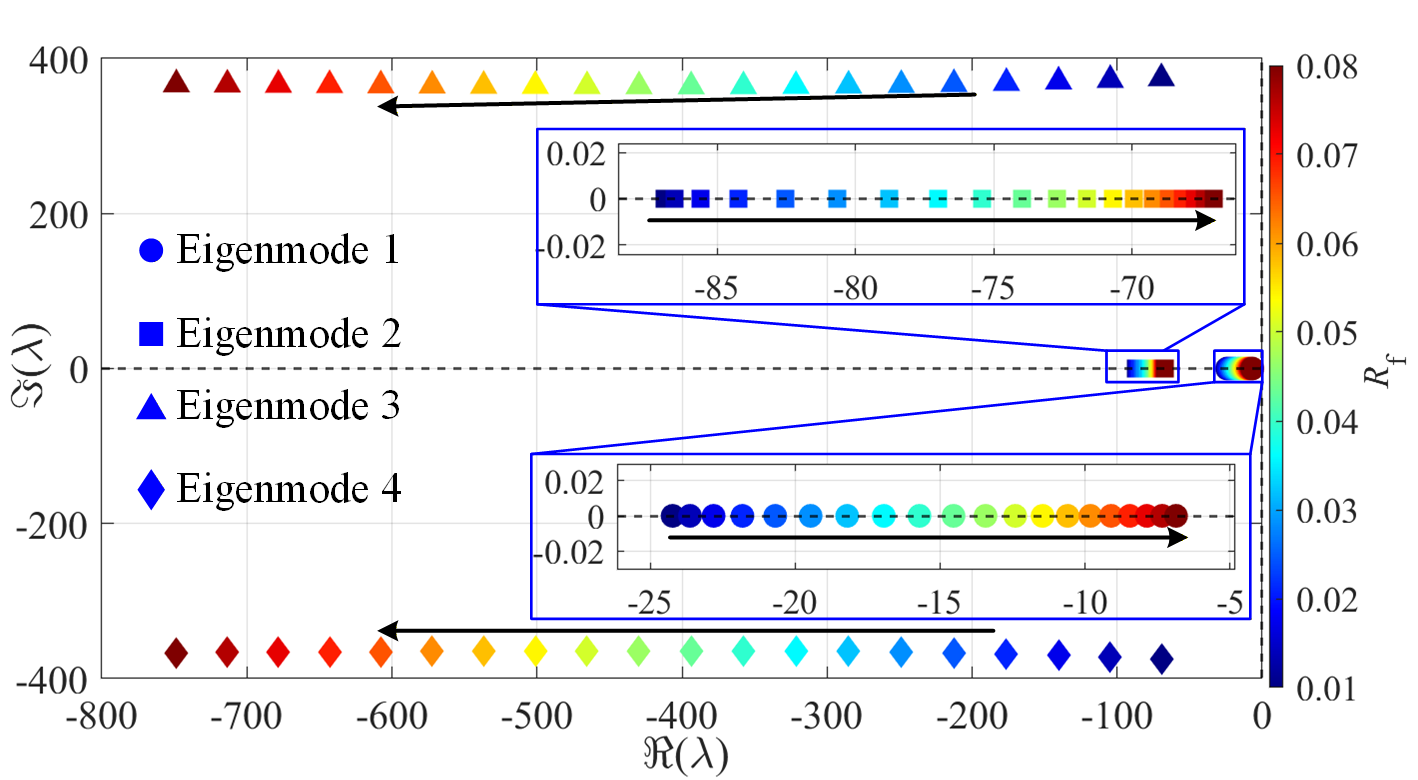}
    \caption{Eigenmode trajectories under varying LC filter resistance $R_\tf$.}
    \label{fig:Rf001008}
\end{figure}

\begin{figure}[hb]
    \centering
    \includegraphics[width=0.9\linewidth]{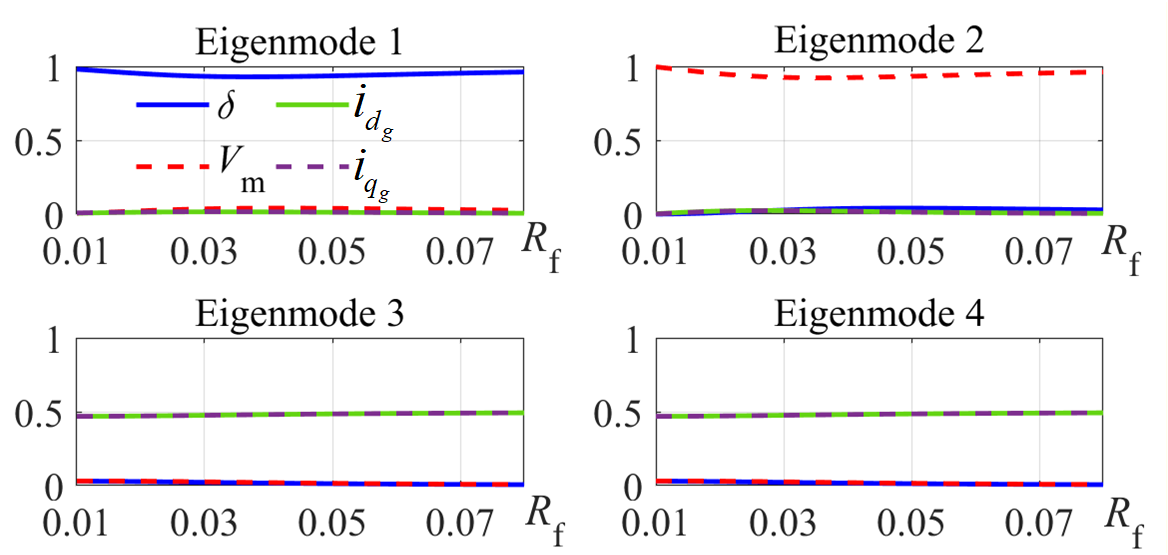}
    \caption{Participation factors under varying LC filter resistance $R_\tf$.}
    \label{fig:PFRf}
\end{figure}

To investigate the impact of filter damping on system stability, the LC filter series resistance $R_\tf$ is increased while all other parameters are kept unchanged with $\eta_1=\eta_2=1, \mu=30, $ and $X_\tf=0.04$ pu. 
For each value of $R_\tf$, the steady-state operating point is recomputed, and the system is linearized around the corresponding equilibrium. The resulting eigenvalue trajectories are illustrated in Figure~\ref{fig:Rf001008}, 
and the corresponding participation factors are shown in Figure~\ref{fig:PFRf}.

 
As $R_\tf$ increases, the oscillatory eigenmodes 3 and 4 move monotonically leftward in the complex plane, indicating enhanced damping due to the increasing line resistance. 
In contrast, eigenmodes 1 and 2, which are mainly associated with $\delta$ and $V_\tm$, exhibit only minor variations and remain well separated from the imaginary axis over the entire range of $R_\tf$.

\subsection{Input-Output Frequency-Domain Characteristics}
\label{SectionIVE:FrequencyDomain}

\begin{figure*}[ht]
    \centering
    \includegraphics[width=1\linewidth]{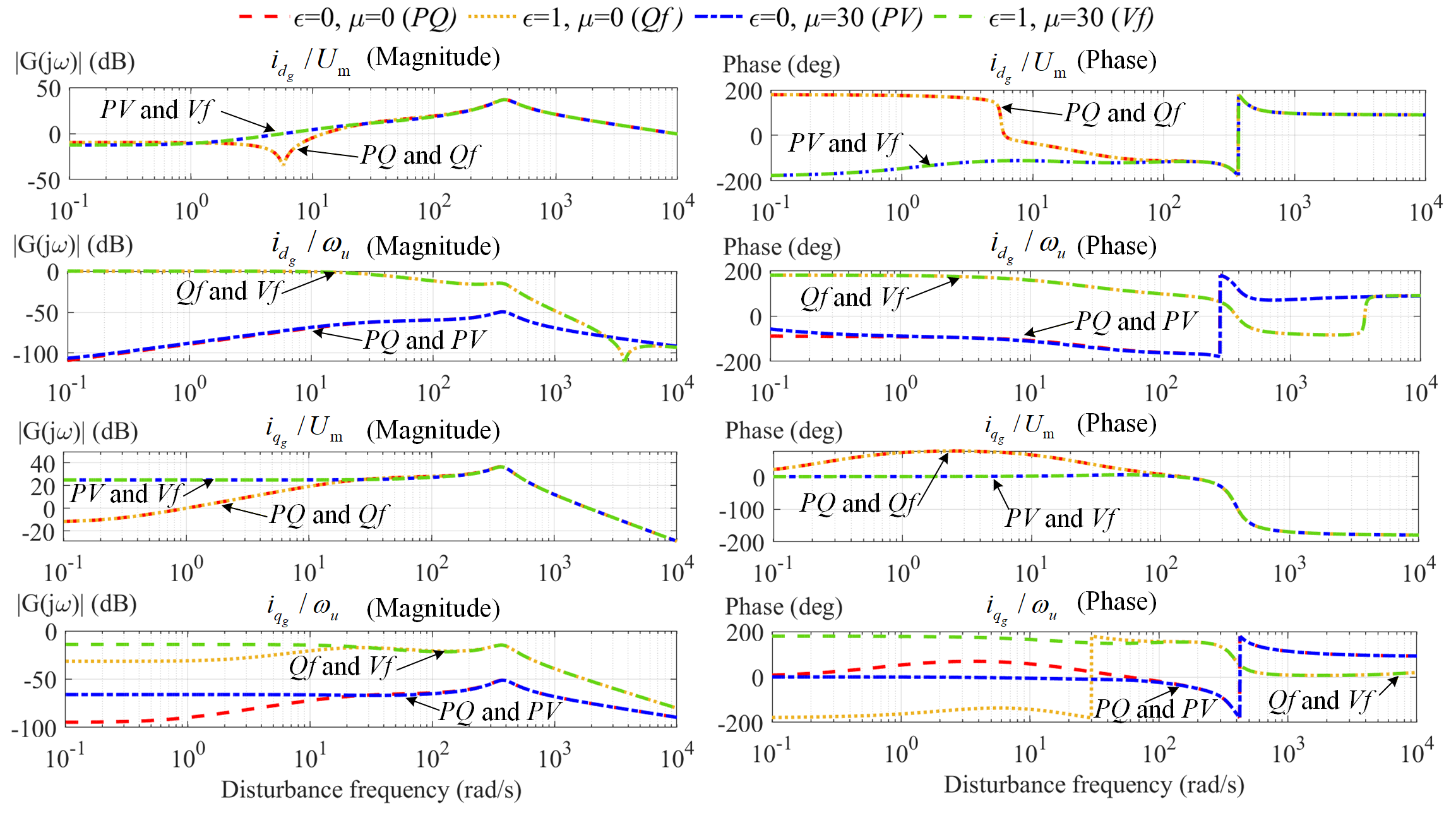}
    \caption{Input-output frequency-domain response. }
    \label{fig:Bode_Gyu}
\end{figure*}

Taking the Laplace transform of the small-signal linearized model \eqref{eq:DynamicsgeneralAxbc} under zero initial conditions yields $\bm{Y}(s)=C(sI-A)^{-1}B\,\bm{U}(s)$. Then, the input-output transfer function matrix from $\bm{u}=[U_\tm,\ \omega_u]^\top$ to $\bm{y}=[i_{d_g},\ i_{q_g}]^\top$ is given by \eqref{eq:Gyu_matrix}:
\begin{equation}  
\bm{G}(s)=C(sI\!-\!A)^{-1}B = \!\begin{bmatrix} \dfrac{i_{d_g}(s)}{U_\tm(s)} & \dfrac{i_{d_g}(s)}{\omega_u(s)}\\[6pt] \dfrac{i_{q_g}(s)}{U_\tm(s)} & \dfrac{i_{q_g}(s)}{\omega_u(s)} \end{bmatrix}, \label{eq:Gyu_matrix} \end{equation}
which captures the frequency-domain behavior of the inverter dynamics. 
Figure~\ref{fig:Bode_Gyu} illustrates the Bode plots of magnitude and phase responses for the entries in \eqref{eq:Gyu_matrix} 
under four representative parameter settings associated with four operating modes: $(\epsilon,\mu)=(0,0)$ for $PQ$ mode, $(1,0)$ for $Qf$ mode, $(0,30)$ for $PV$ mode, and $(1,30)$ for $Vf$ mode.
All operating modes exhibit a pronounced resonant peak around $120\pi$ rad/s,
which is mainly governed by the $RL$ filter dynamics \eqref{eq:RLfdynamicinlocalbusframe}. 

The responses $i_{d_g}(s)/U_\tm(s)$ and $i_{q_g}(s)/U_\tm(s)$ characterize how grid voltage magnitude disturbances propagate to the inverter output currents (or active and reactive powers).
The parameter $\mu$ primarily affects the low-frequency gain and phase characteristics.
When $\mu=0$ ($PQ$/$Qf$ modes), the low-frequency magnitude of $i_{q_g}(s)/U_\tm(s)$ is significantly reduced, indicating a stronger voltage-following behavior and a stronger reactive power tracking. 
In contrast, when $\mu=30$ ($PV$/$Vf$ modes), the inverter behaves like voltage-forming and exhibits higher low-frequency gains. In this case, the low-frequency gain of $i_{q_g}(s)/U_\tm(s)$ is affected by the filter impedance and voltage-forming parameter $\mu$, i,e, the $Q$-$U$ damping.


The transfer functions $i_{d_g}(s)/\omega_u(s)$ and $i_{q_g}(s)/\omega_u(s)$ characterize the coupling between grid-frequency disturbances and the inverter output currents, and hence the active and reactive power responses. 
The parameter $\epsilon$ primarily shapes the characteristics of the transfer function $i_{d_g}(s)/\omega_u(s)$, which represents the $f$-$P$ droop behavior. 
When $\epsilon=1$ ($Qf$/$Vf$ modes),
the inverter operates in frequency-forming mode. In this case, the low-frequency gain 0 dB reflects the $f$-$P$ droop coefficient, which is approximately given by \eqref{eq:fPdroopVfmode}:  $D_{f\text{-}P}= V_\tm^2/\eta_2 \approx 1$ pu.
The corresponding negative phase indicates an effective negative damping contribution introduced by the droop mechanism. In contrast, when $\epsilon=0$, the inverter operates in active-power tracking mode. As a result, grid-frequency disturbances have minimal influence on the output current, leading to a negligible low-frequency gain in $i_{d_g}(s)/\omega_u(s)$.

The transfer function $i_{q_g}(s)/\omega_u(s)$ characterizes the coupling between grid frequency variations and reactive-current dynamics, reflecting the interaction between active and reactive power channels. 
When $\epsilon=0$, this coupling is weak, and the low-frequency gain of $i_{q_g}(s)/\omega_u(s)$ remains small, as indicated by the blue and red curves. 
When $\epsilon=1$, corresponding to the orange and green curves, the inverter operates in frequency-forming mode with $P$-$f$ droop.
In this case, grid-frequency disturbances affect not only the active-power response but also the reactive-power response through dynamic cross-coupling. In particular,
for $\epsilon=1$ and $\mu=30$, shown by the green curve, the effect of grid-frequency perturbations on the $q$-axis current is further amplified due to the combined effects of $f$-$P$ and $V$-$Q$ droop control, which strengthen the coupling between the active and reactive power dynamics.

Overall, the Bode plots in Figure~\ref{fig:Bode_Gyu} confirm that the input-output representation \eqref{eq:Gyu_matrix}
provides a unified frequency-domain interpretation of voltage/frequency-following/forming behaviors.
While the high-frequency responses are largely dictated by the physical filter dynamics,
the low-frequency gain and phase characteristics are systematically shaped by the parameters $\mu$ and $\epsilon$,
thus explaining the distinct disturbance-rejection properties across different operating modes.

\section{Electromagnetic Transient Simulation Results}
\label{Section:EMTsimulation}

In this section, high-fidelity EMT simulations are conducted in MATLAB Simulink under three different test cases. Firstly, an infinite-bus system with a passive impedance load is used to illustrate the characteristics of various operating modes. Secondly, a two-bus system is constructed to demonstrate the dynamic features of the $PQ$-$Vf$ and $PV$-$Qf$ inverter pairs. The shift of frequency-forming capability between the two inverters is also studied to explain the hybrid operating mode. Lastly, the IEEE 39-bus system is employed to evaluate the overall system-level performance.
All these Simulink models have been made publicly available as open-source models \cite{wang2026unifiedgfml}.


\subsection{Case 1: Individual Inverter Connected to an Infinite Bus}
\label{subsection:EMTindividual}
\begin{figure}[ht]
    \centering    \includegraphics[width=0.85\linewidth]{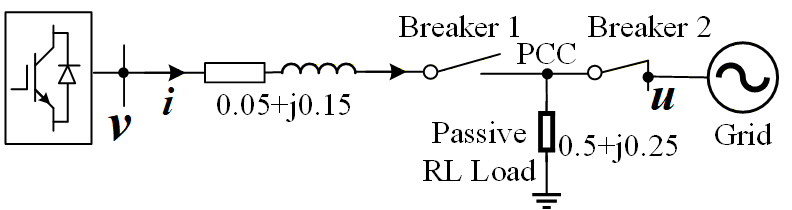}
    \caption{An individual inverter connected to an infinite bus in Case 1.}
    \label{fig:SIMBsystem}
\end{figure}

In this case, the proposed unified GFM–GFL inverter is connected to an infinite bus and a passive RL load, as shown in Figure~\ref{fig:SIMBsystem}. This case is used to demonstrate the dynamic characteristics of different operating modes, as well as the seamless mode transition behaviors. The base power and base voltage are set to 1~MVA and 10~kV, respectively. The reference voltage and angular frequency are set to $V_0=1.075$~pu. and $\omega_0=120\pi$~rad/s, respectively. The filter impedance is given by $R_\tf+\tj X_\tf=0.05+\tj0.15$~pu, and the passive load is set to $0.5+\tj0.25$~pu.

\subsubsection{PQ Control Mode - Voltage and Frequency Following}
\label{section:simu-PQ}

\begin{figure}[ht]
    \centering
    \includegraphics[width=1\linewidth]{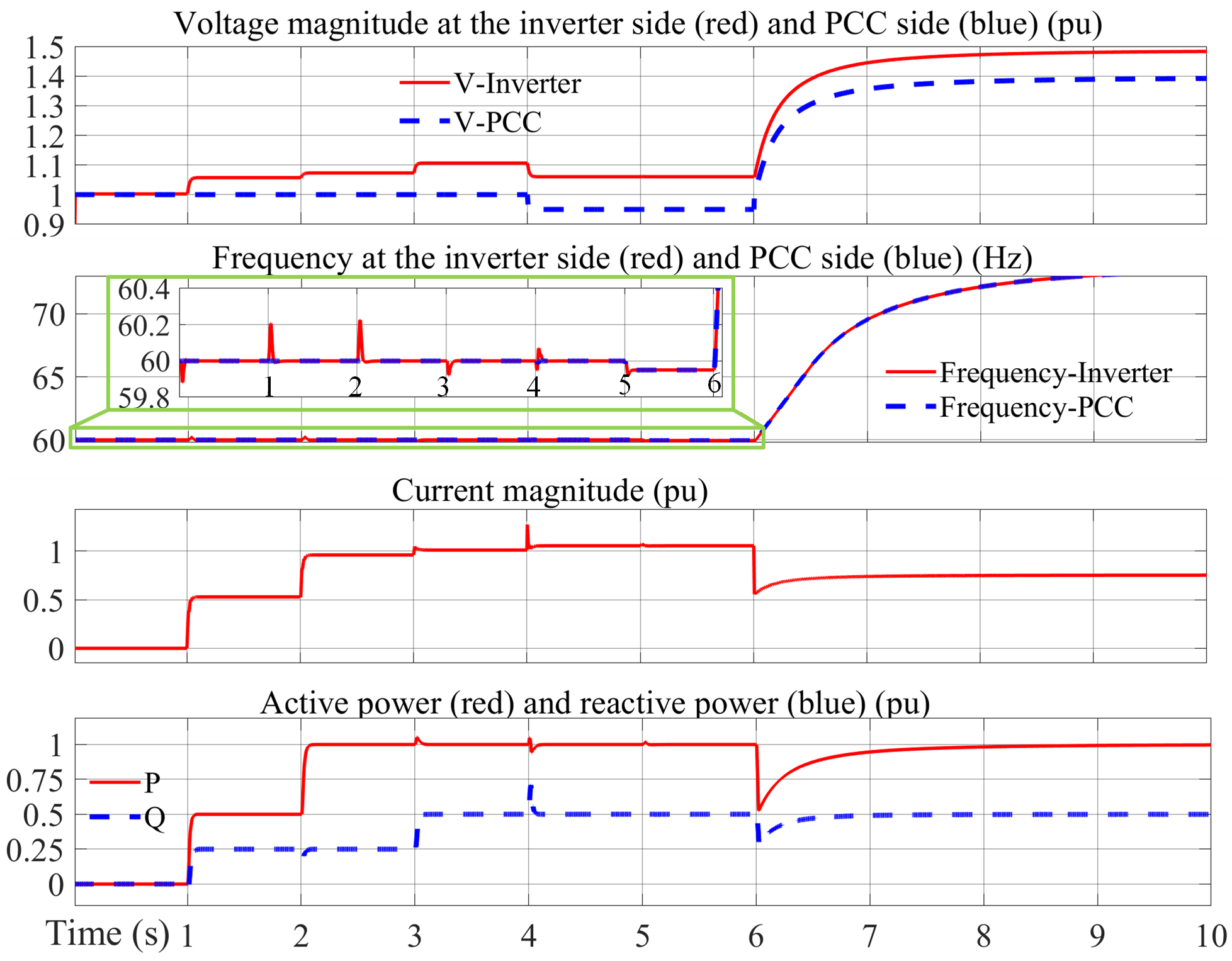}
    \caption{Dynamic performance of $PQ$ mode under disturbances.}
    \label{fig:PQdynamicresu}
\end{figure}

\begin{figure}[ht]
    \centering
    \includegraphics[width=1\linewidth]{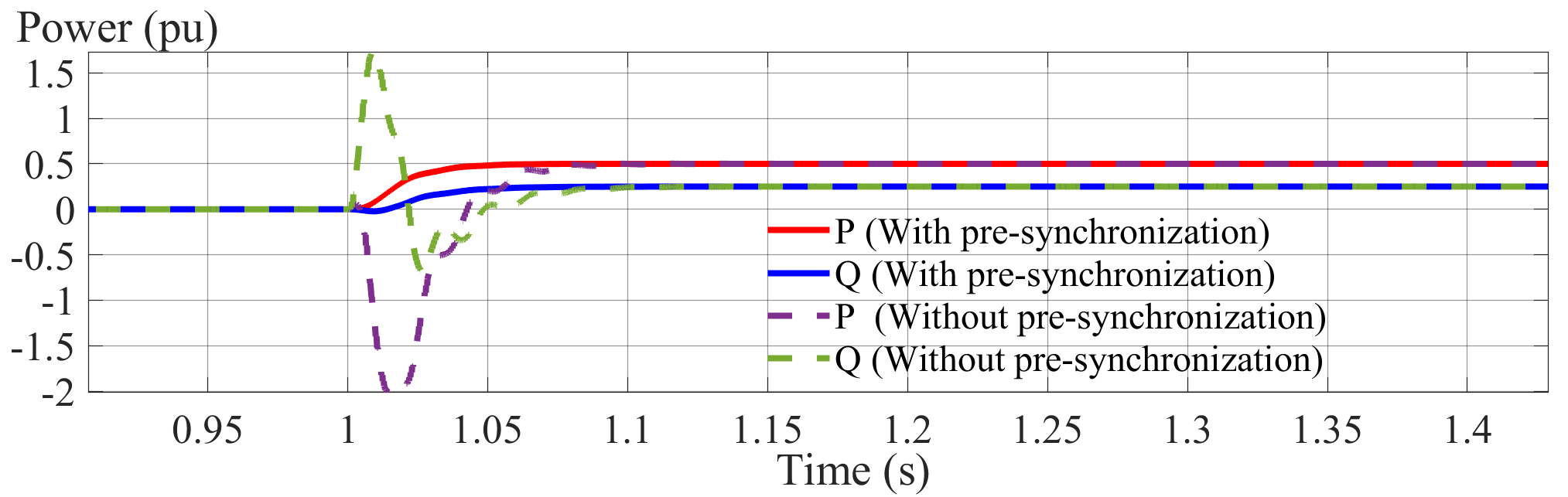}
    \caption{Comparison of initial power responses with and without pre-synchronization.
}
    \label{fig:presyn}
\end{figure}

Setting $\epsilon=0$, $\mu=0$, and $\eta_1=\eta_2=10$, the inverter operates in $PQ$ mode and tracks the prescribed active and reactive power references. The simulation results are shown in Figure~\ref{fig:PQdynamicresu}. Initially, Breaker~1 is open, and Breaker~2 is closed. Before closing Breaker~1, $\gamma$ is set to 1000 and $\tilde{\bv}=\bu$ is enforced to achieve pre-synchronization. At $t=1$~s, Breaker~1 is closed, $\gamma$ is reduced to 0, and the power references are set to $P_0=0.5$~pu and $Q_0=0.25$~pu. 
With the proposed pre-synchronization method, no inrush power or current is observed, and the inverter power gradually converges to the reference. In contrast, without pre-synchronization, significant inrush power and current occur at the instant of grid connection. The comparative results are presented in Figure~\ref{fig:presyn}.

Additionally, a variety of reference changes and grid disturbances are simulated to test performance. Specifically,
at $t=2$~s, the active-power reference $P_0$ is increased to 1.0~pu, and at $t=3$~s, the reactive-power reference $Q_0$ is increased to 0.5~pu. At $t=4$~s, the grid voltage drops to 0.95~pu, and at $t=5$~s, the grid frequency drops to 59.95~Hz. At $t=6$~s, Breaker~1 opens, and the inverter transitions to islanded operation. As shown in Figure~\ref{fig:PQdynamicresu}, the inverter in the $PQ$ grid-following mode can smoothly and rapidly track the commanded power references, even under grid-voltage and grid-frequency disturbances. However, islanded operation is not supported in the $PQ$ grid-following mode; once the inverter is disconnected from the grid, both voltage and frequency collapse.


\subsubsection{PV Control Mode - Voltage Forming and Frequency Following}

\begin{figure}
    \centering
    \includegraphics[width=1\linewidth]{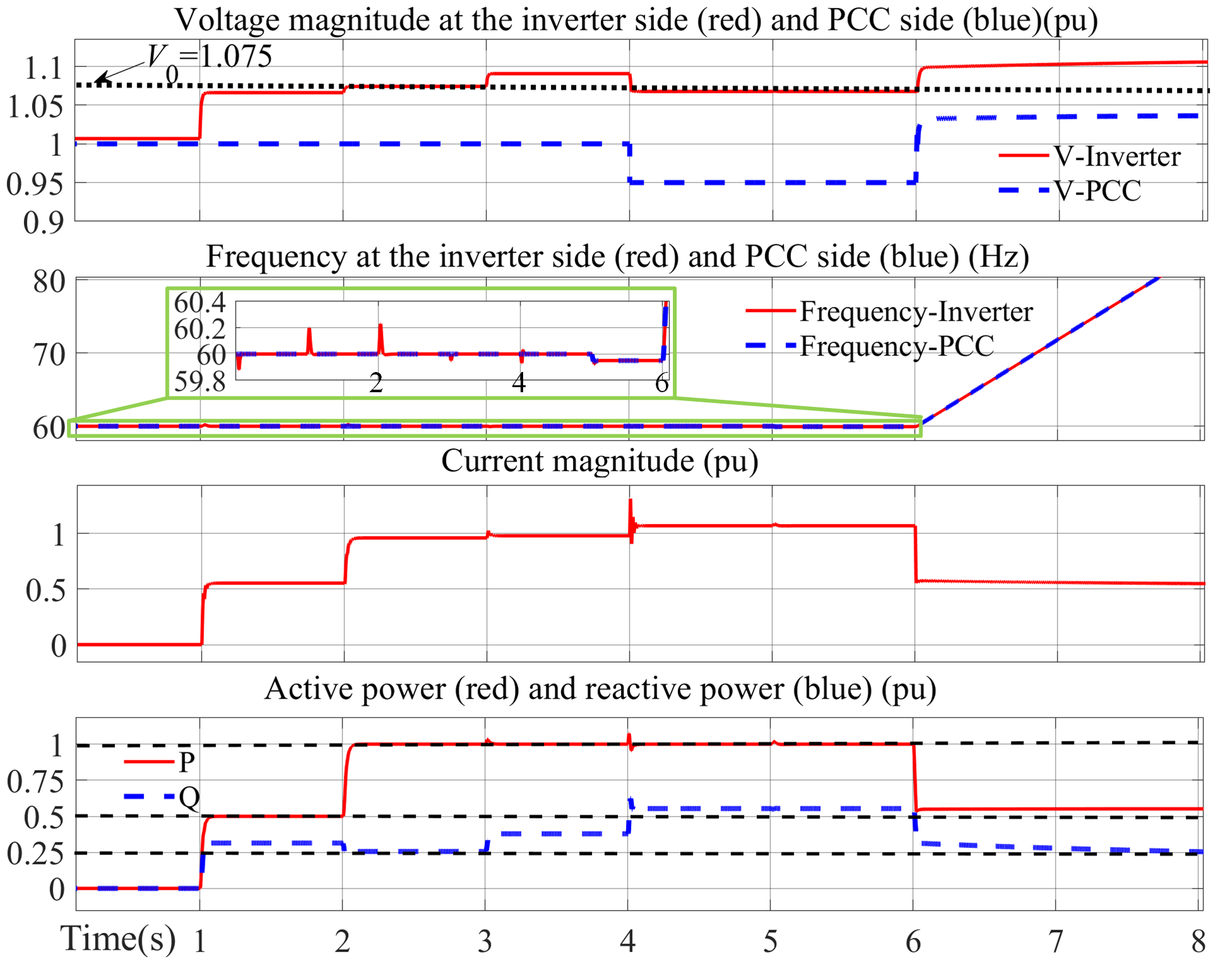}
    \caption{Dynamic performance of $PV$ mode under disturbances.}
    \label{fig:PVdynamicresult}
\end{figure}

Setting $\epsilon=0$, $\mu=30$, and $\eta_1=\eta_2=10$, the inverter operates in $PV$ mode. Before $t=1$~s, the same pre-synchronization procedure as in Section~\ref{section:simu-PQ} is applied. The subsequent disturbances are the same as those used for the $PQ$ mode above. The dynamic response of the $PV$ mode is shown in Figure~\ref{fig:PVdynamicresult}. 
Under grid-voltage and grid-frequency disturbances, the inverter in $PV$ mode accurately tracks the active-power reference. The reactive-power response, however, differs from that in $PQ$ mode. Since the inverter has voltage-forming capability and follows the nonlinear $V$-$Q$ droop \eqref{eq:PVsteadyPQV}, deviations in reactive power are reflected in the inverter voltage magnitude. Specifically, during $t=1$-$2$~s, the reactive power is slightly higher than its reference value of 0.25~pu, and the inverter voltage is correspondingly slightly lower than the reference voltage $V_0=1.075$. During $t=3$-$4$~s, the reactive power is lower than its reference value of 0.5~pu, and the inverter voltage becomes higher than $V_0$. 
Another key characteristic is observed after the inverter is disconnected from the grid at $t=6$~s. Due to the voltage-forming capability, the voltage magnitude does not collapse. Nevertheless, because the inverter lacks frequency-forming capability in $PV$ mode, the frequency cannot be regulated during islanded operation and eventually diverges.

\subsubsection{Qf Control Mode - Frequency Forming and Voltage Following}

\begin{figure}
    \centering
    \includegraphics[width=1\linewidth]{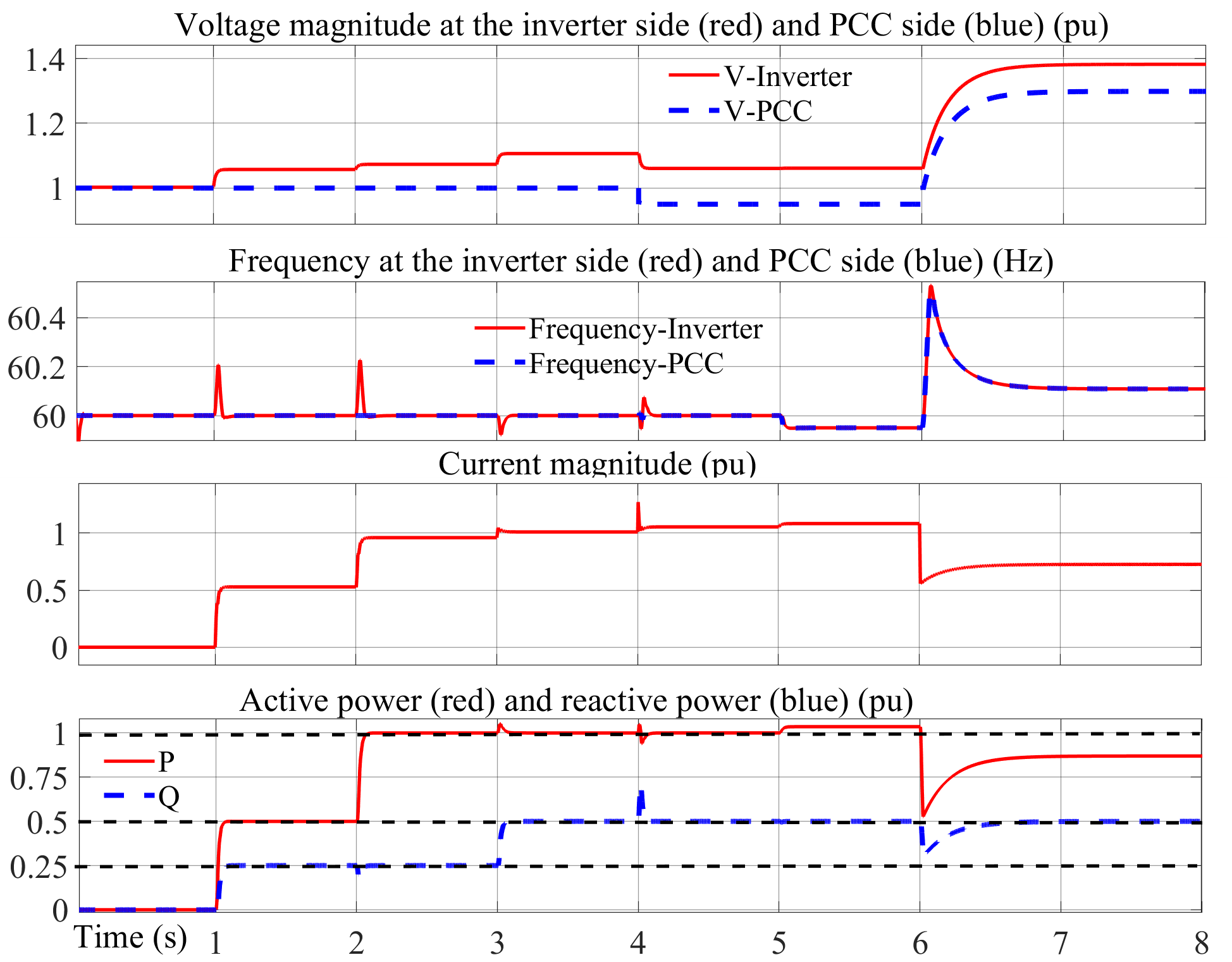}
    \caption{Dynamic performance of $Qf$ mode under disturbances.}
    \label{fig:Qfdynamic}
\end{figure}

Setting $\epsilon=1$ and $\mu=0$, while keeping the other parameters the same as in the previous case, the inverter operates in $Qf$ mode. In this mode, the inverter has frequency-forming capability but does not provide voltage-forming capability. The same disturbance sequence used for the $PQ$ and $PV$ modes is applied, and the simulation results are shown in Figure~\ref{fig:Qfdynamic}.
It is observed that the inverter accurately tracks the reactive-power reference. For the active-power response, the nonlinear $f$-$P$ droop relationship in \eqref{eq:QfthetaV_hatpi2_steadystate} is established. When the grid frequency $\omega_g$ is equal to the reference frequency $\omega_0$, the active power $P$ matches its reference value $P_0$. When the grid frequency drops to 59.5~Hz at $t=5$~s, the inverter active power increases beyond the reference value of 1~pu, indicating that the inverter provides additional power support in response to the frequency deviation. 
After the inverter is disconnected from the grid at $t=6$~s, it supplies only the passive load. Since the $Qf$ mode does not provide voltage-forming capability, the voltage magnitude cannot be maintained and becomes unstable. In attempting to track the reactive-power reference, the controller continues to adjust the voltage magnitude until $Q=Q_0$, leading to voltage instability. In contrast, the frequency remains stable because the $Qf$ mode provides frequency-forming capability. According to the droop relationship, the steady-state frequency settles at approximately 60.1~Hz. This frequency deviation is caused by the active power being lower than its reference value, which induces a frequency shift through the $f$-$P$ droop mechanism.

\subsubsection{Vf Control Mode - Frequency and Voltage Forming}
\label{section:vfmodesim}

\begin{figure}[ht]
    \centering
    \includegraphics[width=1\linewidth]{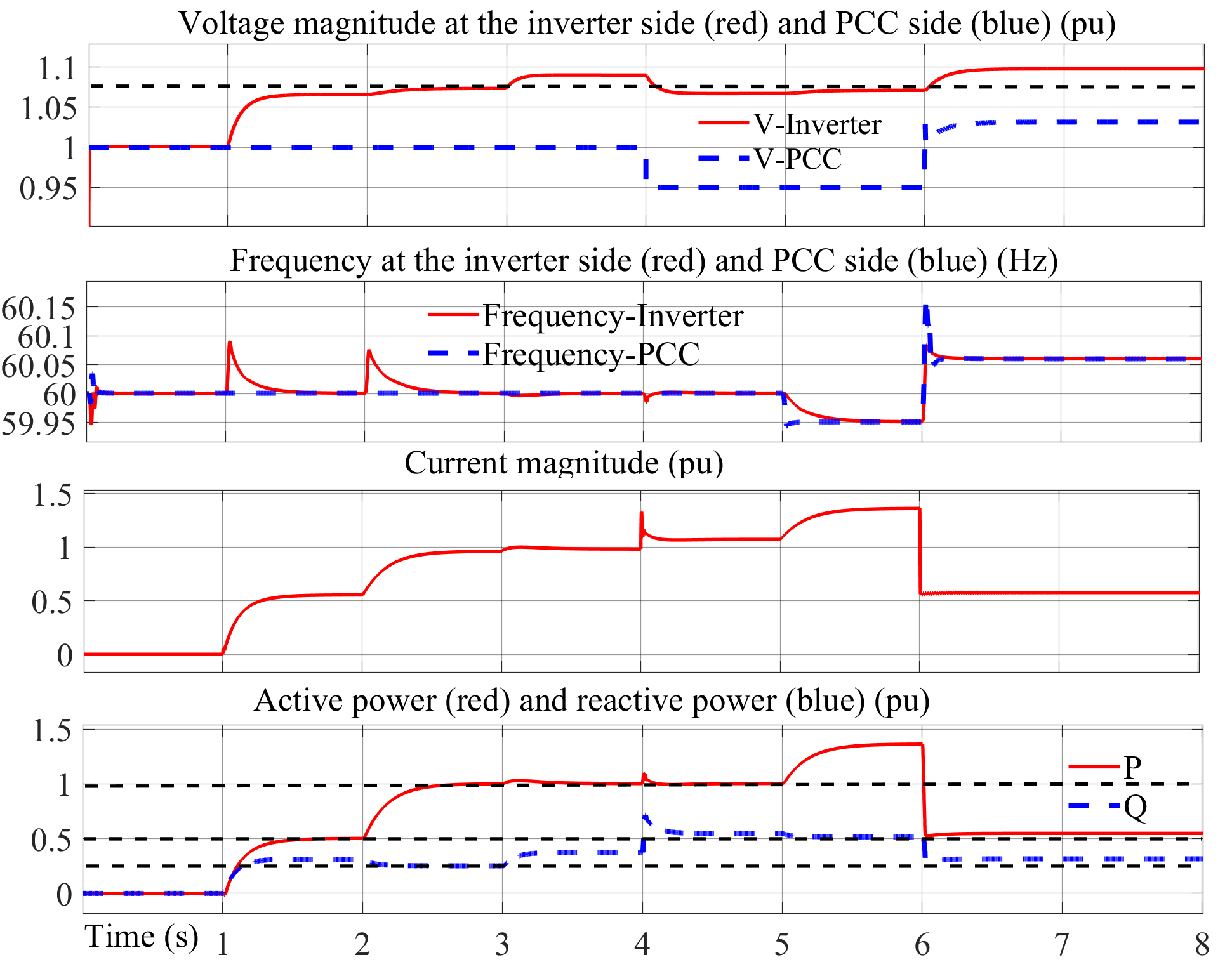}
    \caption{Dynamic performance of $Vf$ mode under disturbances.}
    \label{fig:Vfdynamic}
\end{figure}

Setting $\mu=3$, $\epsilon=1$, and $\eta_1=\eta_2=1$, the inverter operates in $Vf$ mode, which is equivalent to dVOC. The same disturbance sequence as in Section~\ref{section:simu-PQ} is applied, and the simulation results are shown in Figure~\ref{fig:Vfdynamic}. 
When the grid frequency is 60~Hz during $t=1$-$5$~s, the inverter active power satisfies $P=P_0$, corresponding to the nominal operating point of the $f$-$P$ droop characteristic. The reactive power and voltage magnitude follow the nonlinear $V$-$Q$ droop in \eqref{eq:VQdroopVfmode}. During $t=1$-$2$~s, $Q>Q_0=0.25$~pu, and the inverter voltage is slightly lower than the reference value $V_0=1.075$. During $t=2$-$3$~s, $Q=Q_0$, and correspondingly, $V=V_0$. During $t=3$-$4$~s, $Q>Q_0$, and therefore $V<V_0$. At $t=5$~s, the grid frequency drops to 59.95~Hz, and the inverter increases its active-power output according to the nonlinear $f$-$P$ droop in \eqref{eq:fPdroopVfmode} to support the system. 
After the inverter is disconnected from the grid, both voltage and frequency remain stable. Their steady-state values are determined by the passive load and the $f$-$P$ and $V$-$Q$ droop characteristics. Since the passive load is set to $0.5+\tj0.25$~pu, which is lower than the reference power $1+\tj0.5$~pu, both voltage and frequency settle above their reference values. The ability to maintain stable voltage and frequency while supplying a passive load demonstrates the voltage- and frequency-forming capability of the grid-forming inverter, which is fundamentally different from the previous three operating modes.


\begin{figure}
    \centering
    \includegraphics[width=1\linewidth]{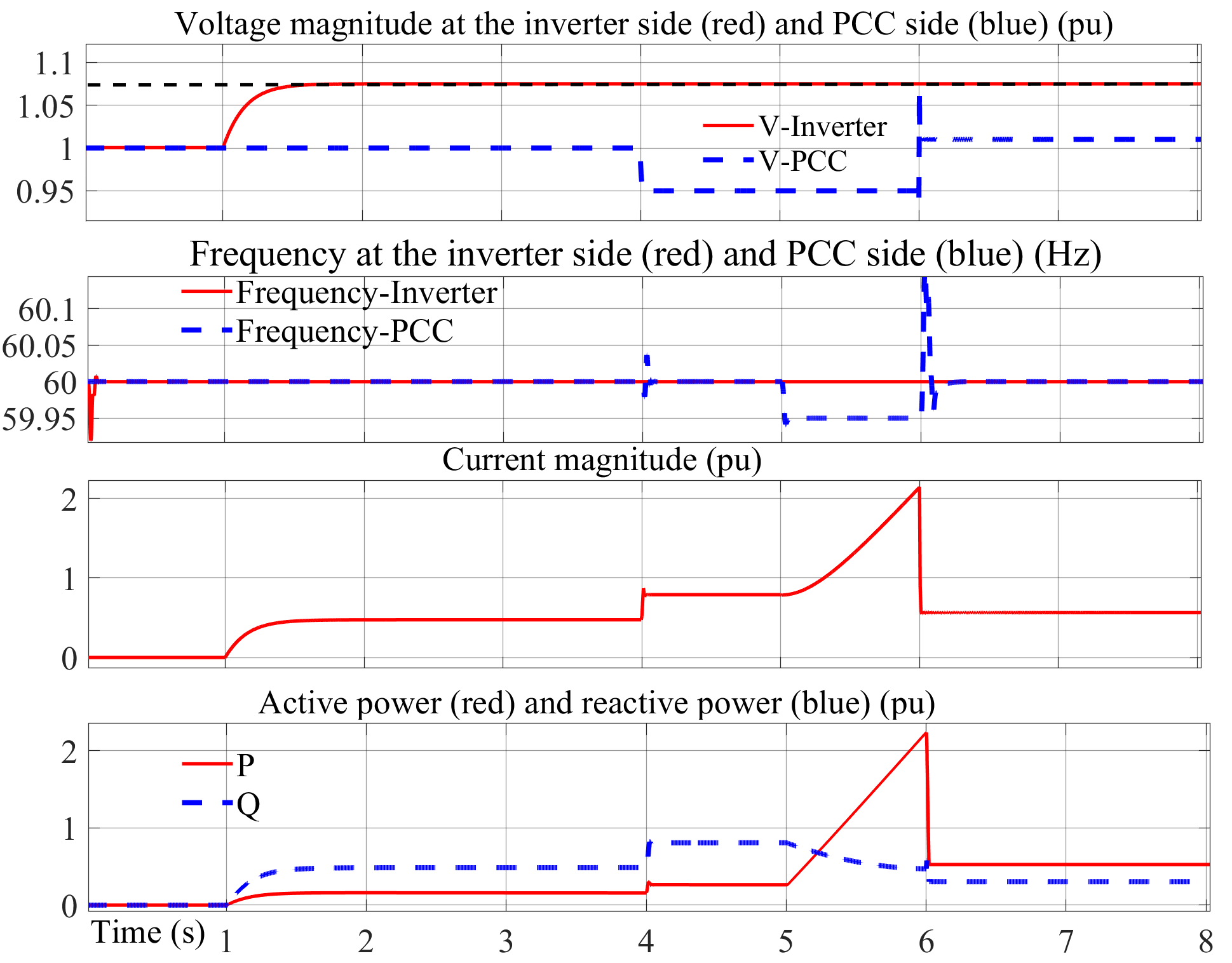}
    \caption{Dynamic performance of $Vf$ mode with constant voltage and frequency under disturbances.}
    \label{fig:constantVf}
\end{figure}

As discussed in Section \ref{section:VfmodeVtheta}, in $Vf$ mode, setting $\eta_1=\eta_2=0$ makes the inverter operate with constant voltage magnitude $V_0$ and constant frequency $\omega_0$. Under the same disturbance sequence as before, the simulation results for constant-$V$ and constant-$f$ operation are shown in Figure~\ref{fig:constantVf}. Before $t=1$~s, pre-synchronization is performed so that the inverter voltage and phase are aligned with the grid. At $t=1$~s, the inverter is connected to the grid, after which its voltage magnitude is regulated to 1.075~pu and its frequency remains at 60~Hz. In this mode, the power output is determined by the network impedance and the voltage difference between the inverter and the grid, and is independent of the power references $P_0$ and $Q_0$. 
At $t=4$~s, the grid voltage drops to 0.95~pu. As a result, both active and reactive power increase due to the larger voltage difference between the inverter and the grid, while the phase difference remains zero. At $t=5$~s, the grid frequency drops to 59.95~Hz. Since the inverter maintains a constant frequency, a frequency difference is introduced, causing the inverter active-power output to increase and provide grid support. At $t=6$~s, Breaker~2 opens and the inverter supplies the passive load independently. The voltage magnitude remains regulated at 1.075~pu.


\subsubsection{Hybrid Operating Mode}

\begin{figure}
    \centering
    \includegraphics[width=1\linewidth]{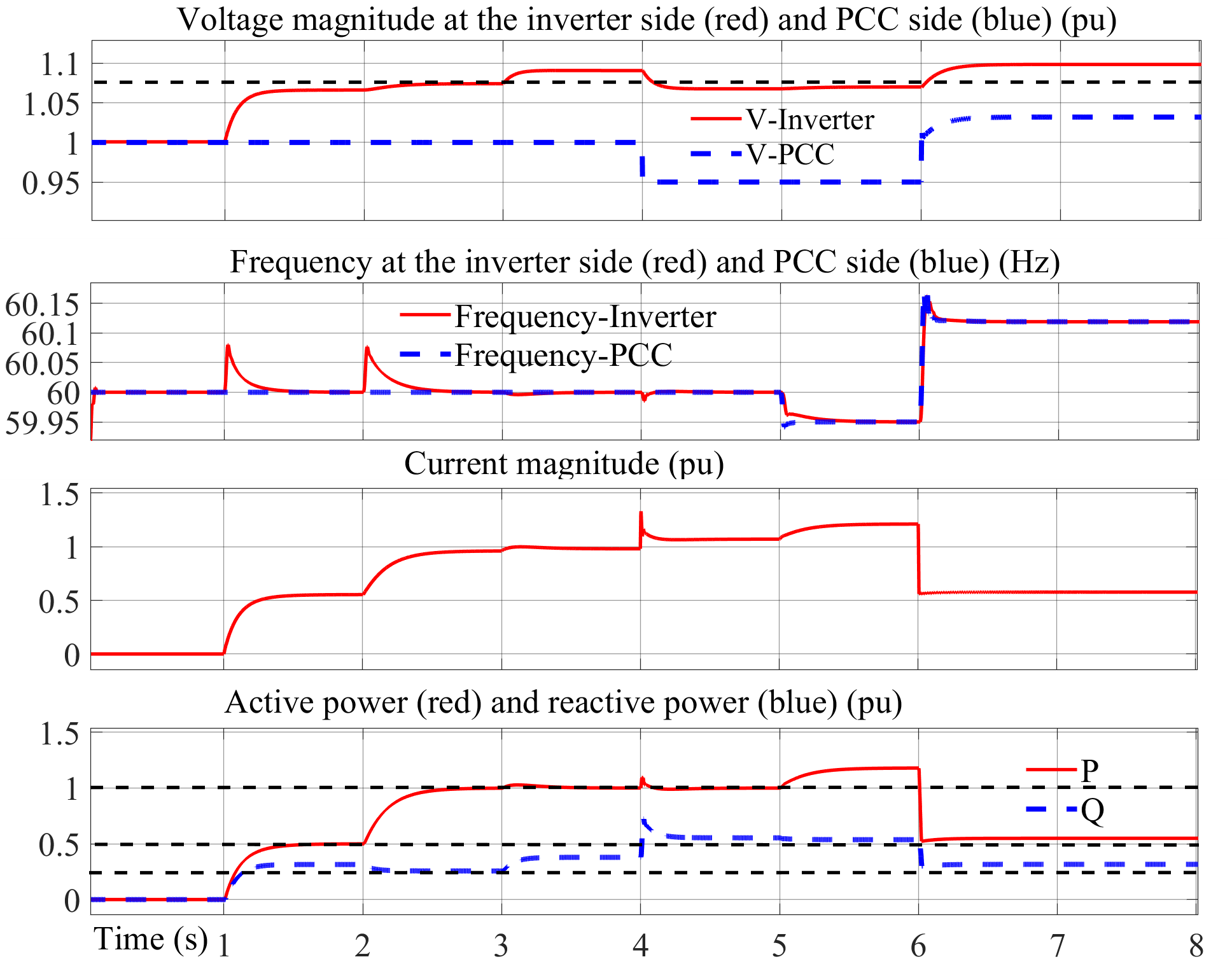}
    \caption{Dynamic performance of hybrid mode  under disturbance.}
    \label{fig:Hybridmode}
\end{figure}

Setting $\epsilon=0.5$, $\mu=3$, and $\eta_1=\eta_2=1$, the inverter operates in a hybrid grid-forming/grid-following mode, which combines voltage- and frequency-forming capability with power-dispatch functionality. The same disturbance sequence as in the previous subsection is applied, and the simulation results are shown in Figure~\ref{fig:Hybridmode}. 
Compared with the $Vf$ mode with $\epsilon=1$ in Figure~\ref{fig:Vfdynamic}, setting $\epsilon=0.5$ strengthens the frequency-following characteristic while reducing the frequency-forming contribution. When the grid frequency drops to 59.95~Hz at $t=5$~s, the inverter frequency follows the grid frequency more rapidly, and the power support provided in response to the frequency deviation is reduced to approximately half of that in the $Vf$ mode. This is because the effective $f$-$P$ droop coefficient is scaled by $\epsilon$, as shown in \eqref{eq:hybridfPdroop}.


\subsubsection{Mode Transitions Under Step Changes in Control Parameters}

\begin{figure}[ht]
    \centering
    \includegraphics[width=0.7\linewidth]{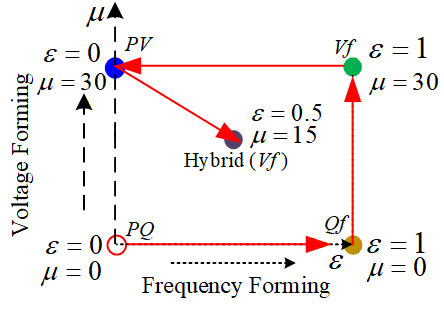}
    \caption{Illustration of the mode transition sequence in simulations.}
    \label{fig:Modetransition}
\end{figure}

\begin{figure}[ht]
    \centering
    \includegraphics[width=1\linewidth]{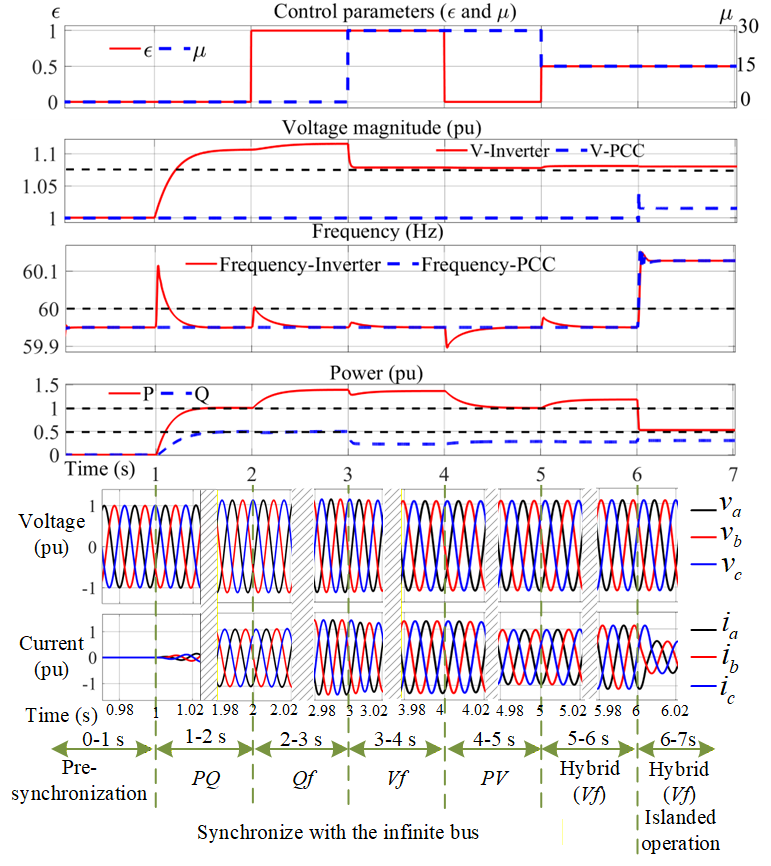}
    \caption{Seamless mode transition under step changes in control parameters.}
    \label{fig:Steptransition}
\end{figure}

To illustrate the seamless mode transition of the proposed unified GFM-GFL control method, this subsection applies step changes to the mode control parameters separately. The corresponding modes and parameters transition trajectories are shown in Figure~\ref{fig:Modetransition}.
All other settings are the same as previous subsections.
Furthermore, the continuous ramp variations are applied to the mode control parameters, and the results are shown in the next subsection \ref{Section:ModeTransitionContinuousChange}. 
The simulation results are shown in Figure~\ref{fig:Steptransition}, where all reference values are indicated by black dotted lines.

For $t < 1$ s, the pre-synchronization method is activated to synchronize the inverter with the grid. At $t = 1$ s, the inverter is connected to the grid and operates in the $PQ$ mode with $\gamma=0$, $\epsilon=0$, and $\mu=0$.
Subsequently, step changes are applied to the control parameters to demonstrate seamless transitions among different operating modes. At $t = 2$ s, $\epsilon$ is changed from 0 to 1, resulting in a transition from the $PQ$ mode to the $Qf$ mode. At $t = 3$ s, $\mu$ is increased to 30, enabling the $Vf$ mode. At $t = 4$ s, $\epsilon$ is reset to zero, and the inverter transitions to the $PV$ mode. Finally, at $t = 5$ s, $\epsilon$ and $\mu$ are adjusted to 0.5 and 15, respectively, yielding a hybrid operating mode.

The results show that all mode transitions are accomplished smoothly despite the step changes in control parameters. The instantaneous voltage and current waveforms shown in Figure~\ref{fig:Steptransition} further confirm that no high-frequency harmonics are introduced during the switching process. No noticeable oscillations, power surges, or waveform distortions are observed during the transitions. 

At $t = 6$ s, the grid breaker Breaker1 is opened to create an islanding condition. Since the inverter remains in the hybrid mode with both voltage-forming and frequency-forming capabilities, stable operation is maintained after islanding. Overall, these results demonstrate that the proposed unified control framework enables continuous and seamless transitions among different operating modes without requiring controller reconfiguration.
\subsubsection{Mode Transition Under Continuous Changes in Control Parameters}
\label{Section:ModeTransitionContinuousChange}

\begin{figure}[ht]
    \centering
    \includegraphics[width=1\linewidth]{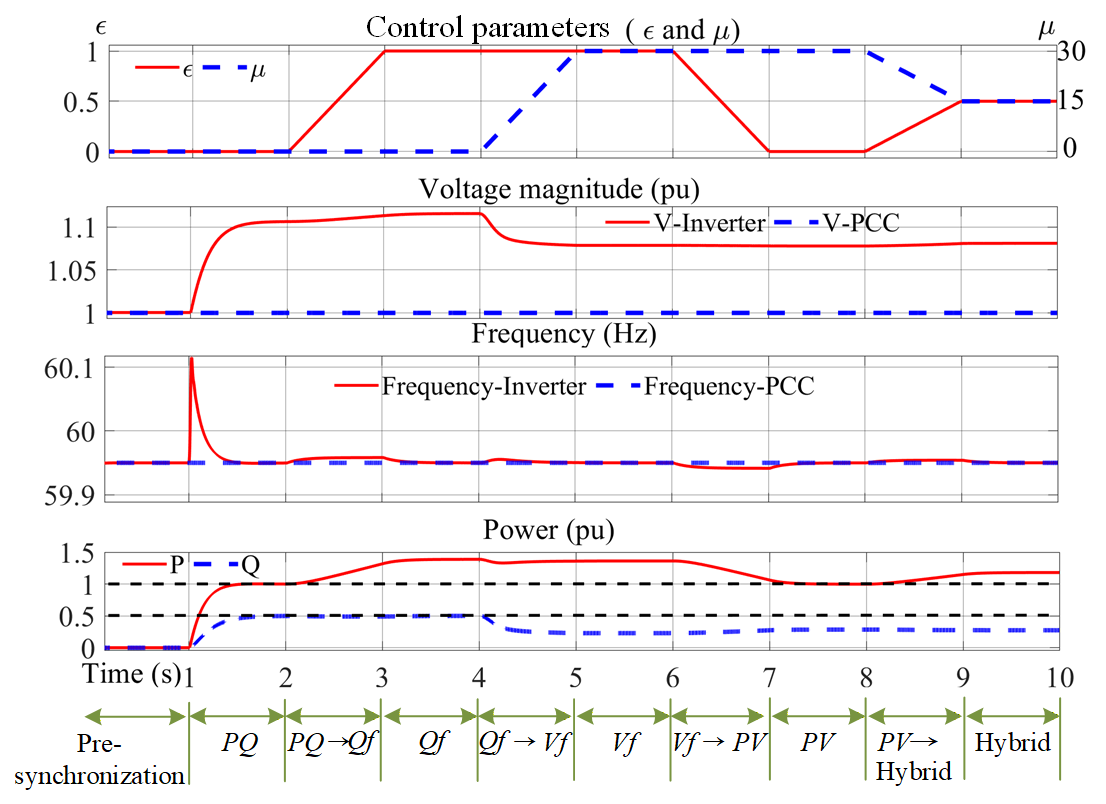}
    \caption{Seamless mode transition under continuous changes in control parameters.}
    \label{fig:Continuoustransition}
\end{figure}


This subsection evaluates the dynamic performance of the proposed controller under continuous variations of the mode parameters. Owing to the unified control structure, smooth transitions can be achieved between frequency-forming and frequency-following behaviors, as well as between voltage-forming and voltage-following characteristics, through continuous parameter adjustment.

The results of the continuous mode transitions are shown in Figure~\ref{fig:Continuoustransition}. The sequence of operating modes is identical to that in the step-change case; however, the control parameters are varied linearly over 1~s instead of being changed instantaneously. Compared with Figure~\ref{fig:Steptransition}, continuous parameter variation results in a smoother migration of the system equilibrium, leading to even more gradual mode transitions while maintaining stable operation throughout the process

\subsection{Case 2: Interconnection of Two Inverters in Different Operating Modes}
\label{Section:TwoInverterSystem}
The $PQ$-$Vf$ and $PV$-$Qf$ pairs can be viewed as dual operating configurations. In this case, we first demonstrate their dynamic responses when two inverters operating in these complementary modes are interconnected and subjected to a sudden load disturbance. Furthermore, to highlight the continuous nature of the proposed framework, we investigate hybrid operating modes in which the frequency-forming capability is continuously distributed between two interconnected inverters by varying $\epsilon$. This allows the system to transition smoothly from a frequency-following/frequency-forming pair to two hybrid-mode inverters with different degrees of frequency-forming capability, thereby illustrating how transient power-sharing and frequency-support responsibilities can be continuously reallocated among interconnected inverters.

\subsubsection{PQ-mode Inverter Connected to a Vf-mode Inverter}
\begin{figure}[ht]
    \centering
    \includegraphics[width=1\linewidth]{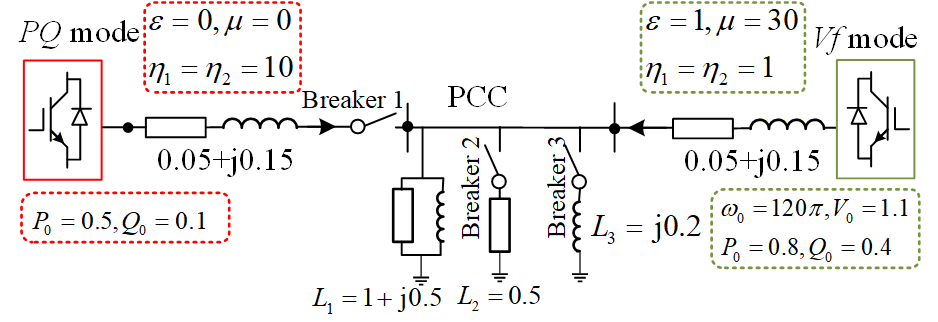}
    \caption{Two-inverter test system with a $PQ$-mode inverter and a $Vf$-mode inverter connected at the PCC.}
    \label{fig:PQandVf}
\end{figure}

\begin{figure}[ht]
\vspace{-10pt}
    \centering
    \subfloat[$PQ$-mode inverter.]{%
        \includegraphics[width=0.48\linewidth]{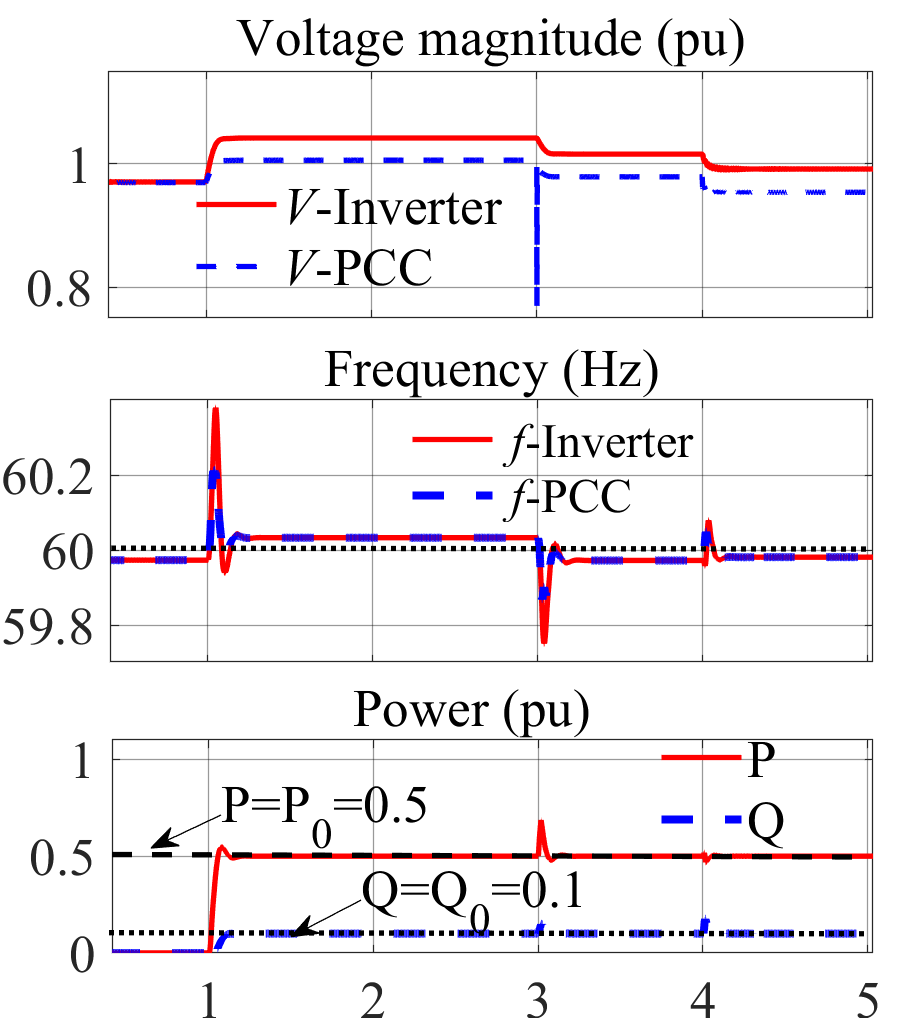}
        \label{fig:PQinPQVf}
    }
    \hfill
    \subfloat[$Vf$-mode inverter.]{%
        \includegraphics[width=0.48\linewidth]{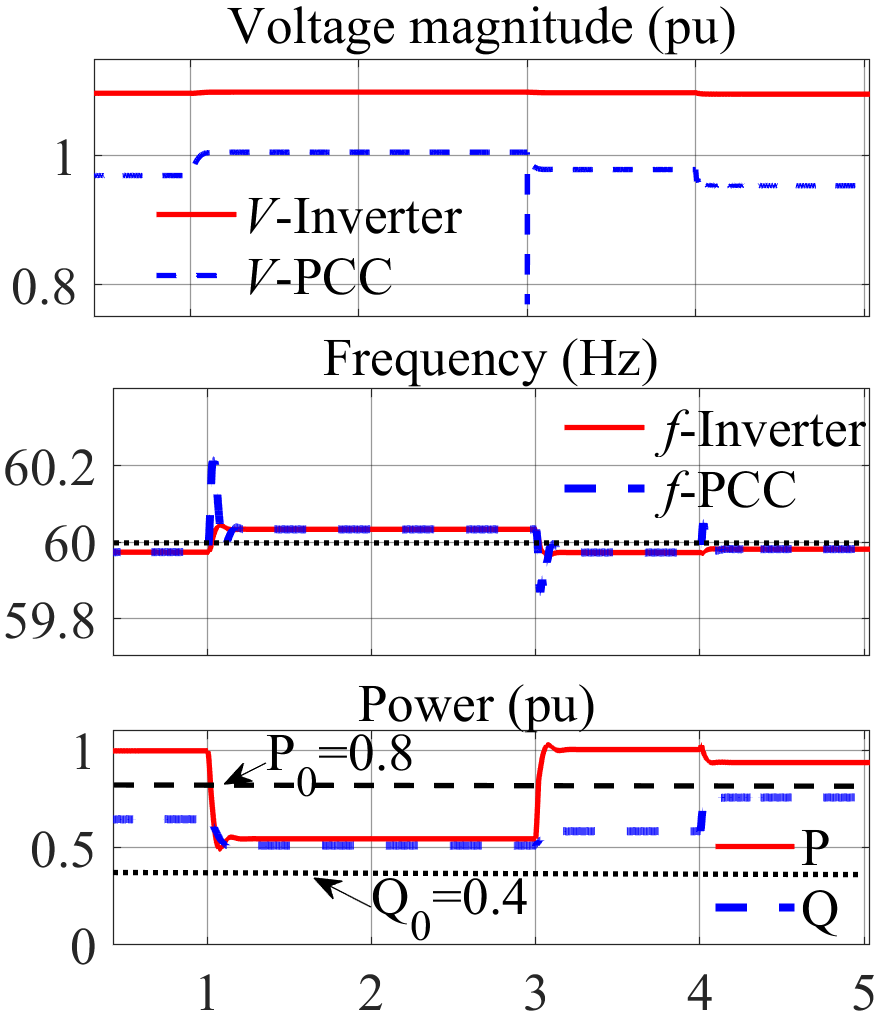}
        \label{fig:VfinPQVf}
    }
    \caption{Dynamic performance comparison for the case $PQ$-mode inverter connecting with $Vf$-mode inverter. (x-axis: Time (s))}
    \label{fig:PQ_Vf_comparison}
\end{figure}

The architecture and parameters are shown in Figure~\ref{fig:PQandVf}. 
At $t=1$~s, Breaker1 is closed and the $PQ$-mode inverter is connected to the PCC. Owing to the pre-synchronization process, the interconnection is achieved without noticeable inrush current or power transients. After connection, the $PQ$-mode inverter maintains its active and reactive power references, while the output power of the $Vf$-mode inverter decreases correspondingly.

At $t=3$~s, load $L_2$ is connected to the PCC. The $PQ$-mode inverter continues to track its power references despite the disturbance, whereas the additional power demand is supplied by the $Vf$-mode inverter. At $t=4$~s, reactive load $L_3$ is connected, and the additional reactive power is similarly absorbed by the $Vf$-mode inverter.

These results demonstrate the compatibility of grid-following and grid-forming operation within the proposed unified framework. The $PQ$-mode inverter maintains accurate power tracking, while the $Vf$-mode inverter automatically balances system power and regulates the grid condition. The seamless interconnection and disturbance response further verify the effectiveness of the proposed control strategy.

\subsubsection{PV-mode Inverter Connected to a Qf-mode Inverter}
\begin{figure}[ht]
    \centering
    \includegraphics[width=1\linewidth]{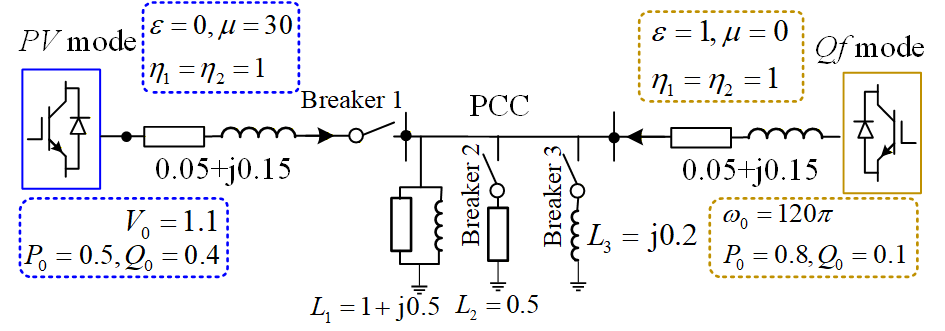}
    \caption{Two-inverter test system consisting of a $PV$-mode inverter and a $Qf$-mode inverter connected through the PCC.}
    \label{fig:PVandQf}
\end{figure}

\begin{figure}[ht]
    \centering
    \subfloat[$PV$-mode inverter.]{%
        \includegraphics[width=0.48\linewidth]{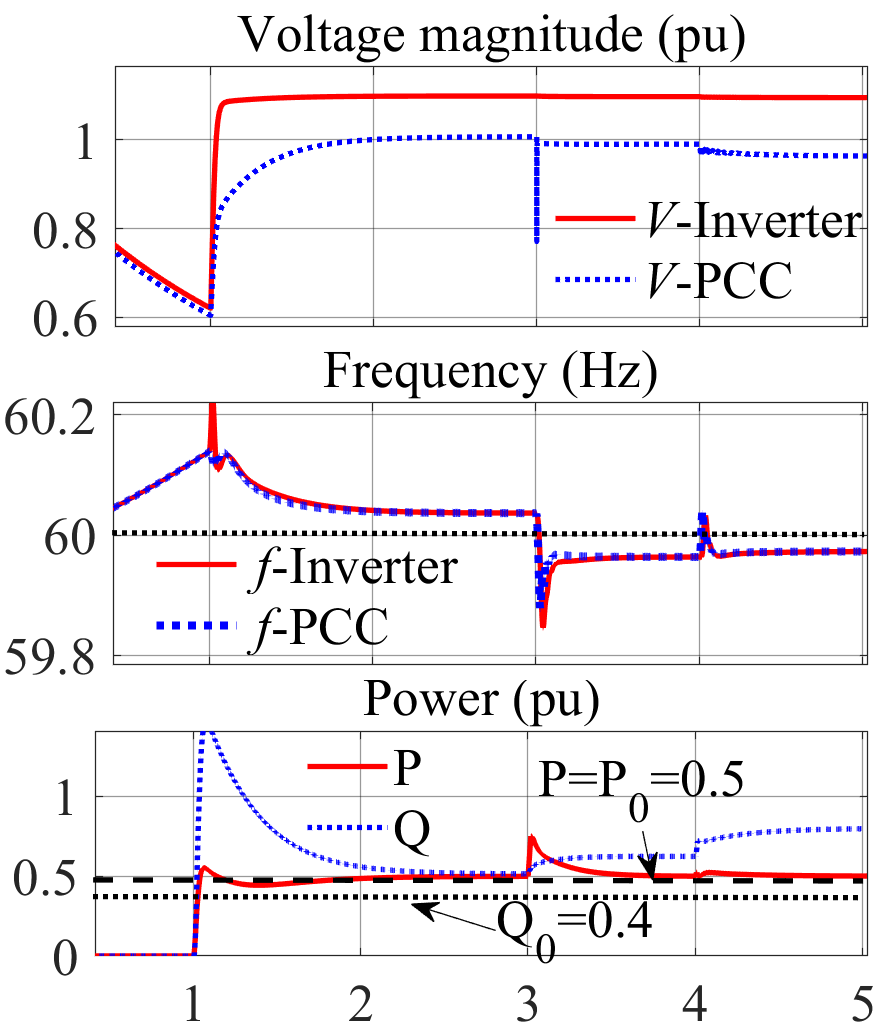}
        \label{fig:PVinPVQf}
    }
    \hfill
    \subfloat[$Qf$-mode inverter.]{%
        \includegraphics[width=0.48\linewidth]{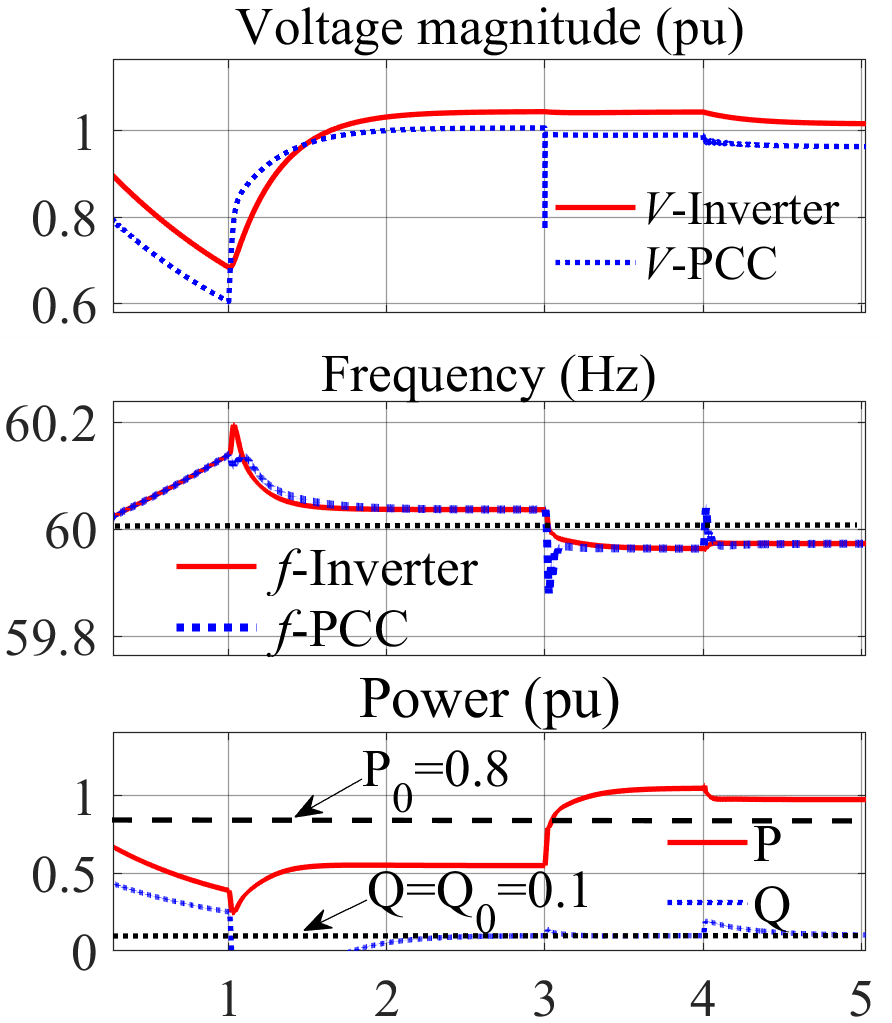}
        \label{fig:QfinPVQf}
    }
    \caption{Dynamic performance comparison for the case $PV$ mode connecting with $Qf$ mode inverter. (x-axis: Time (s))}
    \label{fig:PV_Qf_comparison}
\end{figure}

The $PV$ and $Qf$ modes constitute another complementary pair of operating modes. The test system and corresponding responses are shown in Figure~\ref{fig:PVandQf} and Figure \ref{fig:PV_Qf_comparison}.

Unlike the $PQ$-$Vf$ case, the system initially lacks voltage-forming capability because the standalone $Qf$-mode inverter regulates frequency but not voltage magnitude. As a result, the PCC voltage remains at a low level before interconnection. Prior to $t=1$~s, the $PV$-mode inverter performs pre-synchronization. When Breaker1 is closed at $t=1$~s, the $PV$-mode inverter establishes the PCC voltage, leading to a transient reactive-power surge before the system settles to a new equilibrium.

Subsequently, additional active and reactive loads are connected at $t=3$~s and $t=4$~s, respectively. The results show that the $PV$-mode inverter primarily responds to changes in voltage and reactive-power demand, whereas the $Qf$-mode inverter accommodates the active-power imbalance and regulates system frequency. These results further demonstrate the compatibility of the proposed unified framework for interconnecting inverters operating in different modes and highlight the duality between the $PV$-$Qf$ and $PQ$-$Vf$ operating pairs.

\subsubsection{Continuous Allocation of Frequency-Forming Capability}
Unlike the previous two cases, which demonstrate the compatibility of complementary operating modes, this subsection highlights a unique feature of the proposed framework: the ability to continuously distribute frequency-forming capability among multiple interconnected inverters. By adjusting the parameter $\epsilon$, an inverter can be smoothly transitioned from frequency-following operation ($\epsilon=0$) to frequency-forming operation ($\epsilon=1$), with hybrid operating modes occupying the intermediate region.
\begin{figure}[ht]
    \centering
    \includegraphics[width=1\linewidth]{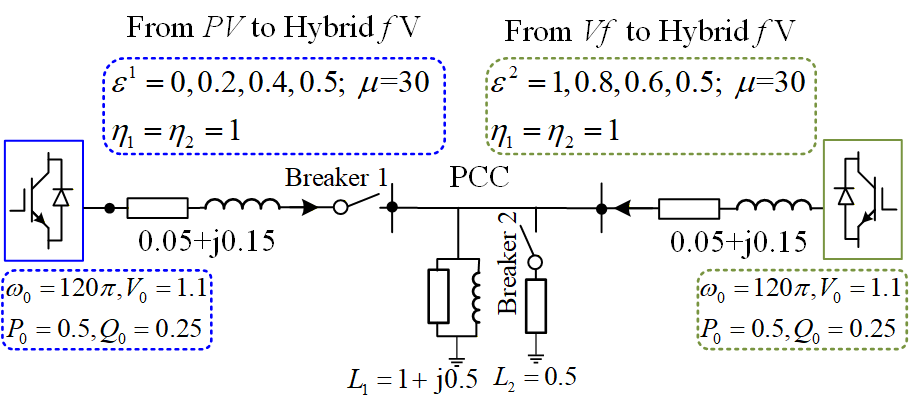}
    \caption{Two-inverter test system for illustrating transitions between frequency-following, hybrid, and frequency-forming modes.}
    \label{fig:FromPVtoHybridfV}
\end{figure}
To illustrate this concept, a two-inverter system is considered, as shown in Figure~\ref{fig:FromPVtoHybridfV}. The parameter pairs $(\epsilon^1,\epsilon^2)$ are chosen as $(0,1)$, $(0.2,0.8)$, $(0.4,0.6)$, and $(0.5,0.5)$ while maintaining $\epsilon^1+\epsilon^2=1$. Consequently, the overall frequency-support capability of the system remains unchanged, whereas its distribution between the two inverters is continuously reallocated. At $t=4$~s, load $L_2$ is connected to the PCC to excite the system dynamics.

\begin{figure}
    \centering
    \includegraphics[width=1\linewidth]{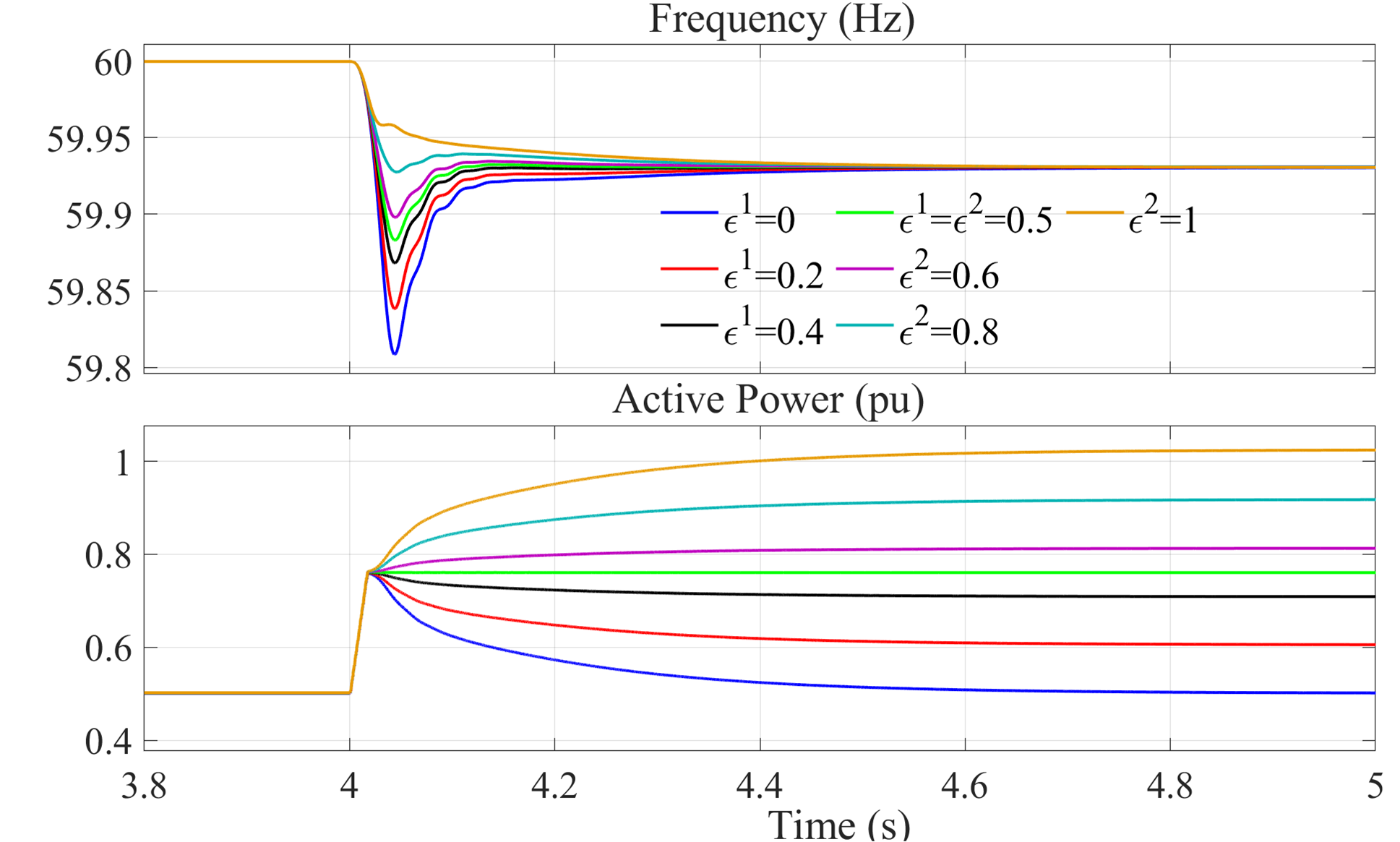}
    \caption{Frequency and power responses under different allocations of frequency-forming capability.}
    \label{fig:resultFromPVtoHybridfVNodroop}
\end{figure}

The responses in Figure~\ref{fig:resultFromPVtoHybridfVNodroop} show that the initial power imbalance is shared according to the network electrical distance and is therefore identical for all cases. As the control dynamics evolve, the subsequent power redistribution depends on $\epsilon$. Inverters with larger $\epsilon$ provide stronger frequency support and contribute a larger share of the disturbance power.

\begin{figure}
    \centering
    \includegraphics[width=1\linewidth]{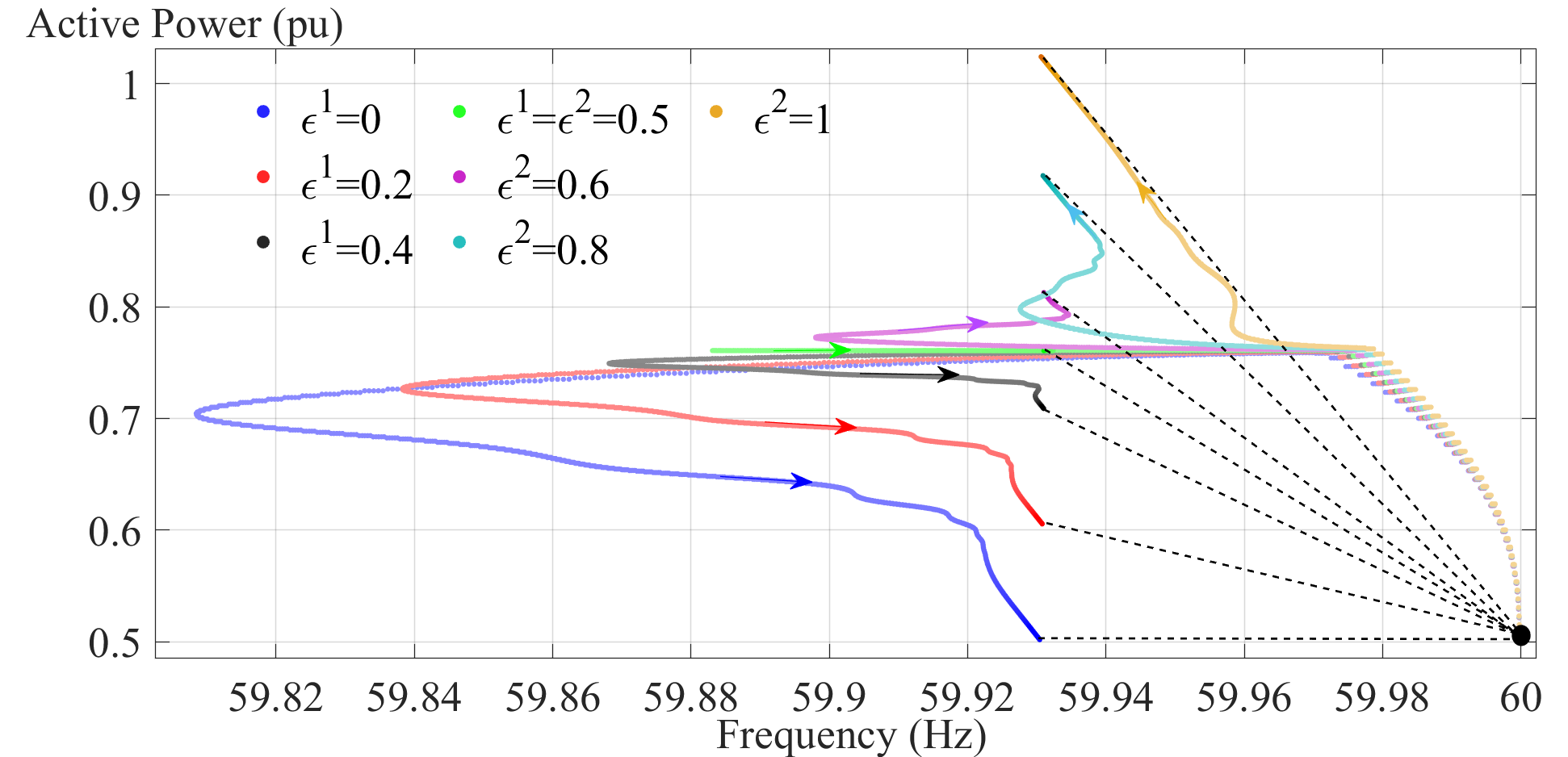}
    \caption{Power-frequency trajectories under different allocations of frequency-forming capability ($\epsilon^1+\epsilon^2=1$).}
    \label{fig:PowerandFrequencyNodroopHybrid}
\end{figure}

The corresponding power-frequency trajectories in Figure~\ref{fig:PowerandFrequencyNodroopHybrid} reveal a continuous transition from frequency-following to frequency-forming behavior. For small $\epsilon$, the trajectories are primarily attracted toward the active-power reference, reflecting power-tracking behavior. As $\epsilon$ increases, the attraction gradually shifts toward the frequency equilibrium, and the trajectories increasingly follow the droop characteristic. When $\epsilon^1=\epsilon^2=0.5$, the system becomes symmetric, and both inverters exhibit identical dynamic responses.




As discussed in Remark \ref{reamrk2}, an additional outer-loop $P$-$f$ droop controller can be incorporated into the power-tracking term to adjust the active-power reference according to frequency deviations. 
Here, this supplementary outer-loop droop is introduced such that the overall steady-state $P$-$f$ droop coefficient remains unchanged as $\epsilon$ varies. The outer-loop droop coefficient is selected as:
\begin{equation}
D_{P\text{-}f}=\frac{3\times V_\tm^2}{2}\frac{(1-\epsilon)}{\eta_2},
\end{equation}
Combined with the intrinsic  $f$-$P$ droop characteristic of the hybrid mode in \eqref{eq:hybridfPdroop}, this choice yields a constant overall $P$-$f$ droop coefficient that is independent of $\epsilon$.

\begin{figure}[ht]
    \centering
    \includegraphics[width=1\linewidth]{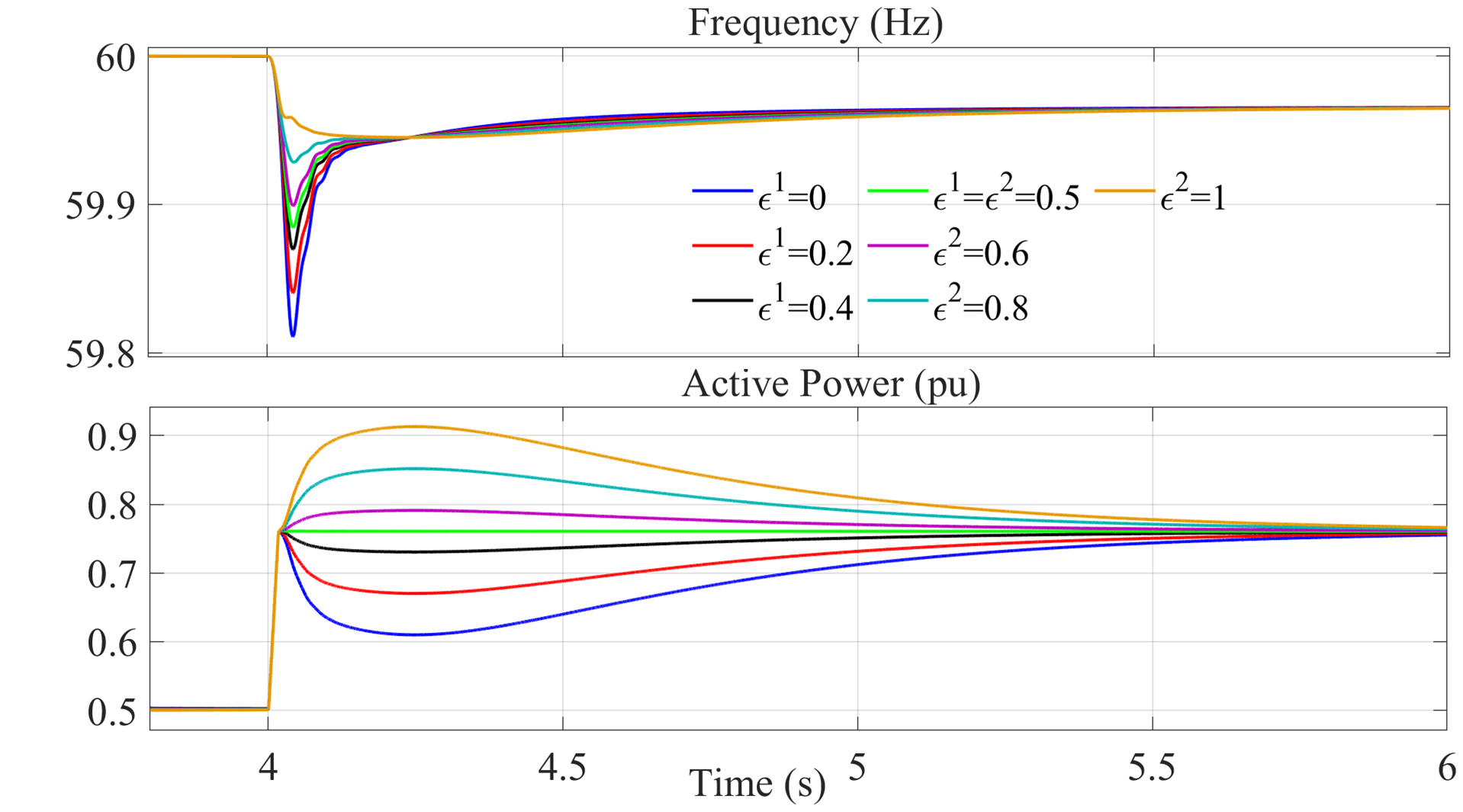}
    \caption{Frequency and power trajectories in time domain under different frequency-forming capabilities, where additional droop-based primary control is introduced to maintain the same equilibrium point as $\epsilon$ varies.}
    \label{fig:FreandPtimeHybridandDroop}
\end{figure}

\begin{figure}[ht]
    \centering
    \includegraphics[width=1\linewidth]{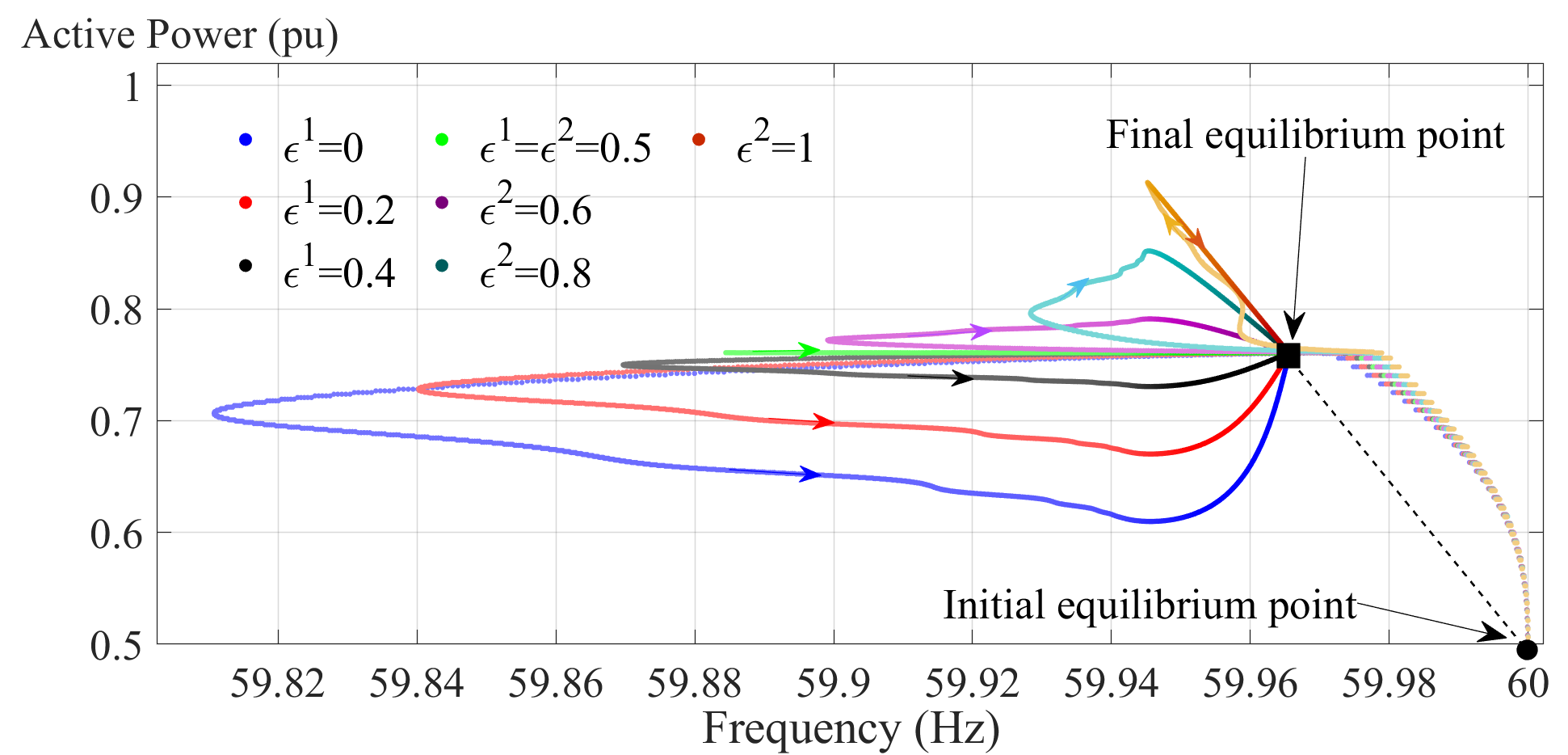}
    \caption{Power-frequency trajectories under different frequency-forming capabilities with additional droop control to maintain the same equilibrium point as $\epsilon$ varies.}
    \label{fig:PowerandFrequencyHybridDroop}
\end{figure}

The resulting responses are shown in Figure~\ref{fig:FreandPtimeHybridandDroop} and Figure~\ref{fig:PowerandFrequencyHybridDroop}. Although all cases converge to the same steady-state equilibrium, their transient trajectories remain markedly different. Larger $\epsilon$ values produce stronger frequency support, resulting in larger initial power responses and smaller frequency deviations. These results indicate that frequency-following and frequency-forming control mainly differ in their transient equilibrium-attraction mechanisms. By continuously adjusting $\epsilon$, the proposed framework enables smooth redistribution of frequency-support responsibility among interconnected inverters while preserving the desired steady-state operating point.

\subsection{Case 3: IEEE 39-Bus System with a Mix of Multiple Inverters and Synchronous Generators}
\begin{figure}[ht]
    \centering
    \includegraphics[width=1\linewidth]{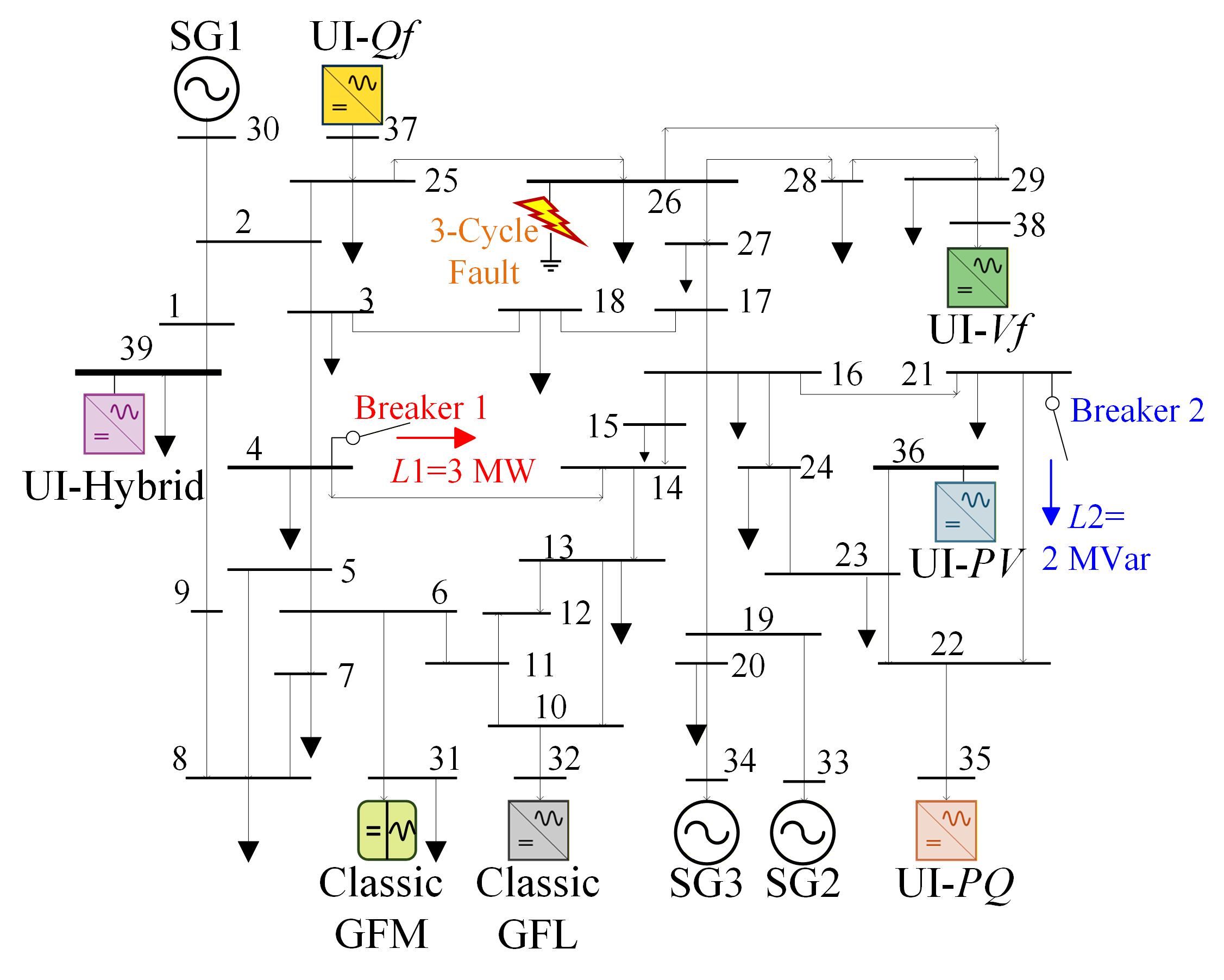}
    \caption{IEEE 39-bus system with proposed unified GFM and GFL control inverter (UI-$PQ$, UI-$PV$, UI-$Qf$, UI-$Vf$, and UI-Hybrid ), classic PI-based inverter (Classic GFM and GFL), and synchronous generator.}
    \label{fig:IEEE39UGFML}
\end{figure}

\begin{figure}[ht]
    \centering
    \includegraphics[width=1\linewidth]{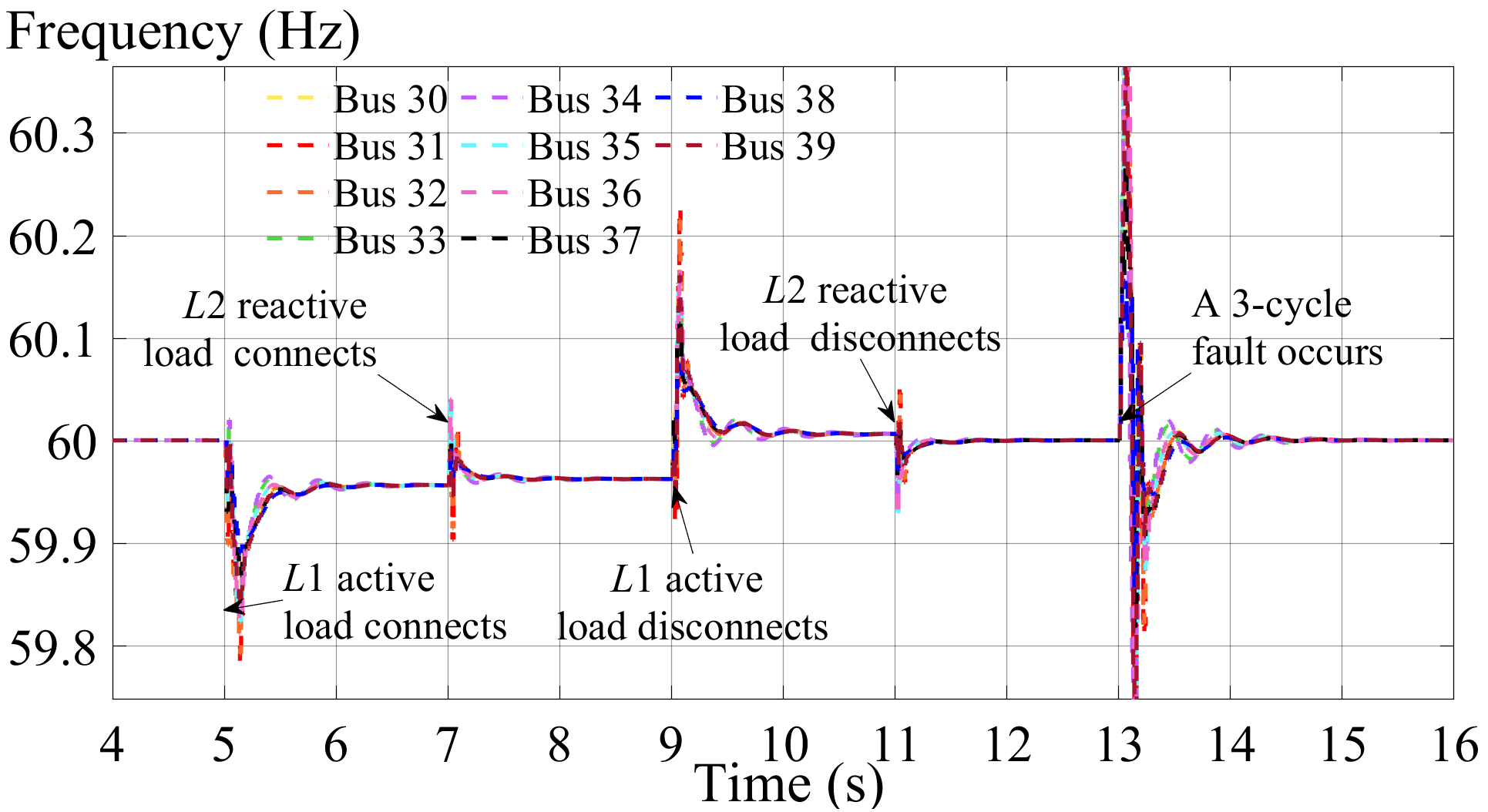}
    \caption{Dynamic frequency responses of inverter and generator buses under disturbance in IEEE 39-bus system test case.}
    \label{fig:FreIEEE39}
\end{figure}

In this subsection, the IEEE 39-bus system is employed to develop a high-fidelity EMT model to evaluate the performance of the proposed unified GFM/GFL inverter control strategy in MATLAB Simulink. The system configuration is illustrated in Figure~\ref{fig:IEEE39UGFML}. The network integrates the proposed unified inverter (UI) operating in different modes, together with conventional GFM and GFL inverters and synchronous generators (SGs). The base power, base voltage, and nominal frequency are set to 1~MVA, 22~kV, and 60~Hz, respectively. Transmission lines are modeled using $\pi$-equivalent circuits, and loads are represented by passive RLC impedances.

\begin{figure}[t]
    \centering
    \includegraphics[width=1\linewidth]{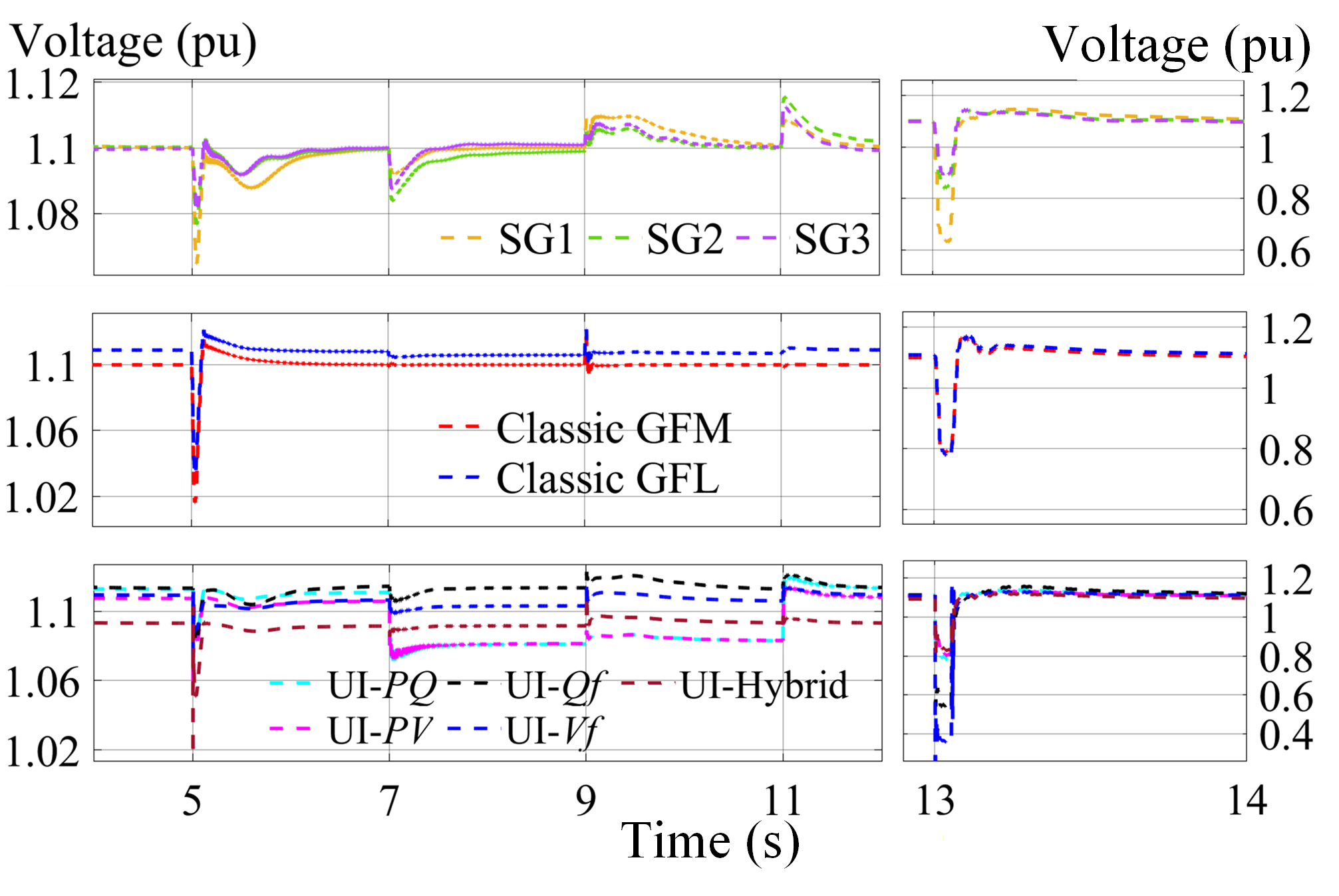}
    \caption{Voltage response of inverters and generators under disturbance in IEEE 39-bus system test case.}
    \label{fig:VolIEEE39}
\end{figure}

\begin{figure}[ht]
    \centering
    \includegraphics[width=1\linewidth]{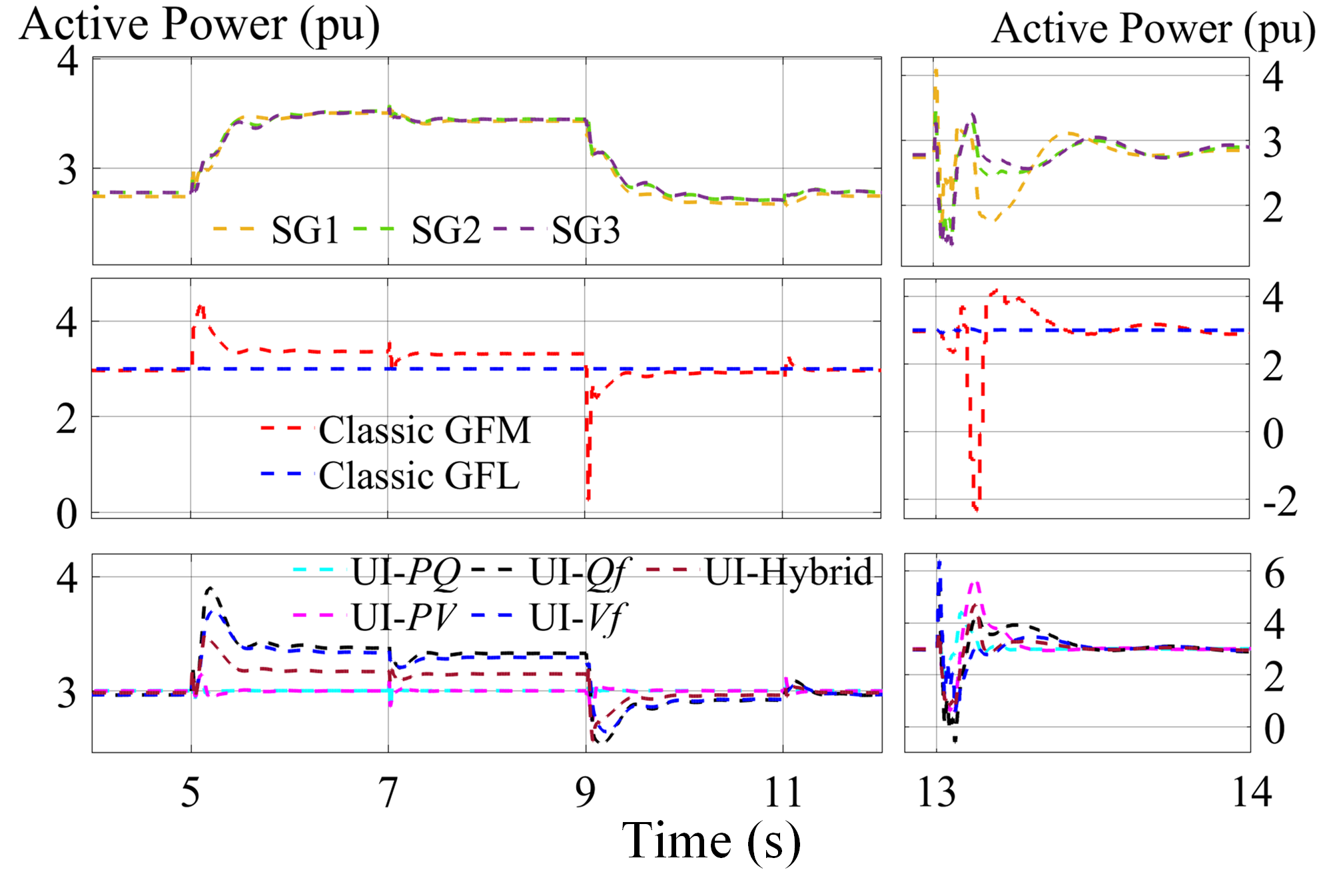}
    \caption{Active power response of inverters and generators under disturbance in IEEE 39-bus system test case. }
    \label{fig:PowerIEEE39}
\end{figure}

\begin{figure}[ht]
    \centering
    \includegraphics[width=1\linewidth]{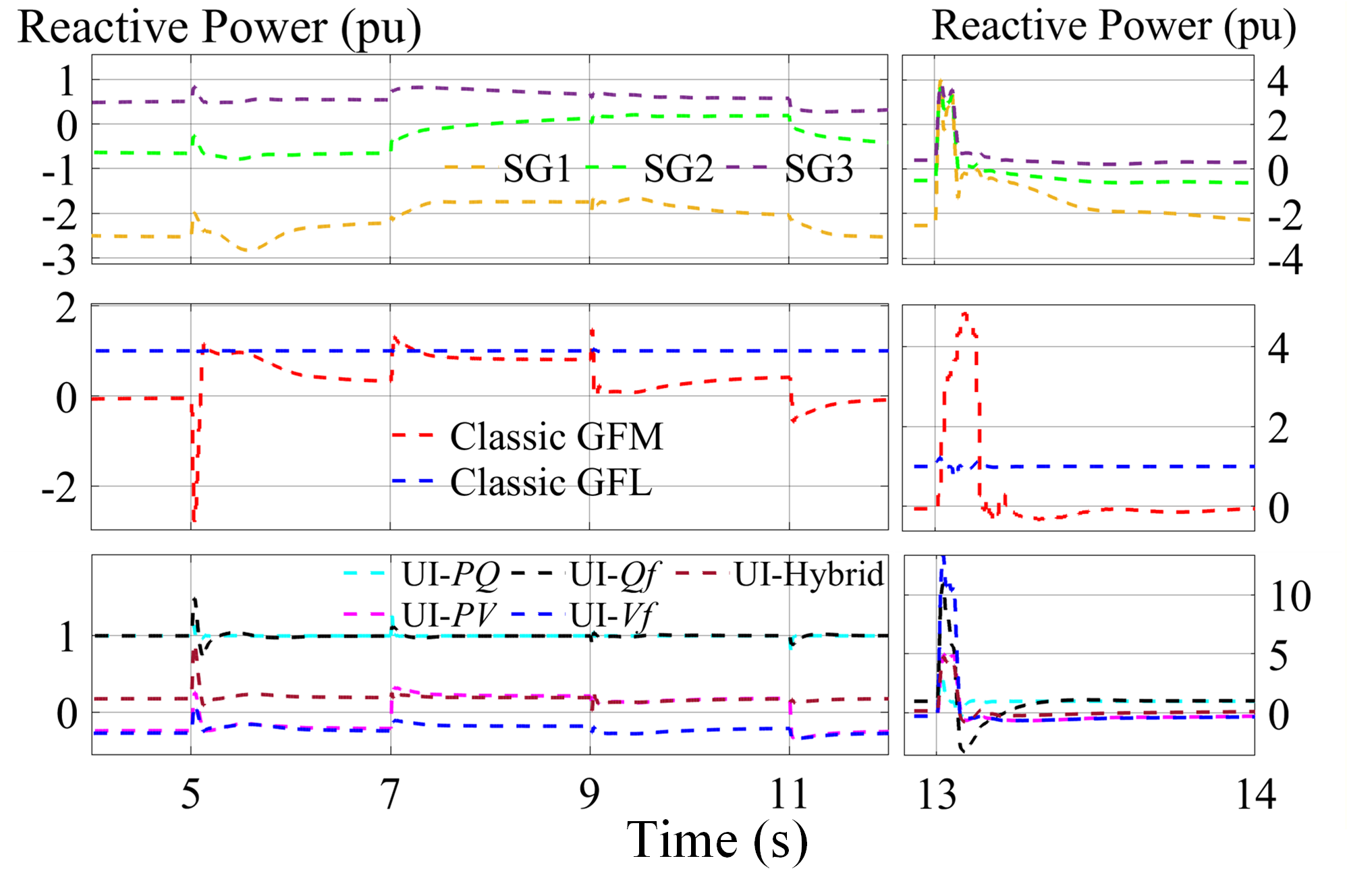}
    \caption{Reactive power response of inverters and generators under disturbance in IEEE 39-bus system test case.}
    \label{fig:QIEEE39}
\end{figure}

For the proposed unified GFM/GFL control inverters, the parameters are the same as those used in Section~\ref{subsection:EMTindividual} with $P_0=3$~pu and $Q_0=1$~pu. 
The conventional GFM inverter employs a PI-based voltage-current dual-loop controller with virtual inertia and $f$-$P$ droop control. The active power reference is $P_0=3$~pu, while the terminal voltage is regulated to a constant reference of $V_0=1.1$~pu. The virtual inertia constant is $H=0.2$~s.
The conventional GFL inverter adopts outer-loop $PQ$ control with $P_0=3$~pu and $Q_0=1$~pu, without primary frequency support. 
All synchronous generators are represented by full-order machine models. The governor provides $P$-$f$ droop control with a mechanical power reference of $P_0=3$~pu, while the excitation system regulates the terminal voltage to a constant reference value. The inertia constant of each SG is set to $H=1$~s.

Sudden load connection/disconnection events and a three-phase-to-ground fault are applied to evaluate the system dynamics. The corresponding frequency, voltage, active-power, and reactive-power responses are shown in Figure~\ref{fig:FreIEEE39} to \ref{fig:QIEEE39}.

Before $t=5$~s, the system operates at steady state with a total load of approximately 30~MW and a nominal frequency of 60~Hz. 
At $t=5$~s, an active load of $L_1=3$~pu is connected at Bus~4, causing the system frequency to decrease to approximately 59.95~Hz. The SGs, conventional GFM inverter, and unified-control inverters operating in $Qf$, $Vf$, and hybrid modes increase their active power outputs to support the frequency disturbance. In contrast, the conventional GFL inverter and the unified-control inverters operating in $PQ$ and $PV$ modes maintain nearly constant active power. The hybrid-mode inverter exhibits an intermediate response between the power-tracking and frequency-forming modes.

At $t=7$~s, a reactive load of $L_2=2$~pu is connected at Bus~21. The SGs, conventional GFM inverter, and unified-control inverters operating in $PV$, $Vf$, and hybrid modes respond to the increased reactive-power demand, whereas the conventional GFL inverter and the unified-control inverters operating in $PQ$ and $Qf$ modes maintain nearly constant reactive power. These results are consistent with the voltage-forming or voltage-following characteristics of each operating mode.

At $t=9$~s and $t=11$~s, $L_1$ and $L_2$ are disconnected, respectively. The frequency and voltage recover toward their pre-disturbance values, and the responses of the SGs and inverters follow the opposite trends of the corresponding load-connection events.

At $t=13$~s, a three-phase-to-ground fault is applied at Bus~26 for three cycles and then cleared. During the fault, bus voltages decrease significantly, reducing the active-power transfer capability of the network. The SGs, conventional GFM inverter, and unified-control inverters exhibit transient power reductions followed by post-fault oscillations. In contrast to the conventional GFL inverter, which maintains nearly constant active power through fast decoupled $dq$-frame $PQ$ control, the unified-control inverter operating in $PQ$ mode exhibits a noticeable power response during the fault.

Overall, the IEEE 39-bus EMT results demonstrate that the proposed unified control framework preserves the expected dynamic characteristics of different operating modes in a large-scale network. It also enables stable coexistence and coordinated dynamic responses among unified-control inverters, conventional inverters, and synchronous generators.

\section{Hardware-in-the-Loop Test Results}
\label{Section:HIL}

\begin{figure*}[ht]
\centering
\setlength{\tabcolsep}{1.5pt}
\renewcommand{\arraystretch}{1.0}

\begin{tabular}{@{}c c c c c@{}}
& {\scriptsize PQ} 
& {\scriptsize PV} 
& {\scriptsize Qf} 
& {\scriptsize Vf} \\

\adjustbox{valign=c}{\scriptsize (a) }
& \adjustbox{valign=c}{\includegraphics[width=0.24\textwidth]{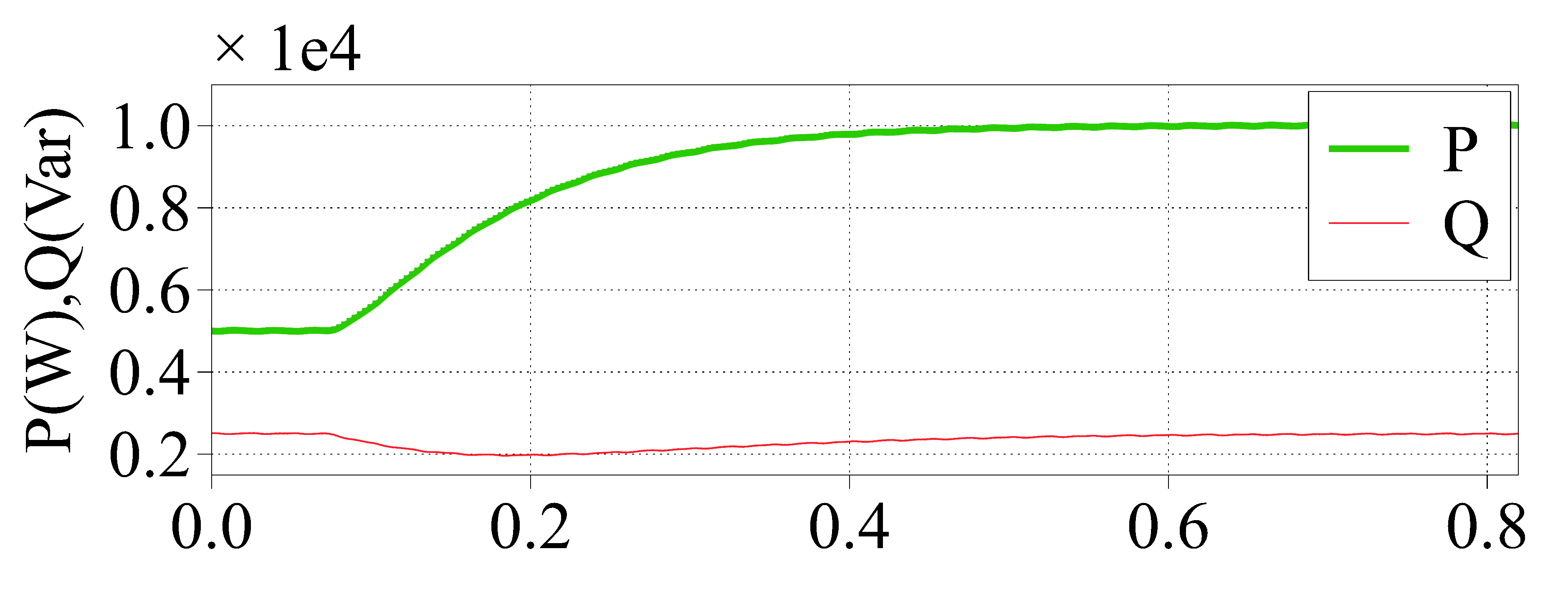}}
& \adjustbox{valign=c}{\includegraphics[width=0.24\textwidth]{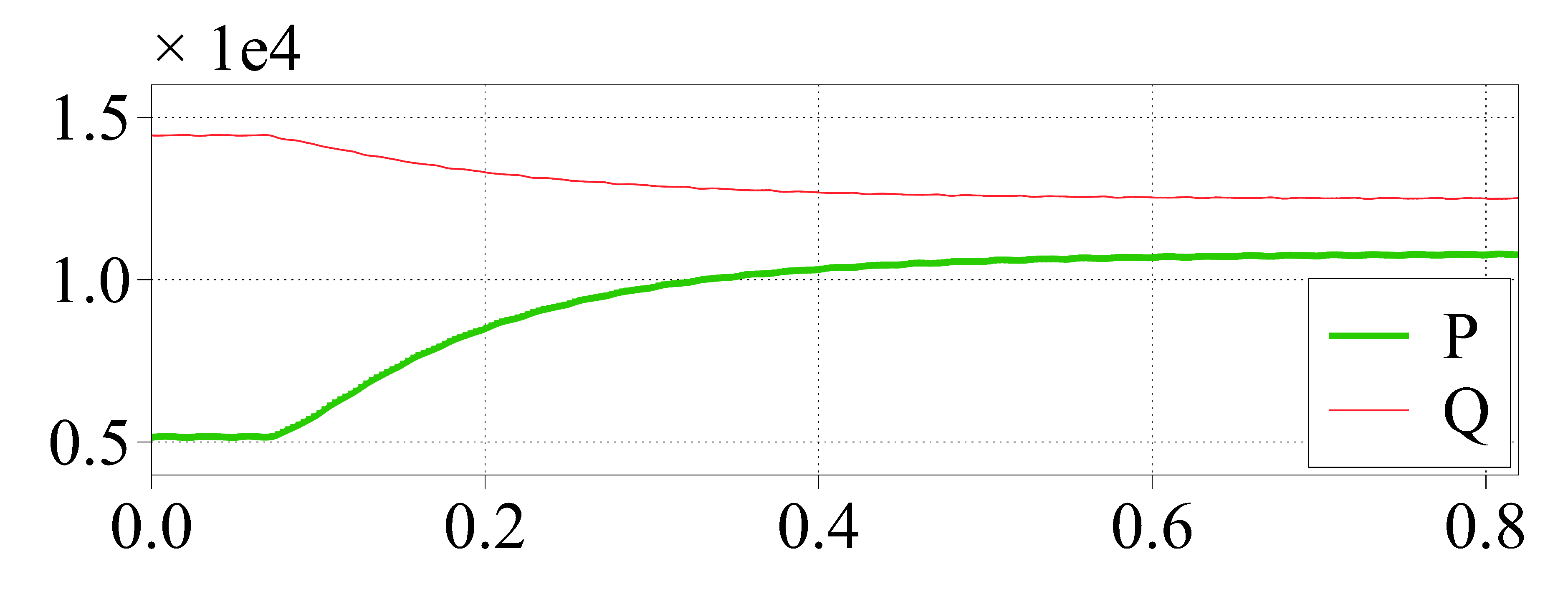}}
& \adjustbox{valign=c}{\includegraphics[width=0.24\textwidth]{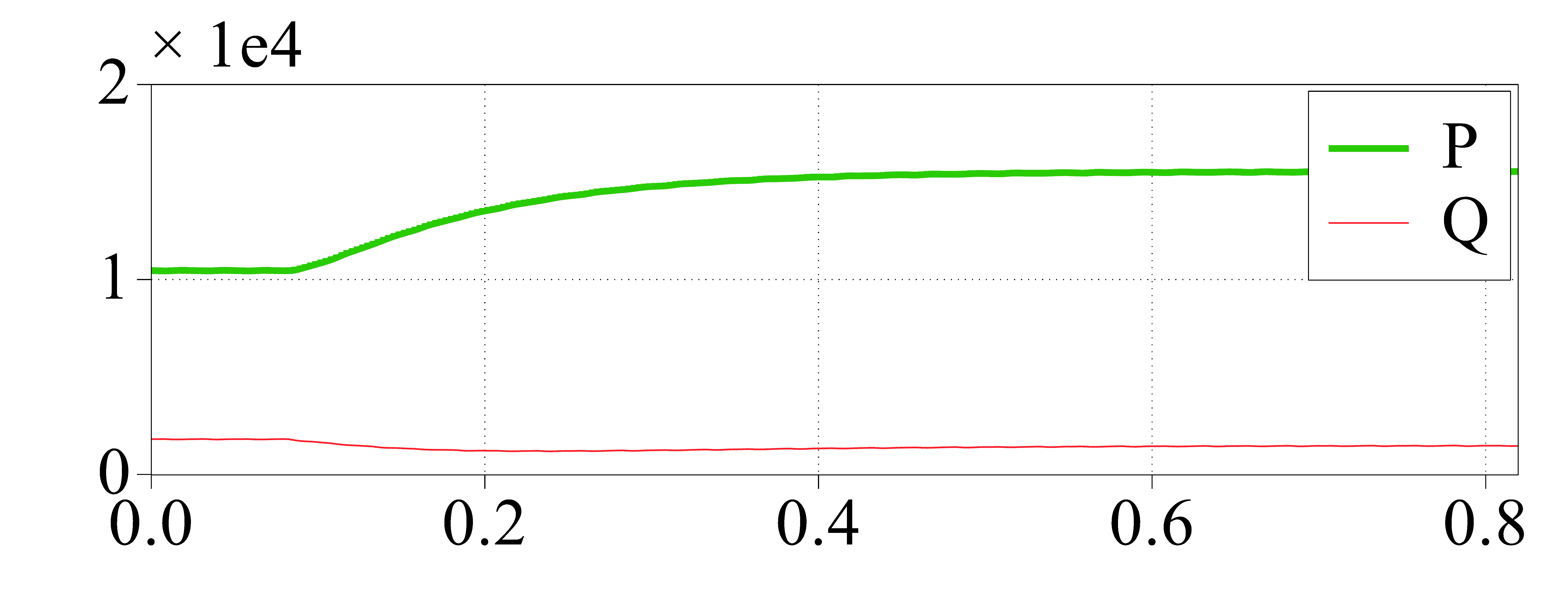}}
& \adjustbox{valign=c}{\includegraphics[width=0.24\textwidth]{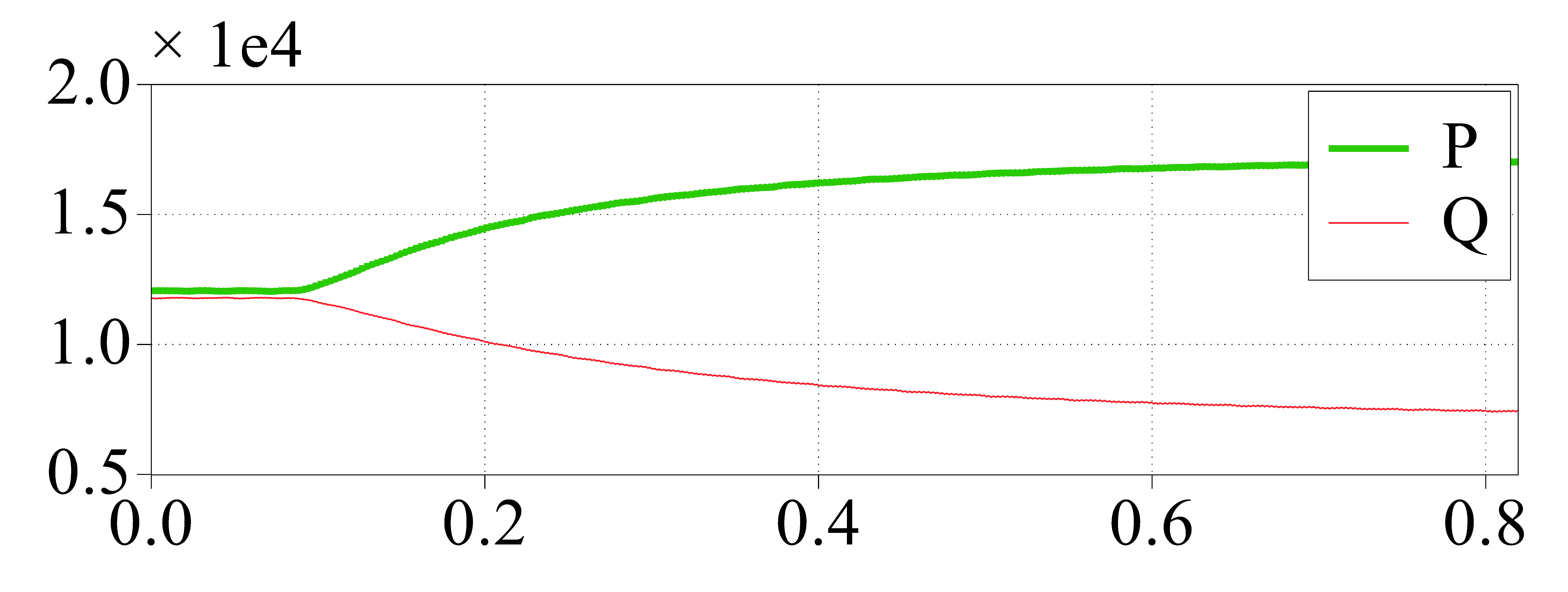}} \\

\adjustbox{valign=c}{\scriptsize (b)}
& \adjustbox{valign=c}{\includegraphics[width=0.24\textwidth]{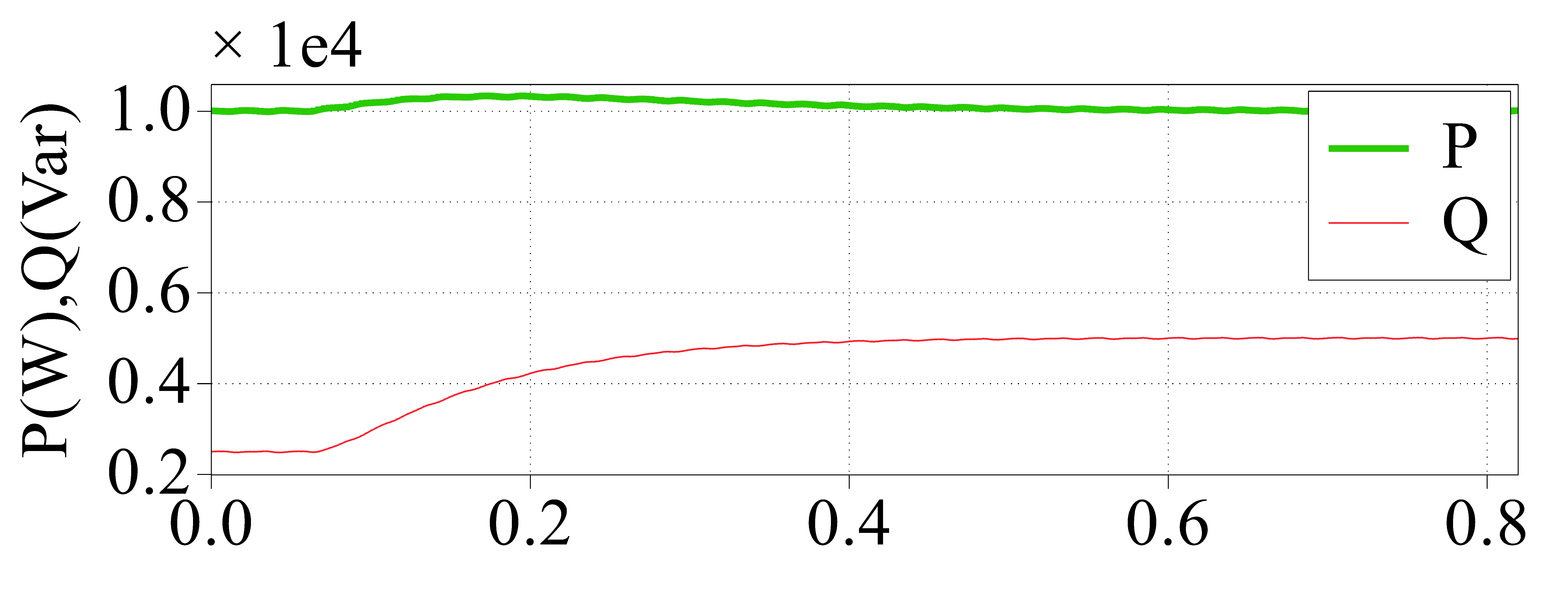}}
& \adjustbox{valign=c}{\includegraphics[width=0.24\textwidth]{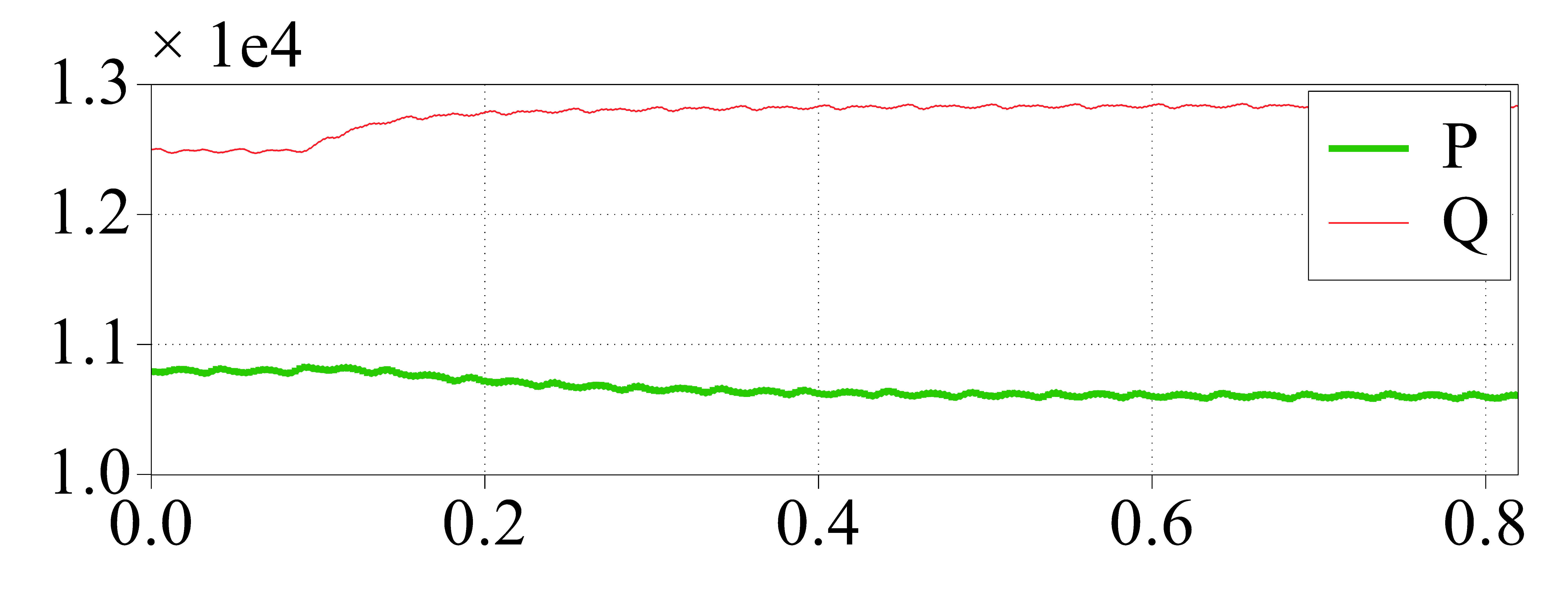}}
& \adjustbox{valign=c}{\includegraphics[width=0.24\textwidth]{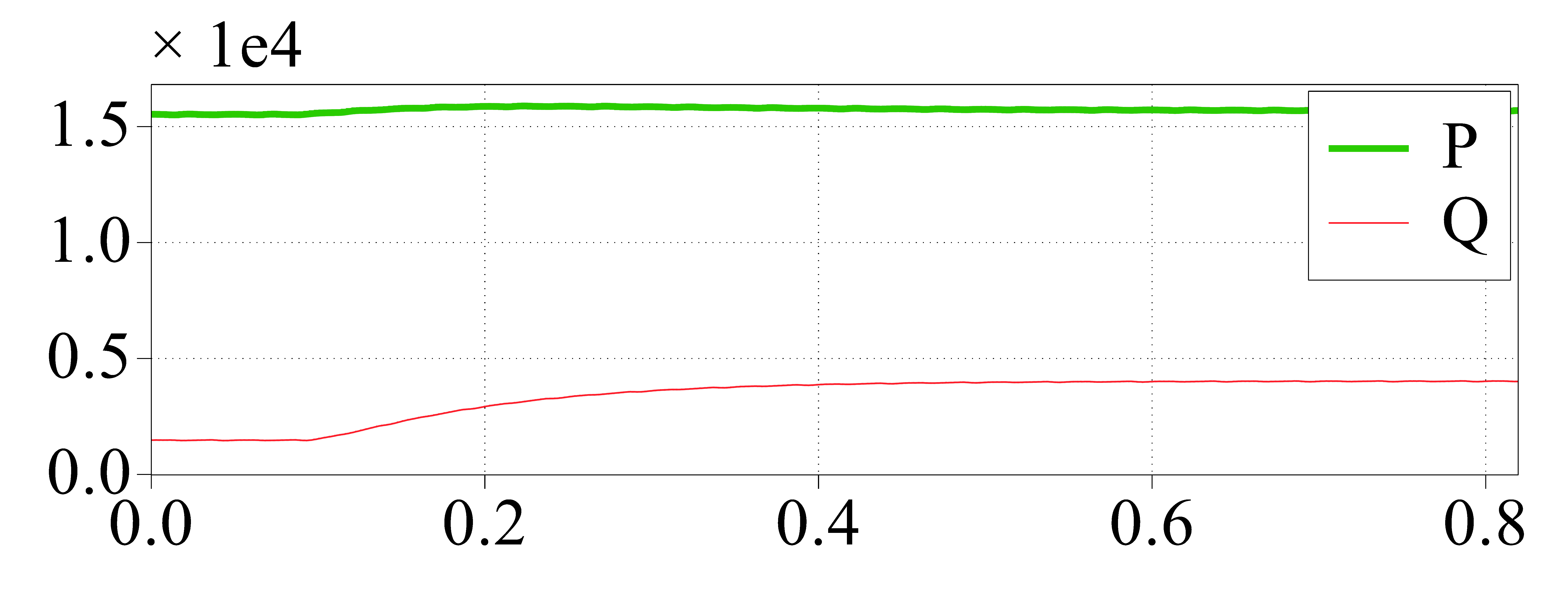}}
& \adjustbox{valign=c}{\includegraphics[width=0.24\textwidth]{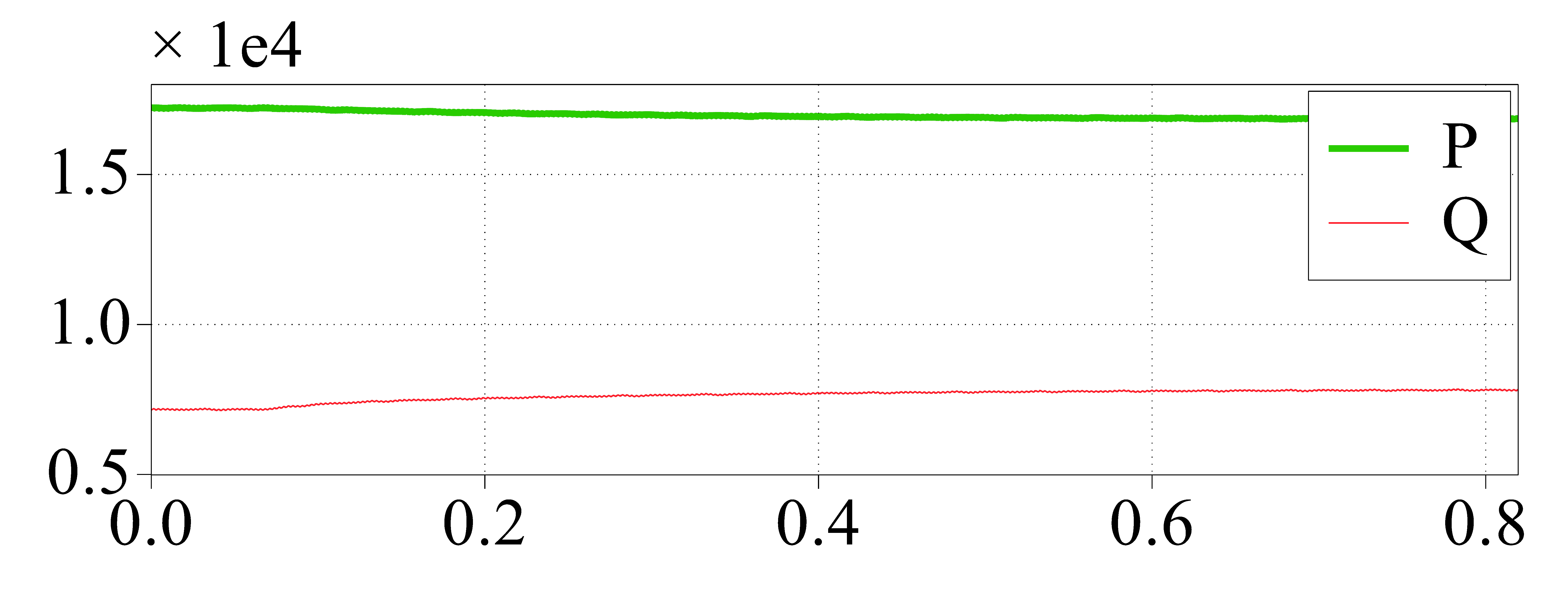}} \\

\adjustbox{valign=c}{\scriptsize (c)}
& \adjustbox{valign=c}{\includegraphics[width=0.24\textwidth]{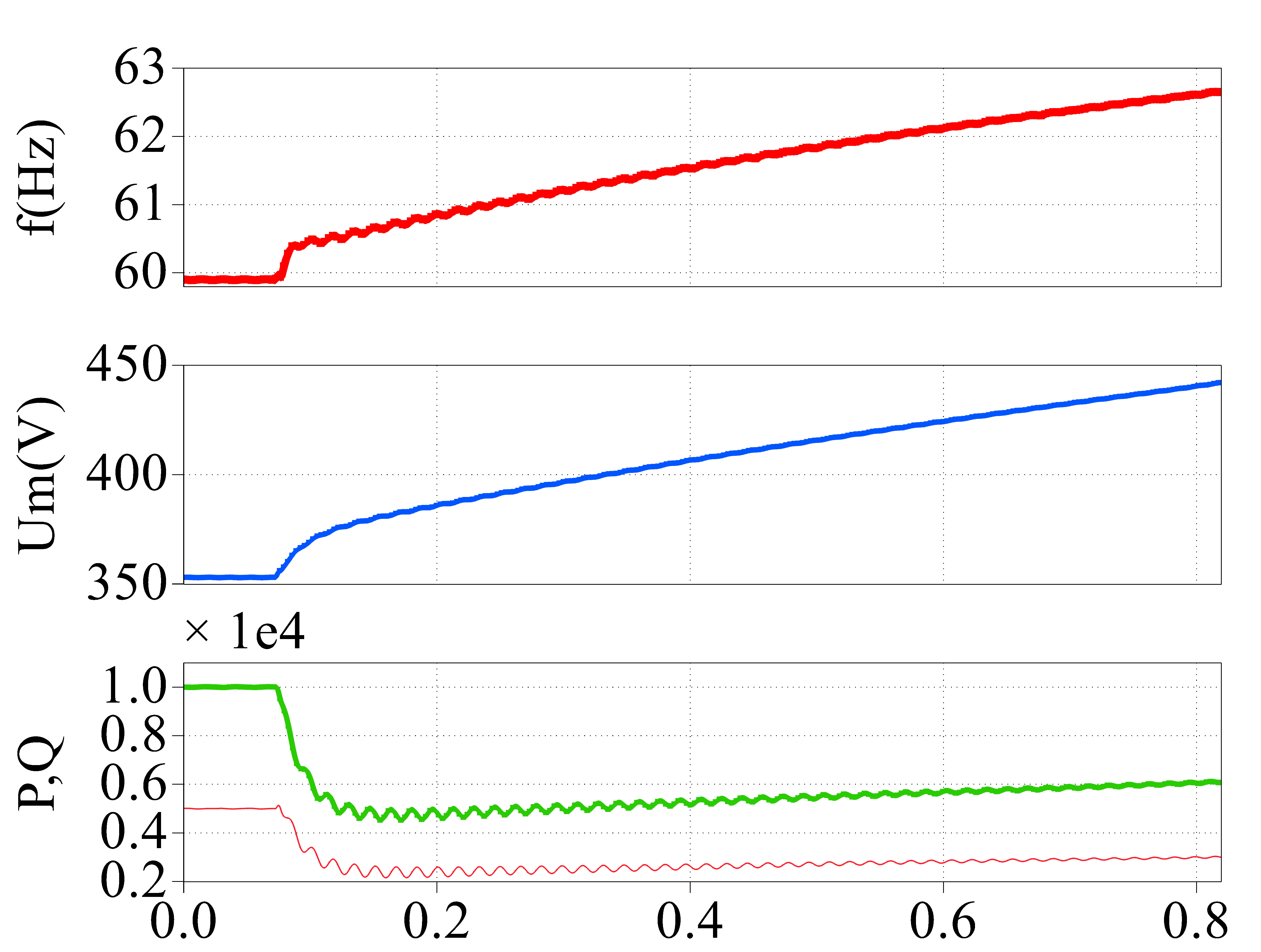}}
& \adjustbox{valign=c}{\includegraphics[width=0.24\textwidth]{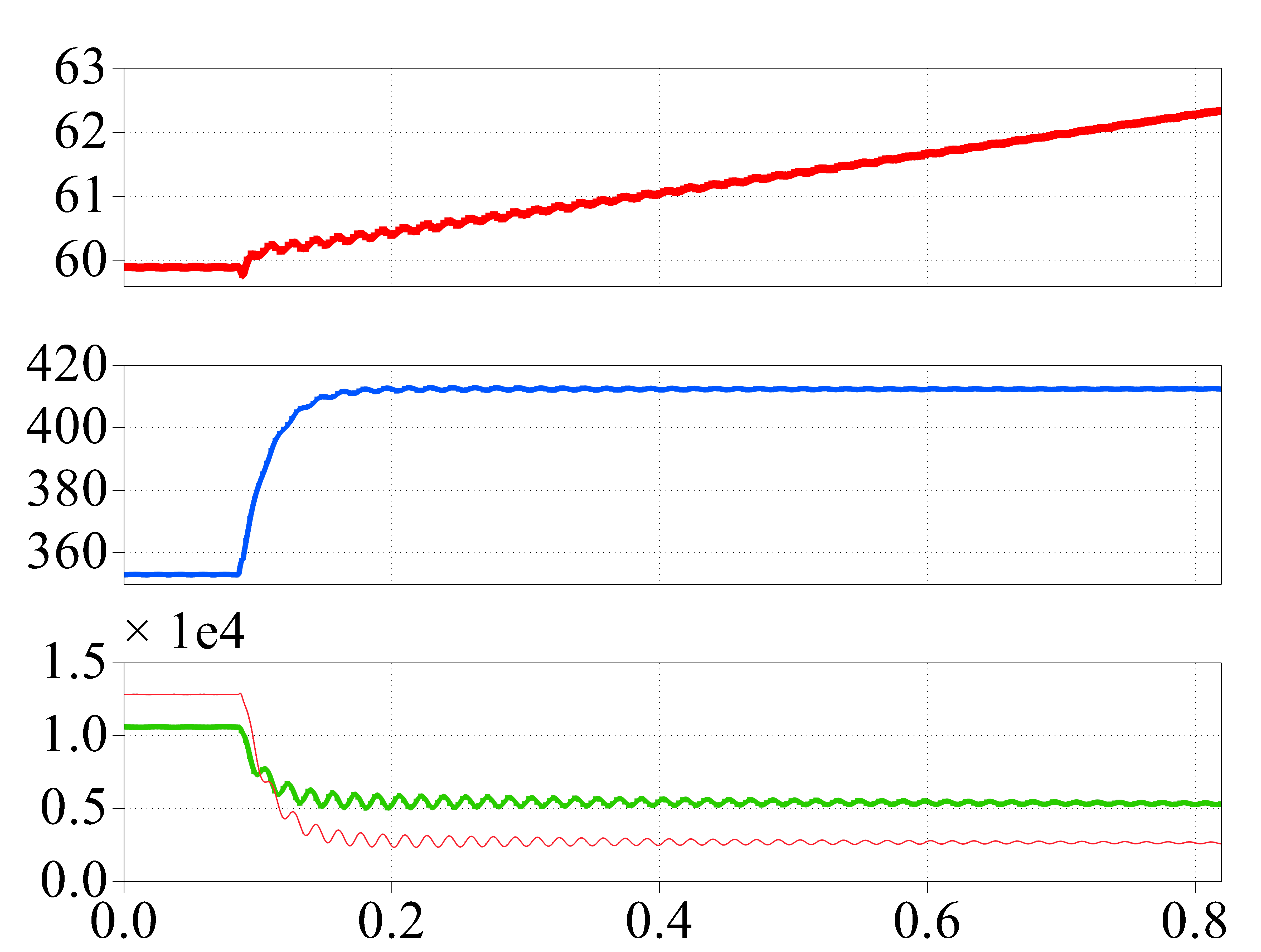}}
& \adjustbox{valign=c}{\includegraphics[width=0.24\textwidth]{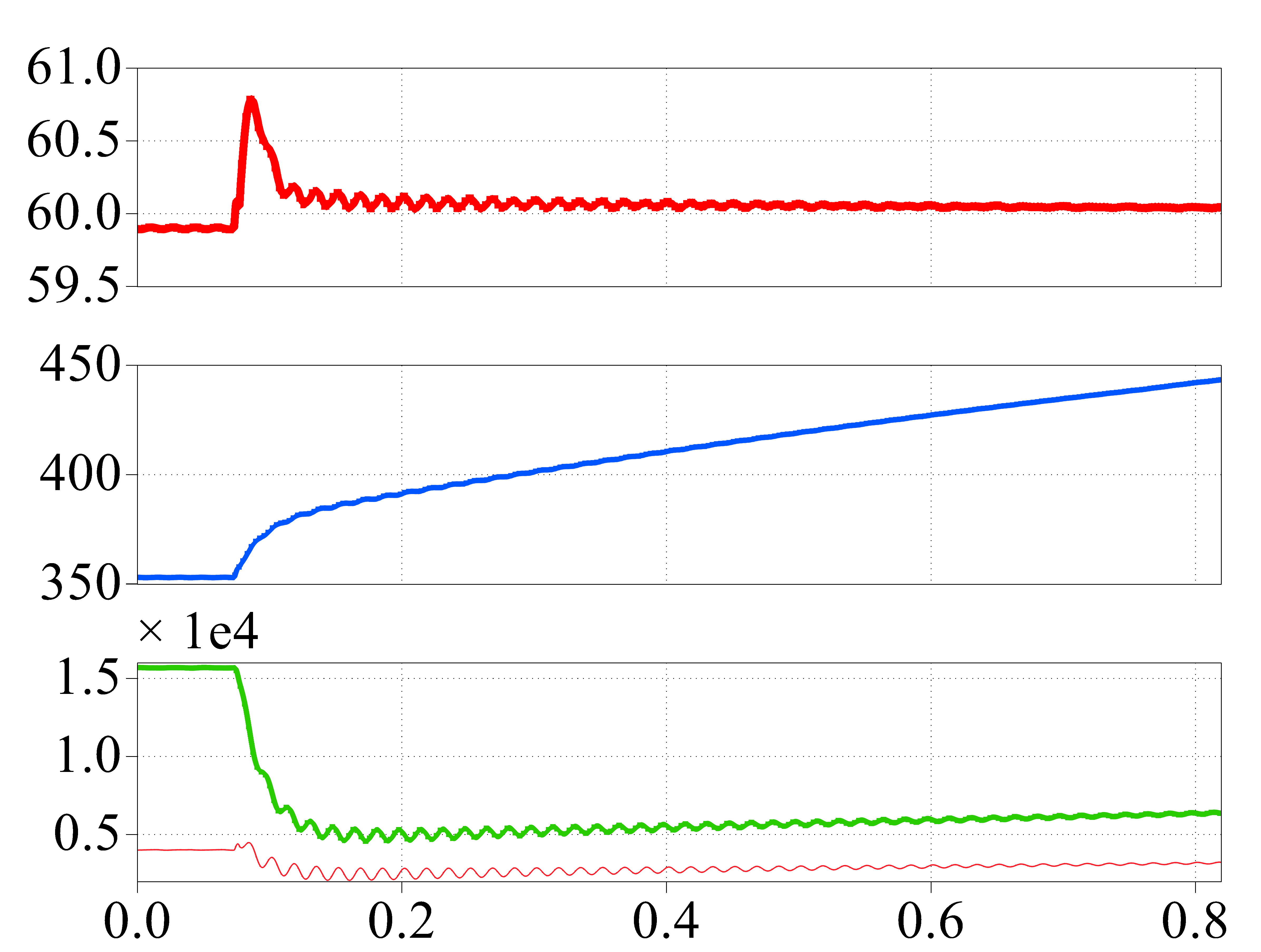}}
& \adjustbox{valign=c}{\includegraphics[width=0.24\textwidth]{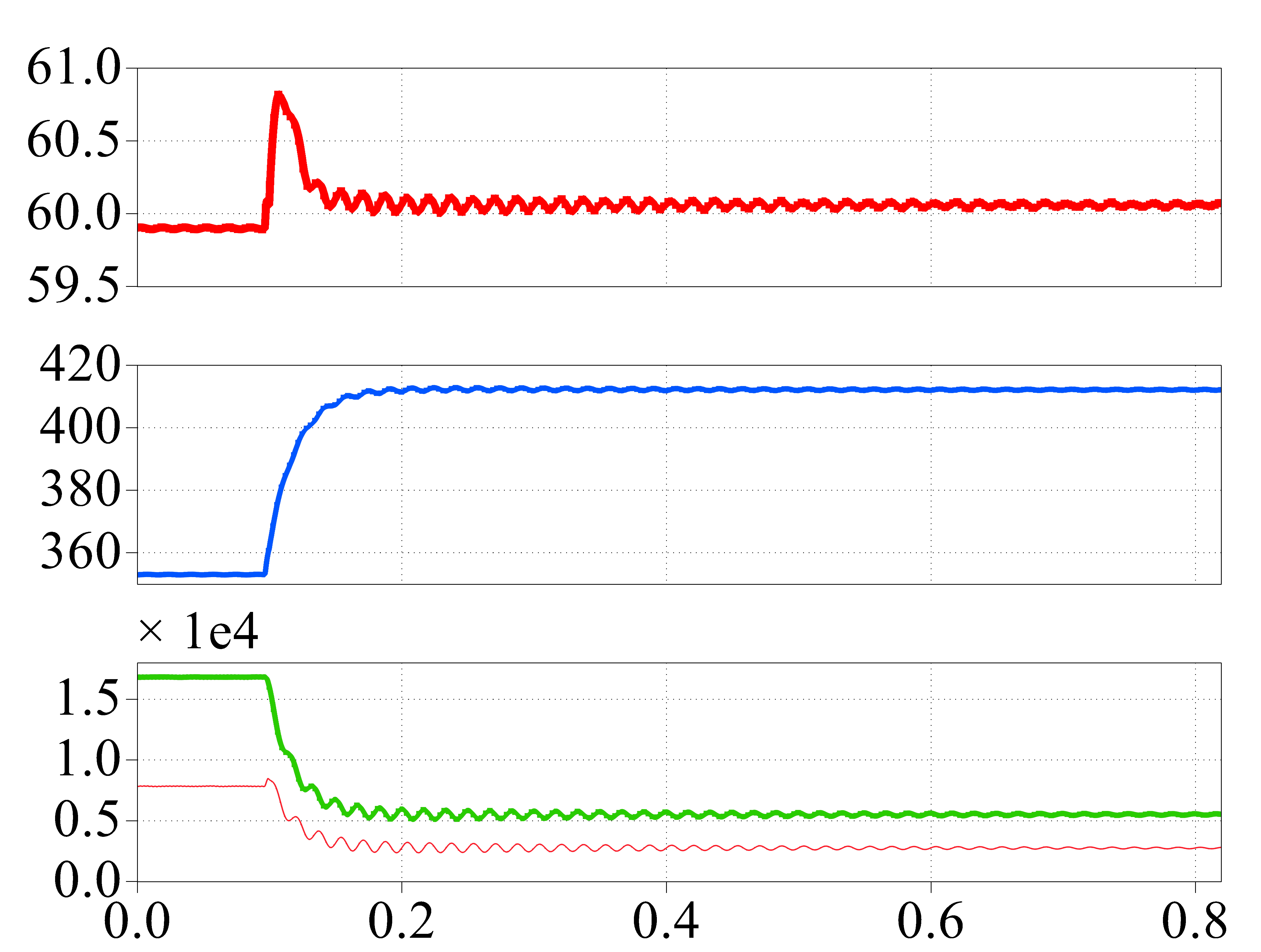}} \\
\end{tabular}

\caption{Dynamic responses under different events and operating modes in the hardware-in-the-loop tests. The columns correspond to the $PQ$, $PV$, $Qf$, and $Vf$ modes, respectively. The rows correspond to: (a) an increase in the active-power reference from 0.5 to 1.0~pu, (b) an increase in the reactive-power reference from 0.25 to 0.5~pu, and (c) grid disconnection followed by islanded operation. The $x$-axis represents time in seconds. The oscilloscope is triggered at $t=0$, when the $P$ or $Q$ reference command or the breaker-status command is issued. A 0.05~s delay is applied before command execution.}
\label{fig:event_mode_grid}
\end{figure*}

\begin{figure}[ht]
    \centering
    \includegraphics[width=1\linewidth]{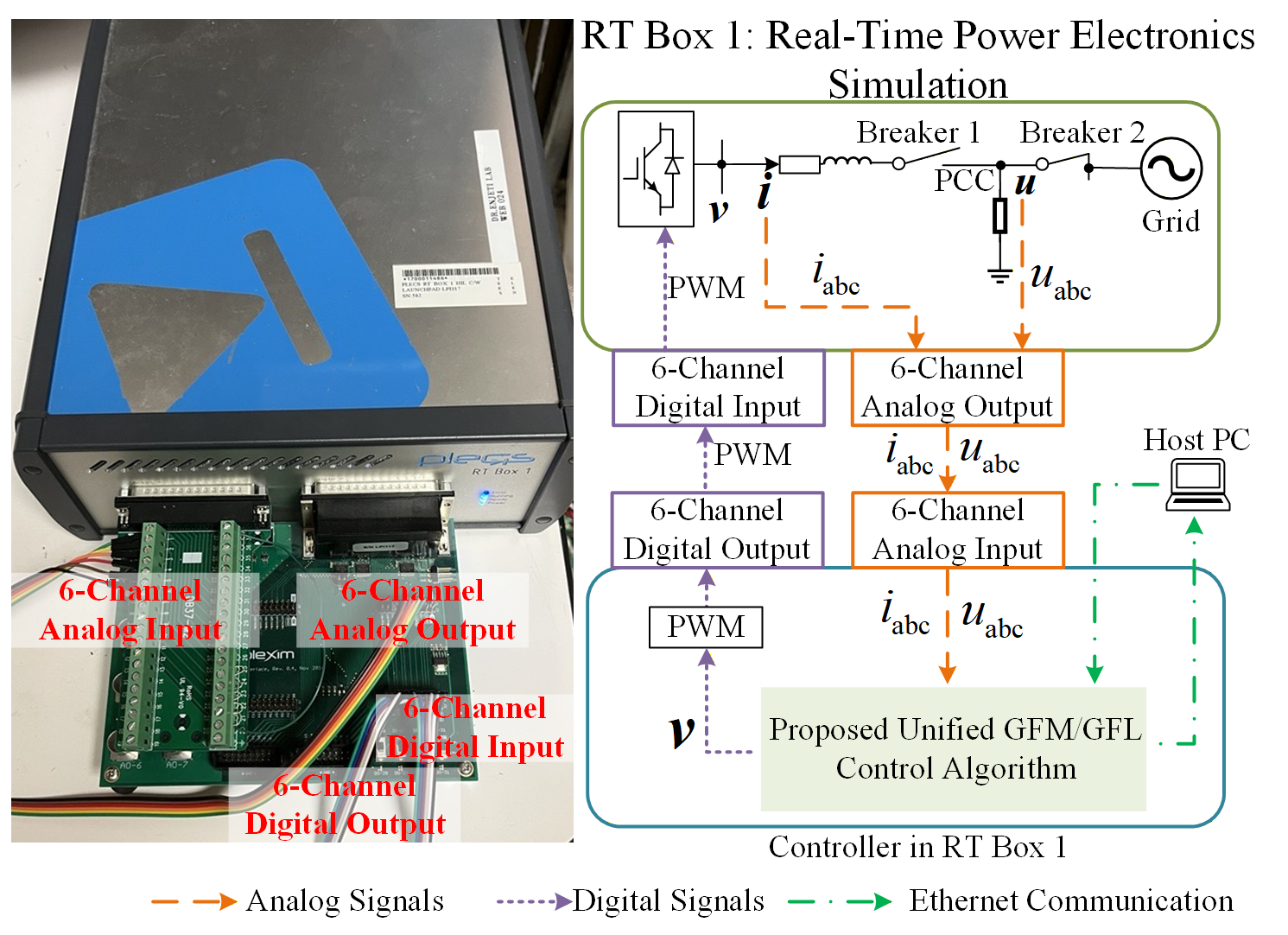}
    \caption{Hardware-in-the-loop implementation of the proposed unified GFM-GFL controller in the PLECS RT box.}
    \label{fig:HILRTbox}
\end{figure}

The effectiveness of the proposed unified GFM-GFL control framework is further verified through hardware-in-the-loop (HIL) tests. A single inverter connected to an infinite bus system in Figure~\ref{fig:SIMBsystem} is implemented on a PLECS RT Box~1 platform \cite{PLEXIM_RTBox}. The real-time inverter simulation, controller implementation, and signal interface are illustrated in Figure~\ref{fig:HILRTbox}. The HIL simulation model is also open-sourced in \cite{wang2026unifiedgfml}.
The main simulation time step is set to 10~$\mu$s, while the IGBT switching process is resolved using a 7.5~ns nano-step. Analog and digital input/output channels are used to establish a closed-loop interface between the simulated plant and controller. 
The controller generates 20~kHz PWM signals that are fed back to the RT Box for real-time switching-level simulation. A host PC communicates with the RT Box via Ethernet for parameter configuration, reference updates, and real-time data acquisition.
The inverter is rated at 10~kW and 480~V with an 800~V DC-link voltage. The grid voltage and frequency are set to 0.9~pu and 59.9~Hz, respectively, and the inverter output filter impedance is $0.03+\tj0.1$~pu. The active-power, reactive-power, voltage, and frequency references are initialized as $P_0=0.5$~pu, $Q_0=0.25$~pu, $V_0=1.1$~pu, and $f_0=60$~Hz.


To evaluate the controller performance, the following events are applied sequentially: (i) $P_0$ is increased from 0.5~pu to 1.0~pu; (ii) $Q_0$ is increased from 0.25~pu to 0.5~pu; and (iii) Breaker~1 is opened to disconnect the inverter from the grid and transition to islanded operation. The HIL test results are shown in Figure~\ref{fig:event_mode_grid}.

The HIL results are consistent with the EMT simulations. In $PQ$ mode, the inverter accurately tracks the active- and reactive-power references but loses both voltage and frequency regulation after islanding.


In $PV$ mode, the inverter regulates the voltage magnitude and tracks the active-power reference. A steady-state active-power tracking error is observed, consistent with the analytical results in \eqref{eq:PVhaterror} and \eqref{eq:PVhaterrorestimate}, resulting in the quasi-$PV$ behavior discussed in Appendix~\ref{Appendix:A-CquasiPV}. After islanding, the voltage remains close to 1.1~pu due to the voltage-forming capability, whereas the frequency collapses because frequency-forming capability is absent.


In $Qf$ mode, the inverter provides frequency-forming capability and tracks the reactive-power reference. After islanding, the frequency remains stable, while the voltage collapses due to the lack of voltage-forming capability.

Among all operating modes, only the $Vf$ mode maintains both voltage and frequency after islanding, confirming its simultaneous voltage-forming and frequency-forming capability. Overall, the HIL results closely match the theoretical analysis and EMT simulations, demonstrating that all operating modes can be realized within a unified controller structure without controller reconfiguration.

\section{Conclusion}\label{sec:conclusion}

This paper proposed a unified inverter control framework that enables hybrid GFM and GFL operation within a single controller structure. By combining self-sustained virtual oscillator control with reference-following synchronization, the proposed framework bridges the conventional GFM and GFL control paradigms. Through a small set of continuous control parameters, the controller can shape voltage- and frequency-forming/following behaviors and realize multiple operating modes, including $PQ$, $PV$, $Qf$, $Vf$, and hybrid modes, without discrete controller switching or structural reconfiguration.
The small-signal stability and input-output frequency-domain characteristics of the proposed controller were analyzed under different parameter settings. Electromagnetic transient simulations and hardware-in-the-loop experiments further validated its effectiveness under grid-connected, islanded, and multi-inverter operating conditions. The results demonstrated seamless mode transitions, stable operation across different modes, and continuous allocation of GFM and GFL capabilities among multiple interconnected inverters. More broadly, the proposed framework shows that GFM and GFL control can be interpreted as two synchronization extremes within a unified virtual-oscillator-based control structure, providing useful insights for the analysis, design, and coordination of inverter controls in future inverter-dominated power systems.

\appendices



\section{Unified GFM-GFL Inverter Control with Power Reference Defined after the Filter Inductor}
\label{Appendix:MultiModedefinedbehindinductor}

\begin{figure}[ht]
    \centering
    \includegraphics[width=0.7\linewidth]{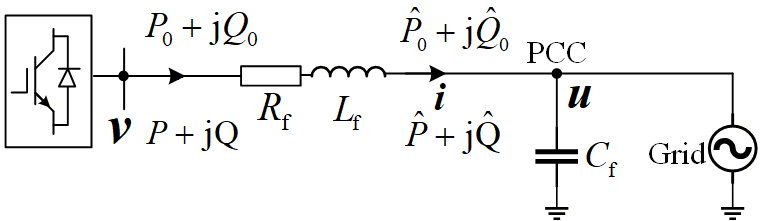}
    \caption{Reference power defined before and after the filter inductor.}
    \label{fig:TwopointsPQrefer}
\end{figure}

The active and reactive power references can be defined either at the inverter terminal (i.e., before the filter inductor) or at the PCC (i.e., after the filter inductor), as shown in Figure~\ref{fig:TwopointsPQrefer}. In contrast to Section \ref{Section:UnifiedController}, this appendix presents the proposed unified GFM-GFL control framework and the corresponding operating modes for power references $\hat{P}_0, \hat{Q}_0$ defined \emph{after} the filter inductor.

Given the reference power $\hat{P}_0+\tj \hat{Q}_0$ defined after the filter inductor, the reference current $\hat{\bi}_0$ is calculated using \eqref{eq:referencePQhat}: 
\begin{equation}
 \label{eq:referencePQhat}
    \hat{\bi}_0=\frac{2}{3}\frac{\hat{P}_0-\tj \hat{Q}_0}{\bar{\bu}}.
\end{equation}
Denote the current tracking error as $\hat{\be}_i=\hat{\bi}_0-{\bi}$. The unified GFM-GFL inverter control dynamics become \eqref{eq:hybridDynamics_ihati}:
\begin{align}\label{eq:hybridDynamics_ihati}
\dot{\bv}=\tj \omega_\epsilon \bv +\mu e_V \bv + \eta \hat{\be}_{\bi}e^{\tj \phi}.
\end{align}

\subsection{Polar Dynamics of Proposed Unified GFM-GFL Control}

As shown in Figure~\ref{fig:TwopointsPQrefer}, the PCC voltage vector $\bu$ can be represented by the inverter terminal voltage $\bv$ as \eqref{eq:u_v}:
\begin{align}\label{eq:u_v}
    \bu=\frac{||\bu||}{||\bv||} e^{-\tj \delta} \bv,
\end{align} 
where $\delta = \theta_v-\theta_u$ is the phase difference between $\bv$ and $\bu$.
Denote $\hat{\be}_S\coloneqq\hat{e}_P+\tj \hat{e}_Q=\frac{2}{3}(\hat{P}_0-\hat{P})+\tj \frac{2}{3}(\hat{Q}_0-\hat{Q})$. Then,
\begin{align}
    \hat{\be}_i=\frac{\bar{\hat{\be}}_S}{\bar{\bu}}=\frac{\bar{\hat{\be}}_S}{||\bu||\cdot||\bv||} e^{-\tj \delta} \bv.
\end{align}

The control dynamics \eqref{eq:hybridDynamics_ihati} can be rewritten as:
 \begin{align} \label{eq:dynamicvbye_hat}
     \dot{\bv}=&\tj \Big[\omega_\epsilon + \eta_2\frac{\hat{e}_P \sin (\phi-\delta) - \hat{e}_Q \cos (\phi-\delta) }{||\bu||\cdot||\bv||}\Big] \bv \nonumber\\ &+\Big[\mu e_V + \eta_1 \frac{\hat{e}_P \cos (\phi-\delta)  +\hat{e}_Q \sin (\phi-\delta) }{||\bu||\cdot||\bv||} \Big] \bv.
 \end{align}
Similar to \eqref{eq:hybridthetaV}, the polar dynamics of $\theta_v$ and $V_\tm$ are derived as \eqref{eq:hybridthetaV_hat}:
\begin{subequations} \label{eq:hybridthetaV_hat}
    \begin{align}
    \dot{V}_\tm=&\mu V_\tm e_V + \eta_1 \frac{ \hat{e}_P \cos (\phi-\delta) + \hat{e}_Q \sin (\phi-\delta)}{U_\tm},\\
        \dot{\theta}_v=&\omega=\omega_\epsilon + \eta_2\frac{\hat{e}_P \sin (\phi-\delta) - \hat{e}_Q \cos (\phi-\delta)}{V_\tm \cdot U_\tm}.
    \end{align}
\end{subequations}
When setting the rotation factor $\phi=\frac{\pi}{2}$, the polar dynamics \eqref{eq:hybridthetaV_hat} reduce to \eqref{eq:hybridthetaV_hatpi2}:
\begin{subequations} \label{eq:hybridthetaV_hatpi2}
    \begin{align}
    \dot{V}_\tm=&\mu V_\tm e_V + \eta_1 \frac{ \hat{e}_P \sin \delta + \hat{e}_Q \cos \delta}{U_\tm}, \label{eq:hybridthetaV_hatpi2:V}\\
        \dot{\theta}_v=&\omega=\omega_\epsilon + \eta_2\frac{\hat{e}_P \cos \delta - \hat{e}_Q \sin \delta}{V_\tm \cdot U_\tm}.  \label{eq:hybridthetaV_hatpi2:theta}
    \end{align}
\end{subequations}

Moreover, \eqref{eq:hybridthetaV_hatpi2} admits a projection-based physical interpretation for both the phase and magnitude dynamics. 
Specifically, the phase-channel term $\hat e_P \cos \delta - \hat e_Q \sin \delta$ corresponds to projecting the power error vector $(\hat e_P,\hat e_Q)$ onto the instantaneous power-angle sensitivity direction. 
Under the grid-side power-flow relationship
\eqref{eq:gridPQ}, the power-angle sensitivities satisfy
$[\partial \hat{P}/\partial \delta,\; \partial \hat{Q}/\partial \delta]^\top \propto [\cos\delta,\,-\sin\delta]^\top$, 
which directly explains the structure of the phase correction term in \eqref{eq:hybridthetaV_hatpi2:theta}. 
\begin{align} \label{eq:gridPQ}
\hat{P} = \frac{V_\tm U_\tm \sin \delta}{X_\tf}, \quad 
\hat{Q} = \frac{V_\tm U_\tm \cos \delta - U_\tm^2}{X_\tf}.
\end{align}
\begin{equation}\label{eq:gridJacobian}
\begin{bmatrix}
\partial \hat{P}/\partial \delta & \partial \hat{P}/\partial V_\tm \\
\partial \hat{Q}/\partial \delta & \partial \hat{Q}/\partial V_\tm
\end{bmatrix}
=
\frac{U_\tm}{X_\tf}
\begin{bmatrix}
V_\tm \cos\delta & \sin\delta \\
- V_\tm\sin\delta & \cos\delta
\end{bmatrix}.
\end{equation}

Similarly, the magnitude-channel term $\hat e_P \sin \delta + \hat e_Q \cos \delta$ can be interpreted as a projection of the same power error vector onto the power-voltage sensitivity direction. 
From \eqref{eq:gridPQ} and the associated Jacobian matrix in \eqref{eq:gridJacobian}, the voltage sensitivities satisfy
$[\partial \hat{P}/\partial V_m,\; \partial \hat{Q}/\partial V_m]^\top \propto [\sin\delta,\;\cos\delta]^\top$, 
thereby showing that the voltage-magnitude dynamics \eqref{eq:hybridthetaV_hatpi2:V} explicitly incorporate both $\partial \hat{P}/\partial V_m$ and $\partial \hat{Q}/\partial V_m$ through a structured projection of the P-Q mismatch.

In the following, we present the multiple operating modes of the unified GFM-GFL control when setting $\phi=\frac{\pi}{2}$.

\subsection{Frequency and Voltage Following: PQ Mode}

 By setting $\epsilon=0$ and $\mu=0$,
the unified GFM-GFL control dynamics \eqref{eq:hybridDynamics_ihati} become \eqref{eq:PQvdothat}:
\begin{align} \label{eq:PQvdothat}
      \dot{\bv}=&\tj  \omega_{u} \bv + \eta (\hat{\bi}_0 -\bi)e^{\tj \phi}.
\end{align}
The polar dynamics \eqref{eq:hybridthetaV_hatpi2} become \eqref{eq:PQhybridthetaV_hatpi2}:
\begin{subequations} \label{eq:PQhybridthetaV_hatpi2}
    \begin{align}
            \dot{V}_\tm=&\eta_1 \frac{ \hat{e}_P \sin \delta + \hat{e}_Q \cos \delta}{U_\tm},\\
        \dot{\delta}=& \eta_2\frac{\hat{e}_P \cos \delta - \hat{e}_Q \sin \delta}{V_\tm \cdot U_\tm}.
    \end{align}
\end{subequations}
Similarly, the steady-state solutions of \eqref{eq:PQhybridthetaV_hatpi2} satisfy:
    \begin{align}
        \hat{e}_P=\hat{P}_0-\hat{P}=0,\quad
        \hat{e}_Q=\hat{Q}_0-\hat{Q}=0.
    \end{align}
In this case, the $PQ$ mode yields the same steady-state behavior as the case where the power reference is defined at the inverter terminal, as given in \eqref{eq:PQmode_steadystate}. The difference is that the phase and magnitude adjustments are implemented through a $\delta$-rotated linear combination of $\hat{e}_P$ and $\hat{e}_Q$.

\subsection{Frequency Following \& Voltage Forming: Quasi-PV Mode}
\label{Appendix:A-CquasiPV}
Setting $\epsilon=0$ and $\mu>0$, the polar dynamics \eqref{eq:hybridthetaV_hatpi2} become:
\begin{subequations} \label{eq:PV_hatpi2}
    \begin{align}
            \dot{V}_\tm&=\mu V_\tm e_V + \eta_1 \frac{\hat{e}_P \sin \delta + \hat{e}_Q \cos \delta}{U_\tm},\\
        \dot{\delta}&=\eta_2\frac{\hat{e}_P \cos \delta - \hat{e}_Q \sin \delta}{V_\tm \cdot U_\tm}.
    \end{align}
\end{subequations}
The steady-state solutions of \eqref{eq:PV_hatpi2} satisfy:
\begin{subequations}
\label{eq:PVhaterror}
   \begin{align}
        \hat{e}_P &= -\frac{\mu V_\mathrm{m} U_\mathrm{m} e_V}{\eta_1} \sin \delta, \\
        \hat{e}_Q &= -\frac{\mu V_\mathrm{m} U_\mathrm{m} e_V}{\eta_1} \cos \delta.
   \end{align}
\end{subequations}
It yields $\|\hat{\mathbf{e}}_S\|=\frac{\mu V_\mathrm{m} U_\mathrm{m} e_V}{\eta_1}$, which
indicates an $S$–$V$ droop behavior; that is, the apparent power $S$ and the voltage magnitude $V_\mathrm{m}$ exhibit a droop relation.

When the phase angle $\delta$ between $\bv$ and $\bu$ is small, the steady-state solutions \eqref{eq:PVhaterror} lead to:
   \begin{align} \label{eq:PVhaterrorestimate}
    \hat{e}_P = \hat{P}_0-\hat{P} \approx 0,\quad 
    \hat{e}_Q = \hat{Q}_0-\hat{Q} \approx -\frac{\mu V_\mathrm{m} U_\mathrm{m} e_V}{\eta_1}.
\end{align} 
Since the active-power tracking error $\hat{e}_P$ cannot be completely eliminated, this mode is referred to as the quasi-$PV$ mode, while 
the inverter still retains frequency-following and voltage-forming capabilities. 

To achieve exact $PV$ operation, one solution is to set the rotation factor
$\phi = \frac{\pi}{2} + \delta$ to offset $\delta$. In this way, \eqref{eq:hybridthetaV_hat} 
yields the same steady-state solutions as \eqref{eq:PVsteadyPQV}, allowing 
the inverter to track $\hat{P} = \hat{P}_0$ exactly while the voltage magnitude 
is regulated through the nonlinear $V$-$Q$ droop.
The phase offset $\delta$ can be obtained either analytically from  \eqref{eq:gridPQ} as:
\begin{align} \label{eq:deltaana}
     \delta = \arcsin\!\left( \frac{\hat{P}_0 X_\tf}{V_\tm U_\tm} \right),
\end{align} 
or it can be continuously estimated from a PLL-based 
measurement of the phase difference between $\bv$ and $\bu$.

\subsection{Frequency Forming \& Voltage Following: Quasi-Qf Mode}

Similarly, there is no exact $Qf$-mode when the power reference is defined after the filter inductor. By setting $\epsilon=1$ and $\mu=0$, the polar dynamics \eqref{eq:hybridthetaV_hatpi2} become:
\begin{subequations} \label{eq:Qf_hatpi2}
    \begin{align}
       \dot{V}_\tm=&\eta_1 \frac{ \hat{e}_P \sin \delta + \hat{e}_Q \cos \delta}{U_\tm},\\
        \dot{\theta}_v=&\omega=\omega_0 + \eta_2\frac{\hat{e}_P \cos \delta - \hat{e}_Q \sin \delta}{V_\tm \cdot U_\tm}. 
    \end{align}
\end{subequations}
The steady state solutions of \eqref{eq:Qf_hatpi2} satisfy:
\begin{subequations}
       \begin{align}
        \hat{e}_P &= \frac{V_\mathrm{m} U_\mathrm{m} (\omega-\omega_0)}{\eta_2} \cos \delta, \\
        \hat{e}_Q &= -\frac{V_\mathrm{m} U_\mathrm{m} (\omega-\omega_0)}{\eta_2} \sin \delta,
       \end{align}
\end{subequations}
which leads to $\|\hat{\bm{e}}_S\|=\frac{V_\tm U_\mathrm{m} (\omega-\omega_0)}{\eta_2}$, indicating an $S$–$f$ droop behavior; that is, the apparent power $S$ and the frequency $f$ exhibit a droop relation. 

When the phase angle $\delta$ between $\bm{v}$ and $\bm{u}$ is small, the steady-state solutions \eqref{eq:QuasiQfsteadyState} lead to:
    \begin{align} \label{eq:QuasiQfsteadyState}
    \hat{e}_P= \hat{P}_0\!-\!\hat{P} \approx \frac{V_\mathrm{m} U_\mathrm{m} (\omega\!-\!\omega_0)}{\eta_2}, \ \ 
    \hat{e}_Q= \hat{Q}_0\!-\!\hat{Q} \approx 0.
\end{align}
Since the reactive power tracking error $\hat{e}_Q$ cannot be completely eliminated, this mode is referred to as the quasi-$Qf$ mode, while the inverter still has the capability of frequency forming and voltage following.

Similar to the quasi-$PV$ mode, to achieve exact $Qf$ operation, one can set $\phi = \frac{\pi}{2} + \delta$, where $\delta$ can be measured via a PLL or analytically computed given the inductor parameters.

\subsection{Frequency and Voltage Forming: Vf Mode}

By setting $\epsilon=1$, $\mu>0$, $\eta_1 \geq 0$, and $\eta_2 \geq 0$, the polar dynamics \eqref{eq:hybridthetaV_hatpi2} become:
\begin{subequations} \label{eq:Vfpolar_hatpi2}
    \begin{align}
            \dot{V}_\tm=&\mu V_\tm e_V + \eta_1 \frac{\hat{e}_P \sin \delta + \hat{e}_Q \cos \delta}{U_\tm}\\
        \dot{\theta}_v=&\omega=\omega_0 + \eta_2\frac{\hat{e}_P \cos \delta - \hat{e}_Q \sin \delta}{V_\tm \cdot U_\tm} 
    \end{align}
\end{subequations}
Similar to the case where the power reference is defined at the inverter terminal, i.e., before the inductor, the resulting behavior does not correspond to simple $f$-$P$ and $V$-$Q$ droop characteristics. Instead, the droop action is applied to a $\delta$-rotated linear combination of $P$ and $Q$. By setting $\phi=\pi/2+\delta$, the effect of the phase offset $\delta$ can be compensated, thereby recovering the standard $V$-$Q$ and $f$-$P$ droop relationships. More generally, by tuning $\phi$, different linear combinations of $P$ and $Q$ can be mapped to the voltage- and frequency-droop characteristics.

\section{Impedance-Shaping Operation Modes}
\label{Appendix:Impedance}

This appendix presents the implementation of impedance-mode control and virtual impedance control within the proposed unified GFM-GFL inverter control framework.

\subsection{Impedance-Mode Control}
\label{AppendixB_A:Impedance}

For the proposed unified GFM-GFL control \eqref{eq:Maindynamic},
the current feedback term $\eta(\bm{i}_0-\bm{i})$ can be utilized to regulate the inverter output impedance at the PCC to a prescribed value $R_0+\tj X_0$. This can be achieved by selecting the current reference as:
\begin{align}
\bm{i}_0=\frac{\bm{u}}{R_0+\tj X_0}.
\end{align}
Under this setting, the current-feedback loop drives the inverter terminal behavior toward the desired impedance characteristic. The rotation angle factor $\phi$ may further be used to adjust the coupling between resistance/reactance regulation and voltage magnitude/frequency regulation.

\subsection{Virtual Impedance Control}
\label{AppendixB_B:VirtualImpedance}
Virtual impedance control can be incorporated under any operating mode to improve power sharing accuracy, enhance damping performance, and increase stability margins under weak-grid conditions. Depending on where the impedance is introduced, two implementation approaches can be considered.

\textit{Inner virtual impedance control} is implemented directly in the modulation-voltage generation stage prior to PWM modulation. The modulation signal is modified as
\begin{align}
\bm{v}'=\bm{v}-Z_v\bm{i},
\end{align}
where $Z_v$ denotes the designed virtual impedance.

Alternatively, \textit{outer virtual impedance control} can be implemented through the outer synchronization or power-control loops by modifying the current reference according to
\begin{align}
\bm{i}_0'=\bm{i}_0-Y_v\bm{u},
\end{align}
where $Y_v$ denotes the designed virtual admittance. The modified reference $\bm{i}_0'$ is subsequently used in the current feedback term $\eta(\bm{i}_0'-\bm{i})$.

Both implementations can be integrated with any operating mode of the proposed unified GFM-GFL framework without modifying the underlying control structure.

\bibliographystyle{IEEEtran}
\bibliography{IEEEabrv,mybibfile}

\end{document}